\documentclass[a4paper,twocolumn,11pt,accepted=2023-02-27]{quantumarticle}
\pdfoutput=1
\usepackage[utf8]{inputenc}
\usepackage[english]{babel}
\usepackage[T1]{fontenc}
\usepackage{bm}% bold math
\usepackage{hyperref}% add hypertext capabilities
\usepackage[shortlabels]{enumitem}

\usepackage{amsfonts}
\usepackage{braket}
\usepackage{mathtools}
\usepackage{amsthm}
\usepackage{ulem}
\usepackage{bbold}
\usepackage{subcaption}
\usepackage{stmaryrd}
\usepackage[table]{xcolor}
\usepackage{amsmath}

\usepackage{appendix}
\usepackage{tikzit}
\usepackage{tikz, circuitikz}
\usepackage{lipsum}
\tikzstyle{env}=[copoint,regular polygon rotate=0,minimum width=0.2cm, fill=black]

\tikzstyle{probs}=[shape=semicircle,fill=white,draw=black,shape border rotate=180,minimum width=1.2cm]

\tikzstyle{nudge}=[yshift=0.6mm]

\tikzstyle{every picture}=[baseline=-0.25em,scale=0.5]
\tikzstyle{dotpic}=[] % for backwards-compatibility
\tikzstyle{diredges}=[every to/.style={diredge}]
\tikzstyle{math matrix}=[matrix of math nodes,left delimiter=(,right delimiter=),inner sep=2pt,column sep=1em,row sep=0.5em,nodes={inner sep=0pt},text height=1.5ex, text depth=0.25ex]

\tikzstyle{inline text}=[text height=1.5ex, text depth=0.25ex,yshift=0.5mm]
\tikzstyle{label}=[font=\footnotesize,text height=1.5ex, text depth=0.25ex]
\tikzstyle{left label}=[label,anchor=east,xshift=2mm]
\tikzstyle{right label}=[label,anchor=west,xshift=-2mm]

\tikzstyle{braceedge}=[decorate,decoration={brace,amplitude=2mm,raise=-1mm}]
\tikzstyle{small braceedge}=[decorate,decoration={brace,amplitude=1mm,raise=-1mm}]

\tikzstyle{doubled}=[line width=1.6pt] % set the line width for all doubled (quantum) maps/wires
\tikzstyle{boldedge}=[doubled,shorten <=-0.17mm,shorten >=-0.17mm]
\tikzstyle{boldedgegray}=[doubled,gray,shorten <=-0.17mm,shorten >=-0.17mm]
\tikzstyle{singleedgegray}=[gray]%,shorten <=-0.1mm,shorten >=-0.1mm]

\tikzstyle{semidoubled}=[line width=1.4pt] % set the line width for all doubled (quantum) maps/wires
\tikzstyle{semiboldedgegray}=[semidoubled,gray,shorten <=-0.17mm,shorten >=-0.17mm]

\tikzstyle{boxedge}=[semiboldedgegray]

\tikzstyle{boldedgedashed}=[very thick,dashed,shorten <=-0.17mm,shorten >=-0.17mm]
\tikzstyle{vboldedgedashed}=[doubled,dashed,shorten <=-0.17mm,shorten >=-0.17mm]
\tikzstyle{left hook arrow}=[left hook-latex]
\tikzstyle{right hook arrow}=[right hook-latex]
\tikzstyle{sembracket}=[line width=0.5pt,shorten <=-0.07mm,shorten >=-0.07mm]

\tikzstyle{causal edge}=[->,thick,gray]
\tikzstyle{causal nondir}=[thick,gray]
\tikzstyle{timeline}=[thick,gray, dashed]

\tikzstyle{cedge}=[<->,thick,gray!70!white]

\tikzstyle{empty diagram}=[draw=gray!40!white,dashed,shape=rectangle,minimum width=1cm,minimum height=1cm]
\tikzstyle{empty diagram small}=[draw=gray!50!white,dashed,shape=rectangle,minimum width=0.6cm,minimum height=0.5cm]

\tikzstyle{dot}=[inner sep=0mm,minimum width=2mm,minimum height=2mm,draw,shape=circle]  
\tikzstyle{Wsquare}=[white dot, shape=regular polygon, rounded corners=0.8 mm, minimum size=3.3 mm, regular polygon sides=3, outer sep=-0.2mm]
\tikzstyle{Wsquareadj}=[white dot, shape=regular polygon, rounded corners=0.8 mm, minimum size=3.3 mm, regular polygon sides=3, outer sep=-0.2mm, regular polygon rotate=180]
\tikzstyle{ddot}=[inner sep=0mm, doubled, minimum width=2.5mm,minimum height=2.5mm,draw,shape=circle]

\tikzstyle{black dot}=[dot,fill=black]
\tikzstyle{white dot}=[dot,fill=white,,text depth=-0.2mm]
\tikzstyle{white Wsquare}=[Wsquare,fill=white,,text depth=-0.2mm]
\tikzstyle{white Wsquareadj}=[Wsquareadj,fill=white,,text depth=-0.2mm]
\tikzstyle{green dot}=[white dot] % for backwards-compatibility
\tikzstyle{gray dot}=[dot,fill=gray!40!white,,text depth=-0.2mm]
\tikzstyle{red dot}=[gray dot] % for backwards-compatibility

\tikzstyle{black ddot}=[ddot,fill=black]
\tikzstyle{white ddot}=[ddot,fill=white]
\tikzstyle{gray ddot}=[ddot,fill=gray!40!white]

\tikzstyle{gray edge}=[gray!60!white]

\tikzstyle{small dot}=[inner sep=0.5mm,minimum width=0pt,minimum height=0pt,draw,shape=circle]

\tikzstyle{small black dot}=[small dot,fill=black]
\tikzstyle{small white dot}=[small dot,fill=white]
\tikzstyle{small gray dot}=[small dot,fill=gray!40!white]

\tikzstyle{causal dot}=[inner sep=0.4mm,minimum width=0pt,minimum height=0pt,draw=white,shape=circle,fill=gray!40!white]

\tikzstyle{phase dimensions}=[minimum size=5mm,font=\footnotesize,rectangle,rounded corners=2.5mm,inner sep=0.2mm,outer sep=-2mm]
\tikzstyle{dphase dimensions}=[minimum size=5mm,font=\footnotesize,rectangle,rounded corners=2.5mm,inner sep=0.2mm,outer sep=-2mm]
\tikzstyle{white phase dot}=[dot,fill=white,phase dimensions]
\tikzstyle{white phase ddot}=[ddot,fill=white,dphase dimensions]

\tikzstyle{white rect ddot}=[draw=black,fill=white,doubled,minimum size=5mm,font=\footnotesize,rectangle,rounded corners=2.5mm,inner sep=0.2mm]
\tikzstyle{gray rect ddot}=[draw=black,fill=gray!40!white,doubled,minimum size=6mm,font=\footnotesize,rectangle,rounded corners=3mm]

\tikzstyle{gray phase dot}=[dot,fill=gray!40!white,phase dimensions]
\tikzstyle{gray phase ddot}=[ddot,fill=gray!40!white,dphase dimensions]
\tikzstyle{grey phase dot}=[gray phase dot]
\tikzstyle{grey phase ddot}=[gray phase ddot]

\tikzstyle{small phase dimensions}=[minimum size=4mm,font=\tiny,rectangle,rounded corners=2mm,inner sep=0.2mm,outer sep=-2mm]
\tikzstyle{small dphase dimensions}=[minimum size=4mm,font=\tiny,rectangle,rounded corners=2mm,inner sep=0.2mm,outer sep=-2mm]

\tikzstyle{small gray phase dot}=[dot,fill=gray!40!white,small phase dimensions]
\tikzstyle{small gray phase ddot}=[ddot,fill=gray!40!white,small dphase dimensions]

\tikzstyle{small map}=[draw,shape=rectangle,minimum height=4mm,minimum width=4mm,fill=white]

\tikzstyle{cnot}=[fill=white,shape=circle,inner sep=-1.4pt]

\tikzstyle{asym hadamard}=[fill=white,draw,shape=NEbox,inner sep=0.6mm,font=\footnotesize,minimum height=4mm]
\tikzstyle{asym hadamard conj}=[fill=white,draw,shape=NWbox,inner sep=0.6mm,font=\footnotesize,minimum height=4mm]
\tikzstyle{asym hadamard dag}=[fill=white,draw,shape=SEbox,inner sep=0.6mm,font=\footnotesize,minimum height=4mm]

\tikzstyle{hadamard}=[fill=white,draw,inner sep=0.6mm,font=\footnotesize,minimum height=4mm,minimum width=4mm]
\tikzstyle{small hadamard}=[fill=white,draw,inner sep=0.6mm,minimum height=1.5mm,minimum width=1.5mm]
\tikzstyle{small hadamard rotate}=[small hadamard,rotate=45]
\tikzstyle{dhadamard}=[hadamard,doubled]
\tikzstyle{small dhadamard}=[small hadamard,doubled]
\tikzstyle{small dhadamard rotate}=[small hadamard rotate,doubled]
\tikzstyle{antipode}=[white dot,inner sep=0.3mm,font=\footnotesize]

\tikzstyle{scalar}=[diamond,draw,inner sep=0.5pt,font=\small]
\tikzstyle{dscalar}=[diamond,doubled, draw,inner sep=0.5pt,font=\small]

\tikzstyle{small box}=[rectangle,inline text,fill=white,draw,minimum height=5mm,yshift=-0.5mm,minimum width=5mm,font=\small]
\tikzstyle{small gray box}=[small box,fill=gray!30]
\tikzstyle{medium box}=[rectangle,inline text,fill=white,draw,minimum height=5mm,yshift=-0.5mm,minimum width=8mm,font=\small]
\tikzstyle{square box}=[small box] % for backwards-compatibility
\tikzstyle{medium gray box}=[small box,fill=gray!30]
\tikzstyle{semilarge box}=[rectangle,inline text,fill=white,draw,minimum height=5mm,yshift=-0.5mm,minimum width=12.5mm,font=\small]
\tikzstyle{large box}=[rectangle,inline text,fill=white,draw,minimum height=5mm,yshift=-0.5mm,minimum width=15mm,font=\small]
\tikzstyle{large gray box}=[small box,fill=gray!30]

\tikzstyle{Bayes box}=[rectangle,fill=black,draw, minimum height=3mm, minimum width=3mm]

\tikzstyle{gray square point}=[small box,fill=gray!50]

\tikzstyle{dphase box white}=[dhadamard]
\tikzstyle{dphase box gray}=[dhadamard,fill=gray!50!white]
\tikzstyle{phase box white}=[hadamard]
\tikzstyle{phase box gray}=[hadamard,fill=gray!50!white]

\tikzstyle{point}=[regular polygon,regular polygon sides=3,draw,scale=0.75,inner sep=-0.5pt,minimum width=9mm,fill=white,regular polygon rotate=180]
\tikzstyle{copoint}=[regular polygon,regular polygon sides=3,draw,scale=0.75,inner sep=-0.5pt,minimum width=9mm,fill=white]
\tikzstyle{dpoint}=[point,doubled]
\tikzstyle{dcopoint}=[copoint,doubled]

\tikzstyle{wide copoint}=[fill=white,draw,shape=isosceles triangle,shape border rotate=90,isosceles triangle stretches=true,inner sep=0pt,minimum width=1.5cm,minimum height=6.12mm]
\tikzstyle{wide point}=[fill=white,draw,shape=isosceles triangle,shape border rotate=-90,isosceles triangle stretches=true,inner sep=0pt,minimum width=1.5cm,minimum height=6.12mm,yshift=-0.0mm]
\tikzstyle{wide point plus}=[fill=white,draw,shape=isosceles triangle,shape border rotate=-90,isosceles triangle stretches=true,inner sep=0pt,minimum width=1.74cm,minimum height=7mm,yshift=-0.0mm]

\tikzstyle{wide dpoint}=[fill=white,doubled,draw,shape=isosceles triangle,shape border rotate=-90,isosceles triangle stretches=true,inner sep=0pt,minimum width=1.5cm,minimum height=6.12mm,yshift=-0.0mm]

\tikzstyle{tinypoint}=[regular polygon,regular polygon sides=3,draw,scale=0.55,inner sep=-0.15pt,minimum width=6mm,fill=white,regular polygon rotate=180] 

\tikzstyle{white point}=[point]
\tikzstyle{white dpoint}=[dpoint]
\tikzstyle{green point}=[white point] % for backwards-compatibility
\tikzstyle{white copoint}=[copoint]
\tikzstyle{gray point}=[point,fill=gray!40!white]
\tikzstyle{gray dpoint}=[gray point,doubled]
\tikzstyle{red point}=[gray point] % for backwards-compatibility
\tikzstyle{gray copoint}=[copoint,fill=gray!40!white]
\tikzstyle{gray dcopoint}=[gray copoint,doubled]

\tikzstyle{white point guide}=[regular polygon,regular polygon sides=3,font=\scriptsize,draw,scale=0.65,inner sep=-0.5pt,minimum width=9mm,fill=white,regular polygon rotate=180]

\tikzstyle{black point}=[point,fill=black,font=\color{white}]
\tikzstyle{black copoint}=[copoint,fill=black,font=\color{white}]

\tikzstyle{tiny gray point}=[tinypoint,fill=gray!40!white]

\tikzstyle{diredge}=[->]
\tikzstyle{ddiredge}=[<->]
\tikzstyle{rdiredge}=[<-]
\tikzstyle{thickdiredge}=[->, very thick]
\tikzstyle{pointer edge}=[->,very thick,gray]
\tikzstyle{pointer edge part}=[very thick,gray]
\tikzstyle{dashed edge}=[dashed]
\tikzstyle{thick dashed edge}=[very thick,dashed]
\tikzstyle{thick gray dashed edge}=[thick dashed edge,gray!40]
\tikzstyle{thick map edge}=[very thick,|->]

\makeatletter
\newcommand{\boxshape}[3]{%
\pgfdeclareshape{#1}{
\inheritsavedanchors[from=rectangle] % this is nearly a rectangle
\inheritanchorborder[from=rectangle]
\inheritanchor[from=rectangle]{center}
\inheritanchor[from=rectangle]{north}
\inheritanchor[from=rectangle]{south}
\inheritanchor[from=rectangle]{west}
\inheritanchor[from=rectangle]{east}
\backgroundpath{% this is new
\southwest \pgf@xa=\pgf@x \pgf@ya=\pgf@y
\northeast \pgf@xb=\pgf@x \pgf@yb=\pgf@y

\@tempdima=#2
\@tempdimb=#3

\pgfpathmoveto{\pgfpoint{\pgf@xa - 5pt + \@tempdima}{\pgf@ya}}
\pgfpathlineto{\pgfpoint{\pgf@xa - 5pt - \@tempdima}{\pgf@yb}}
\pgfpathlineto{\pgfpoint{\pgf@xb + 5pt + \@tempdimb}{\pgf@yb}}
\pgfpathlineto{\pgfpoint{\pgf@xb + 5pt - \@tempdimb}{\pgf@ya}}
\pgfpathlineto{\pgfpoint{\pgf@xa - 5pt + \@tempdima}{\pgf@ya}}
\pgfpathclose
}
}}

\boxshape{NEbox}{0pt}{5pt}
\boxshape{SEbox}{0pt}{-5pt}
\boxshape{NWbox}{5pt}{0pt}
\boxshape{SWbox}{-5pt}{0pt}
\boxshape{EBox}{-3pt}{3pt}
\boxshape{WBox}{3pt}{-3pt}
\makeatother

\tikzstyle{cloud}=[shape=cloud,draw,minimum width=1.5cm,minimum height=1.5cm]

\tikzstyle{map}=[draw,shape=rectangle,inner sep=2pt,minimum height=6mm,minimum width=7mm,fill=white]
\tikzstyle{dashedmap}=[draw,dashed,shape=NEbox,inner sep=2pt,minimum height=6mm,fill=white]
\tikzstyle{mapdag}=[draw,shape=SEbox,inner sep=2pt,minimum height=6mm,fill=white]
\tikzstyle{mapadj}=[draw,shape=SEbox,inner sep=2pt,minimum height=6mm,fill=white]
\tikzstyle{maptrans}=[draw,shape=SWbox,inner sep=2pt,minimum height=6mm,fill=white]
\tikzstyle{mapconj}=[draw,shape=NWbox,inner sep=2pt,minimum height=6mm,fill=white]

\tikzstyle{medium map}=[draw,shape=NEbox,inner sep=2pt,minimum height=6mm,fill=white,minimum width=7mm]
\tikzstyle{medium map dag}=[draw,shape=SEbox,inner sep=2pt,minimum height=6mm,fill=white,minimum width=7mm]
\tikzstyle{medium map adj}=[draw,shape=SEbox,inner sep=2pt,minimum height=6mm,fill=white,minimum width=7mm]
\tikzstyle{medium map trans}=[draw,shape=SWbox,inner sep=2pt,minimum height=6mm,fill=white,minimum width=7mm]
\tikzstyle{medium map conj}=[draw,shape=NWbox,inner sep=2pt,minimum height=6mm,fill=white,minimum width=7mm]
\tikzstyle{semilarge map}=[draw,shape=rectangle,inner sep=2pt,minimum height=6mm,fill=white,minimum width=10.5mm]
\tikzstyle{semilarge map trans}=[draw,shape=SWbox,inner sep=2pt,minimum height=6mm,fill=white,minimum width=9.5mm]
\tikzstyle{semilarge map adj}=[draw,shape=SEbox,inner sep=2pt,minimum height=6mm,fill=white,minimum width=9.5mm]
\tikzstyle{semilarge map dag}=[draw,shape=SEbox,inner sep=2pt,minimum height=6mm,fill=white,minimum width=9.5mm]
\tikzstyle{semilarge map conj}=[draw,shape=NWbox,inner sep=2pt,minimum height=6mm,fill=white,minimum width=9.5mm]
\tikzstyle{large map}=[draw,shape=rectangle,inner sep=2pt,minimum height=6mm,fill=white,minimum width=16mm]
\tikzstyle{large map conj}=[draw,shape=NWbox,inner sep=2pt,minimum height=6mm,fill=white,minimum width=12mm]
\tikzstyle{very large map}=[draw,shape=rectangle,inner sep=2pt,minimum height=6mm,fill=white,minimum width=20mm]
\tikzstyle{large map dag}=[draw,shape=SEbox,inner sep=2pt,minimum height=6mm,fill=white,minimum width=12mm]

\tikzstyle{medium dmap}=[draw,doubled,shape=NEbox,inner sep=2pt,minimum height=6mm,fill=white,minimum width=7mm]
\tikzstyle{medium dmap dag}=[draw,doubled,shape=SEbox,inner sep=2pt,minimum height=6mm,fill=white,minimum width=7mm]
\tikzstyle{medium dmap adj}=[draw,doubled,shape=SEbox,inner sep=2pt,minimum height=6mm,fill=white,minimum width=7mm]
\tikzstyle{medium dmap trans}=[draw,doubled,shape=SWbox,inner sep=2pt,minimum height=6mm,fill=white,minimum width=7mm]
\tikzstyle{medium dmap conj}=[draw,doubled,shape=NWbox,inner sep=2pt,minimum height=6mm,fill=white,minimum width=7mm]
\tikzstyle{semilarge dmap}=[draw,doubled,shape=NEbox,inner sep=2pt,minimum height=6mm,fill=white,minimum width=9.5mm]
\tikzstyle{semilarge dmap trans}=[draw,doubled,shape=SWbox,inner sep=2pt,minimum height=6mm,fill=white,minimum width=9.5mm]
\tikzstyle{semilarge dmap adj}=[draw,doubled,shape=SEbox,inner sep=2pt,minimum height=6mm,fill=white,minimum width=9.5mm]
\tikzstyle{semilarge dmap dag}=[draw,doubled,shape=SEbox,inner sep=2pt,minimum height=6mm,fill=white,minimum width=9.5mm]
\tikzstyle{semilarge dmap conj}=[draw,doubled,shape=NWbox,inner sep=2pt,minimum height=6mm,fill=white,minimum width=9.5mm]
\tikzstyle{large dmap}=[draw,doubled,shape=NEbox,inner sep=2pt,minimum height=6mm,fill=white,minimum width=12mm]
\tikzstyle{large dmap conj}=[draw,doubled,shape=NWbox,inner sep=2pt,minimum height=6mm,fill=white,minimum width=12mm]
\tikzstyle{large dmap trans}=[draw,doubled,shape=SWbox,inner sep=2pt,minimum height=6mm,fill=white,minimum width=12mm]
\tikzstyle{large dmap adj}=[draw,doubled,shape=SEbox,inner sep=2pt,minimum height=6mm,fill=white,minimum width=12mm]
\tikzstyle{large dmap dag}=[draw,doubled,shape=SEbox,inner sep=2pt,minimum height=6mm,fill=white,minimum width=12mm]
\tikzstyle{very large dmap}=[draw,doubled,shape=NEbox,inner sep=2pt,minimum height=6mm,fill=white,minimum width=19.5mm]

\tikzstyle{muxbox}=[draw,shape=rectangle,minimum height=3mm,minimum width=3mm,fill=white]
\tikzstyle{dmuxbox}=[muxbox,doubled]

\tikzstyle{box}=[draw,shape=rectangle,inner sep=2pt,minimum height=6mm,minimum width=6mm,fill=white]
\tikzstyle{dbox}=[draw,doubled,shape=rectangle,inner sep=2pt,minimum height=6mm,minimum width=6mm,fill=white]
\tikzstyle{dmap}=[draw,doubled,shape=NEbox,inner sep=2pt,minimum height=6mm,fill=white]
\tikzstyle{dmapdag}=[draw,doubled,shape=SEbox,inner sep=2pt,minimum height=6mm,fill=white]
\tikzstyle{dmapadj}=[draw,doubled,shape=SEbox,inner sep=2pt,minimum height=6mm,fill=white]
\tikzstyle{dmaptrans}=[draw,doubled,shape=SWbox,inner sep=2pt,minimum height=6mm,fill=white]
\tikzstyle{dmapconj}=[draw,doubled,shape=NWbox,inner sep=2pt,minimum height=6mm,fill=white]

\tikzstyle{ddmap}=[draw,doubled,dashed,shape=NEbox,inner sep=2pt,minimum height=6mm,fill=white]
\tikzstyle{ddmapdag}=[draw,doubled,dashed,shape=SEbox,inner sep=2pt,minimum height=6mm,fill=white]
\tikzstyle{ddmapadj}=[draw,doubled,dashed,shape=SEbox,inner sep=2pt,minimum height=6mm,fill=white]
\tikzstyle{ddmaptrans}=[draw,doubled,dashed,shape=SWbox,inner sep=2pt,minimum height=6mm,fill=white]
\tikzstyle{ddmapconj}=[draw,doubled,dashed,shape=NWbox,inner sep=2pt,minimum height=6mm,fill=white]

\boxshape{sNEbox}{0pt}{3pt}
\boxshape{sSEbox}{0pt}{-3pt}
\boxshape{sNWbox}{3pt}{0pt}
\boxshape{sSWbox}{-3pt}{0pt}
\tikzstyle{smap}=[draw,shape=sNEbox,fill=white]
\tikzstyle{smapdag}=[draw,shape=sSEbox,fill=white]
\tikzstyle{smapadj}=[draw,shape=sSEbox,fill=white]
\tikzstyle{smaptrans}=[draw,shape=sSWbox,fill=white]
\tikzstyle{smapconj}=[draw,shape=sNWbox,fill=white]

\tikzstyle{dsmap}=[draw,dashed,shape=sNEbox,fill=white]
\tikzstyle{dsmapdag}=[draw,dashed,shape=sSEbox,fill=white]
\tikzstyle{dsmaptrans}=[draw,dashed,shape=sSWbox,fill=white]
\tikzstyle{dsmapconj}=[draw,dashed,shape=sNWbox,fill=white]

\boxshape{mNEbox}{0pt}{10pt}
\boxshape{mSEbox}{0pt}{-10pt}
\boxshape{mNWbox}{10pt}{0pt}
\boxshape{mSWbox}{-10pt}{0pt}
\tikzstyle{mmap}=[draw,shape=mNEbox]
\tikzstyle{mmapdag}=[draw,shape=mSEbox]
\tikzstyle{mmaptrans}=[draw,shape=mSWbox]
\tikzstyle{mmapconj}=[draw,shape=mNWbox]

\tikzstyle{mmapgray}=[draw,fill=gray!40!white,shape=mNEbox]
\tikzstyle{smapgray}=[draw,fill=gray!40!white,shape=sNEbox]

\makeatletter

\pgfdeclareshape{cornerpoint}{
\inheritsavedanchors[from=rectangle] % this is nearly a rectangle
\inheritanchorborder[from=rectangle]
\inheritanchor[from=rectangle]{center}
\inheritanchor[from=rectangle]{north}
\inheritanchor[from=rectangle]{south}
\inheritanchor[from=rectangle]{west}
\inheritanchor[from=rectangle]{east}
\backgroundpath{% this is new
\southwest \pgf@xa=\pgf@x \pgf@ya=\pgf@y
\northeast \pgf@xb=\pgf@x \pgf@yb=\pgf@y

\pgfmathsetmacro{\pgf@shorten@left}{\pgfkeysvalueof{/tikz/shorten left}}
\pgfmathsetmacro{\pgf@shorten@right}{\pgfkeysvalueof{/tikz/shorten right}}

\pgfpathmoveto{\pgfpoint{0.5 * (\pgf@xa + \pgf@xb)}{\pgf@ya - 5pt}}
\pgfpathlineto{\pgfpoint{\pgf@xa - 8pt + \pgf@shorten@left}{\pgf@yb - 1.5 * \pgf@shorten@left}}
\pgfpathlineto{\pgfpoint{\pgf@xa - 8pt + \pgf@shorten@left}{\pgf@yb}}
\pgfpathlineto{\pgfpoint{\pgf@xb + 8pt - \pgf@shorten@right}{\pgf@yb}}
\pgfpathlineto{\pgfpoint{\pgf@xb + 8pt - \pgf@shorten@right}{\pgf@yb - 1.5 * \pgf@shorten@right}}
\pgfpathclose
}
}

\pgfdeclareshape{cornercopoint}{
\inheritsavedanchors[from=rectangle] % this is nearly a rectangle
\inheritanchorborder[from=rectangle]
\inheritanchor[from=rectangle]{center}
\inheritanchor[from=rectangle]{north}
\inheritanchor[from=rectangle]{south}
\inheritanchor[from=rectangle]{west}
\inheritanchor[from=rectangle]{east}
\backgroundpath{% this is new
\southwest \pgf@xa=\pgf@x \pgf@ya=\pgf@y
\northeast \pgf@xb=\pgf@x \pgf@yb=\pgf@y

\pgfmathsetmacro{\pgf@shorten@left}{\pgfkeysvalueof{/tikz/shorten left}}
\pgfmathsetmacro{\pgf@shorten@right}{\pgfkeysvalueof{/tikz/shorten right}}

\pgfpathmoveto{\pgfpoint{0.5 * (\pgf@xa + \pgf@xb)}{\pgf@yb + 5pt}}
\pgfpathlineto{\pgfpoint{\pgf@xa - 8pt + \pgf@shorten@left}{\pgf@ya + 1.5 * \pgf@shorten@left}}
\pgfpathlineto{\pgfpoint{\pgf@xa - 8pt + \pgf@shorten@left}{\pgf@ya}}
\pgfpathlineto{\pgfpoint{\pgf@xb + 8pt - \pgf@shorten@right}{\pgf@ya}}
\pgfpathlineto{\pgfpoint{\pgf@xb + 8pt - \pgf@shorten@right}{\pgf@ya + 1.5 * \pgf@shorten@right}}
\pgfpathclose
}
}

\makeatother

\pgfkeyssetvalue{/tikz/shorten left}{0pt}
\pgfkeyssetvalue{/tikz/shorten right}{0pt}

\tikzstyle{kpoint common}=[draw,fill=white,inner sep=1pt,minimum height=4mm]
\tikzstyle{kpoint sc}=[shape=cornerpoint,kpoint common]
\tikzstyle{kpoint adjoint sc}=[shape=cornercopoint,kpoint common]
\tikzstyle{kpoint}=[shape=cornerpoint,shorten left=5pt,kpoint common]
\tikzstyle{kpoint adjoint}=[shape=cornercopoint,shorten left=5pt,kpoint common]
\tikzstyle{kpoint conjugate}=[shape=cornerpoint,shorten right=5pt,kpoint common]
\tikzstyle{kpoint transpose}=[shape=cornercopoint,shorten right=5pt,kpoint common]
\tikzstyle{kpoint symm}=[shape=cornerpoint,shorten left=5pt,shorten right=5pt,kpoint common]

\tikzstyle{black kpoint}=[shape=cornerpoint,shorten left=5pt,kpoint common,fill=black,font=\color{white}]
\tikzstyle{black kpoint adjoint}=[shape=cornercopoint,shorten left=5pt,kpoint common,fill=black,font=\color{white}]
\tikzstyle{black kpointadj}=[shape=cornercopoint,shorten left=5pt,kpoint common,fill=black,font=\color{white}]

\tikzstyle{black dkpoint}=[shape=cornerpoint,shorten left=5pt,kpoint common,fill=black, doubled,font=\color{white}]
\tikzstyle{black dkpoint adjoint}=[shape=cornercopoint,shorten left=5pt,kpoint common,fill=black, doubled,font=\color{white}]
\tikzstyle{black dkpointadj}=[shape=cornercopoint,shorten left=5pt,kpoint common,fill=black, doubled,font=\color{white}] 

\tikzstyle{kpointdag}=[kpoint adjoint]
\tikzstyle{kpointadj}=[kpoint adjoint]
\tikzstyle{kpointconj}=[kpoint conjugate]
\tikzstyle{kpointtrans}=[kpoint transpose]

\tikzstyle{big kpoint}=[kpoint, minimum width=1.2 cm, minimum height=8mm, inner sep=4pt, text depth=3mm]

\tikzstyle{wide kpoint}=[kpoint, minimum width=1 cm, inner sep=2pt]%, text depth=-0.7 mm]
\tikzstyle{wide kpointdag}=[kpointdag, minimum width=1 cm, inner sep=2pt]%, text depth=0.7 mm]
\tikzstyle{wide kpointconj}=[kpointconj, minimum width=1 cm, inner sep=2pt]%, text depth=-0.7 mm]
\tikzstyle{wide kpointtrans}=[kpointtrans, minimum width=1 cm, inner sep=2pt]%, text depth=0.7 mm]

\tikzstyle{gray kpoint}=[kpoint,fill=gray!50!white]
\tikzstyle{gray kpointdag}=[kpointdag,fill=gray!50!white]
\tikzstyle{gray kpointadj}=[kpointadj,fill=gray!50!white]
\tikzstyle{gray kpointconj}=[kpointconj,fill=gray!50!white]
\tikzstyle{gray kpointtrans}=[kpointtrans,fill=gray!50!white]

\tikzstyle{gray dkpoint}=[kpoint,fill=gray!50!white,doubled]
\tikzstyle{gray dkpointdag}=[kpointdag,fill=gray!50!white,doubled]
\tikzstyle{gray dkpointadj}=[kpointadj,fill=gray!50!white,doubled]
\tikzstyle{gray dkpointconj}=[kpointconj,fill=gray!50!white,doubled]
\tikzstyle{gray dkpointtrans}=[kpointtrans,fill=gray!50!white,doubled]

\tikzstyle{white label}=[draw,fill=white,rectangle,inner sep=0.7 mm]
\tikzstyle{gray label}=[draw,fill=gray!50!white,rectangle,inner sep=0.7 mm]
\tikzstyle{black label}=[draw,fill=black,rectangle,inner sep=0.7 mm]

\tikzstyle{dkpoint}=[kpoint,doubled]
\tikzstyle{wide dkpoint}=[wide kpoint,doubled]
\tikzstyle{dkpointdag}=[kpoint adjoint,doubled]
\tikzstyle{wide dkpointdag}=[wide kpointdag,doubled]
\tikzstyle{dkcopoint}=[kpoint adjoint,doubled]
\tikzstyle{dkpointadj}=[kpoint adjoint,doubled]
\tikzstyle{dkpointconj}=[kpoint conjugate,doubled]
\tikzstyle{dkpointtrans}=[kpoint transpose,doubled]

\tikzstyle{kscalar}=[kpoint common, shape=EBox, inner xsep=-1pt, inner ysep=3pt,font=\small]
\tikzstyle{kscalarconj}=[kpoint common, shape=WBox, inner xsep=-1pt, inner ysep=3pt,font=\small]

\tikzstyle{spekpoint}=[kpoint sc,minimum height=5mm,inner sep=3pt]
\tikzstyle{spekcopoint}=[kpoint adjoint sc,minimum height=5mm,inner sep=3pt]

\tikzstyle{dspekpoint}=[spekpoint,doubled]
\tikzstyle{dspekcopoint}=[spekcopoint,doubled]

 \tikzstyle{discard}=[ground,rotate=180,scale=1.5,inner sep=-2mm]
 \tikzstyle{downground}=[circuit ee IEC,thick,ground,rotate=-90,scale=1.5,inner sep=-2mm]

\tikzstyle{maxmix}=[regular polygon,regular polygon sides=3,draw=black,xscale=0.4,yscale=0.3,inner sep=-0.5pt,minimum width=10mm,fill=gray,regular polygon rotate=180]

 \tikzstyle{bigground}=[regular polygon,regular polygon sides=3,draw=gray,scale=0.50,inner sep=-0.5pt,minimum width=10mm,fill=gray]
\tikzstyle{arrs}=[-latex,font=\small,auto]
\tikzstyle{arrow plain}=[arrs]
\tikzstyle{arrow dashed}=[dashed,arrs]
\tikzstyle{arrow bold}=[very thick,arrs]
\tikzstyle{arrow hide}=[draw=white!0,-]
\tikzstyle{arrow reverse}=[latex-]
\tikzstyle{cdnode}=[]

\usepackage{graphicx}[pdftex]
\usepackage{physics}

\begin{document}

\title{Causal structure in the presence of sectorial constraints, with application to the quantum switch}

\author{Nick Ormrod}
\email{nicholas.ormrod@cs.ox.ac.uk}
\affiliation{Quantum Group, Department of Computer Science, University of Oxford}

\author{Augustin Vanrietvelde}
\email{augustin.vanrietvelde@inria.fr}
\affiliation{Quantum Group, Department of Computer Science, University of Oxford}
\affiliation{Department of Physics, Imperial College London}
\affiliation{HKU-Oxford Joint Laboratory for Quantum Information and Computation}

\author{Jonathan Barrett}
\email{jonathan.barrett@cs.ox.ac.uk}
\affiliation{Quantum Group, Department of Computer Science, University of Oxford}

\begin{abstract}
Existing work on quantum causal structure assumes that one can perform arbitrary operations on the systems of interest. But this condition is often not met. Here, we extend the framework for quantum causal modelling to situations where a system can suffer \textit{sectorial constraints}, that is, restrictions on the orthogonal subspaces of its Hilbert space that may be mapped to one another.  Our framework (a) proves that a number of different intuitions about causal relations turn out to be equivalent; (b) shows that quantum causal structures in the presence of sectorial constraints can be represented with a directed graph; and (c) defines a fine-graining of the causal structure in which the individual sectors of a system bear causal relations. As an example, we apply our framework to purported photonic implementations of the quantum switch to show that while their coarse-grained causal structure is cyclic, their fine-grained causal structure is acyclic. We therefore conclude that these experiments realize indefinite causal order only in a weak sense. Notably, this is the first argument to this effect that is not rooted in the assumption that the causal relata must be localized in spacetime.
\end{abstract}

\maketitle
\setcounter{tocdepth}{0}

%\tableofcontents
\newtheorem{theorem}{Theorem}[section]

\newtheorem{remark}{Remark}[section]

\newtheorem{corollary}{Corollary}[section]

\newtheorem{lemma}{Lemma}[section]

\newtheorem{sublemma}{Sublemma}[section]

\newtheorem{proposition}{Proposition}[section]

\newtheorem{definition}{Definition}[section]

\newcommand{\discard}{%
\begin{tikzpicture}%
\draw (0,1ex) -- (5ex,1ex);%
\draw (1ex,2ex) -- (4ex,2ex);%
\draw (2ex,3ex) -- (3ex,3ex);%
\draw (2.5ex, -2ex) -- (2.5ex, 1ex);
\end{tikzpicture}%
}

%%% Calligraphic Alphabet

\newcommand{\ca}{\mathcal A}
\newcommand{\cb}{\mathcal B}
\newcommand{\cc}{\mathcal C}
\newcommand{\cd}{\mathcal D}
\newcommand{\ce}{\mathcal E}
\newcommand{\cf}{\mathcal F}
\newcommand{\cg}{\mathcal G}
\newcommand{\ch}{\mathcal H}
\newcommand{\ci}{\mathcal I}
\newcommand{\cj}{\mathcal J}
\newcommand{\ck}{\mathcal K}
\newcommand{\cl}{\mathcal L}
\newcommand{\cm}{\mathcal M}
\newcommand{\cn}{\mathcal N}
\newcommand{\co}{\mathcal O}
\newcommand{\calp}{\mathcal P}
\newcommand{\cq}{\mathcal Q}
\newcommand{\calr}{\mathcal R}
\newcommand{\cs}{\mathcal S}
\newcommand{\ct}{\mathcal T}
\newcommand{\cu}{\mathcal U}
\newcommand{\cv}{\mathcal V}
\newcommand{\cw}{\mathcal W}
\newcommand{\cx}{\mathcal X}
\newcommand{\cy}{\mathcal Y}
\newcommand{\cz}{\mathcal Z}
\newcommand{\dc}{\rightsquigarrow}

%%%%%%%%%%%%%%%%%%%%%%%%%%%%%%%%%%%%%%%%%%%%%%%%%%%%%%%%%%%%%%%%
\section{Introduction} 
%%%%%%%%%%%%%%%%%%%%%%%%%%%%%%%%%%%%%%%%%%%%%%%%%%%%%%%%%%%%%%%%

Causal structure in quantum theory has recently been the subject of intense study, both theoretically \cite{hardy2007towards, Chiribella_2013, oreshkov2012quantum, araujo2015witnessing, oreshkov2012quantum, barrett2020quantum, paunkovic2020causal, felce2020quantum, barrett2021cyclic, kissinger2017categorical, lorenz2020causal, branciard2015simplest, araujo2014computational, felce2022indefinite} and in the laboratory \cite{procopio2015experimental, rubino2017experimental, goswami2018indefinite, rubino2019experimental, nie2020experimental, cao2021experimental} (see \cite{Goswami_2020} for a review of the experiments). An impetus for this was the discovery that a simple extension of quantum theory allows for \textit{indefinite causal order}  \cite{hardy2007towards, Hardy2009quantum, Chiribella_2009, chiribella2009beyond, Chiribella_2013, chiribella2012perfect, Colnaghi_2012, oreshkov2012quantum, baumeler2016space, baumeler2014maximal, araujo2017purification, vanrietvelde2022consistent}, in which the causal structure of a scenario inherits some of the quantum indeterminacy associated with superpositions of states. Such structures should be important in any theory of quantum gravity in which one can have a superposition of different classical solutions of Einstein's field equations \cite{Hardy2009quantum}.

However, even when the causal order is definite, quantum theory poses significant challenges to classical notions of cause and effect. For example, no common cause can explain the Bell correlations in accordance with Reichenbach's principle\footnote{This states that conditioning on the common cause $C$ should eliminate the correlations between the correlated variables, $A$ and $B$ \cite{reichenbach1956direction}. Formally, $P(AB|C)=P(A|C)P(B|C)$.} unless superluminal, retrocausal, or superdeterminstic causal influences are invoked. Even allowing such exotic influences, any classical causal explanation of the Bell correlations would have to be fine-tuned \cite{Wood_2015}.

One is left with two options: abandon the notions of cause and effect altogether, or update them for a quantum setting. The latter approach was taken in \cite{Allen_2017, barrett2020quantum, barrett2021cyclic} by generalizing classical causal models \cite{pearl2009causality} to create an intrinsically quantum theory of causal structure.\footnote{See \cite{pienaar2015graph, costa2016, pienaar2019time, pienaar2020quantum} for alternative approaches to quantum causal modelling, or \cite{gogioso2022topology} for a theory-independent approach to generalizing classical causal models.} The resulting framework of `quantum causal models' provides answers to the basic foundational questions about causality -- what are the relata of causal relations? what does it mean for one thing to be a direct cause of another? are quantum causal relations time-symmetric? Beyond this, it lays out the implications of a postulated causal structure for the correlations we can actually observe. This approach has also borne fruit for the question of indefinite causal order by rigorously analysing \cite{barrett2021cyclic} the causal structures of the quantum switch \cite{Chiribella_2013} and the Lugano process  \cite{baumeler2016space}.

However, the existing framework makes a highly nontrivial assumption about the scenarios it serves to model. Namely, that any quantum operation on the Hilbert space of one of the systems (and possibly some local ancillas) is a possible intervention. Mathematically, this assumption is manifested in the use of a process matrix to represent the physical scenario, which can be combined with any set of local quantum operations for each system to yield a valid probability distribution. 

The assumption is important since the framework defines causal relations in terms of the possibility of signalling given the underlying unitary process between agents that each have access to one of the relevant systems. But if the interventions that the agents can perform are restricted, then this might limit their ability to signal to one another. Thus restrictions on the possible interventions should change the causal structure. It follows that the existing framework for quantum causal models is inadequate for modelling scenarios in which there are restrictions on the possible interventions on a system.

Some particularly interesting restrictions are \textit{sectorial constraints.} These are restrictions on which sectors (i.e., orthogonal subspaces) of the input space can be mapped to which sectors of the output space by some transformation. They apply to the common quantum optical technique of sending a photon into a superposition of trajectories (in which they arise from the practical impossibility of creating a photon from the vacuum) \cite{Chiribella_2019}; to various superselected systems; to the Aharonov-Bohm effect \cite{aharonov1959significance, Erez_2010}; the del Santo-Dakić protocol \cite{Del_Santo_2018, Carrying, Massa_2019, faleiro2020}; recent alleged implementations \cite{goswami2018indefinite, procopio2015experimental, rubino2017experimental, rubino2019experimental} of the quantum switch \cite{chiribella2009beyond, Chiribella_2013}; and many more quantum protocols.
 More abstractly, sectorial constraints are at the heart of the various powerful theorems and techniques in quantum theory that make use of non-factor $C^*$ algebras, whose structure is specified by the sectorial constraints suffered by their elements. Examples include the Schur-Weyl duality and its generalisations \cite{marvian2014, Harrow2005ApplicationsOC}, or the existence of decoherence-free subspaces \cite{palma1996quantum, duan1997preserving, zanardi1997noiseless, lidar1998decoherence, beige2000quantum, kwiat2000experimental}.

What this shows is that the causal structure of many foundationally interesting and practically significant scenarios cannot be adequately analysed by the existing framework for quantum causal models, which fails to accommodate the sectorial constraints that help define them. The goal of this paper is to present an appropriate framework for assessing the causal structure in these situations. 

Our framework is self-contained, scalable, and applicable to a wide range of quantum-theoretical scenarios. It shows how various intuitions of causal influence turn out to be equivalent in the unitary case. It does not only extend standard quantum causal modelling to situations featuring sectorial constraints; it also shows that in such situations, a more fine-grained account of the causal structure can be given by leveraging on the sectorial structure of the operations. The resulting \textit{sectorized causal structure} possesses a natural conceptual interpretation, and is strictly more detailed than the standard, unsectorized causal structure.

The sectorized causal structure  can shed light on enigmatic features of the more standard sort of causal structure. To illustrate this, we end this paper with a detailed analysis of recent purported photonic implementations \cite{goswami2018indefinite, procopio2015experimental, rubino2017experimental, rubino2019experimental} of the quantum switch \cite{chiribella2009beyond, Chiribella_2013}. Weighing in on a recent debate over whether these experiments realize indefinite causal order \cite{oreshkov2019time, procopio2015experimental, paunkovic2020causal}, we argue that they only do so at the coarse-grained, unsectorized level, but not at the more detailed, sectorized level. We conclude that they are weak realizations of indefinite causal order.
Unlike previous arguments to a similar effect \cite{paunkovic2020causal}, ours is not at all rooted in the assumption the objects that bear causal relations should be localized in spacetime. In fact, our basic assumptions resemble those that are usually taken to motivate the view that the experiments do realize indefinite causal order.

%Beyond implementations of the switch, our framework could similarly be applied to other quantum scenarios which feature sectorial constraints, and whose causal structure has been a subject of particular attention: examples include the Aharonov-Bohm effect \cite{aharonov1959significance}, and the recently proposed del Santo-Dakić protocol \cite{Del_Santo_2018, Carrying, Massa_2019, faleiro2020}.

The structure of the paper is as follows. First, we recap in Section \ref{sec:routed framework} the framework of \textit{routed maps and supermaps}, which was recently developed \cite{vanrietvelde2020routed, vanrietvelde2021universal, composableconstraints} as a way of hardcoding sectorial constraints into models of quantum processes, providing a basis for our analysis. In Section \ref{sec:first order}, we provide a review of the canonical notion of a causal relation between the inputs and outputs of a standard unitary transformation, then show how it can be naturally generalised to the case of \textit{routed unitary transformations}, satisfying sectorial constraints. We then define fine-grainings of causal relations through routed unitaries, and thus define the sectorized causal structure.

Building on this, in Section \ref{sec:higher order} we first summarise how the existing quantum causal models framework treats the causal structure of a more complicated sort of scenario, represented by a higher-order transformation known as a supermap\footnote{Note that supermaps can equivalently be seen as \textit{process matrices}, used elsewhere in the literature on indefinite causal order. More precisely, process matrices are essentially a Choi-Jamiołkowski representation of supermaps.}, then generalise this analysis to \textit{routed} supermaps, in which operations satisfy sectorial constraints. We show that in this case again, a sectorized causal structure can be defined. In Section \ref{sec:switch}, we apply the resulting framework to the alleged photonic implementations of the switch. Finally, in Section \ref{sec:rswitchconsequences}, we fend off possible objections to our analysis of those experiments.

%%%%%%%%%%%%%%%%%%%%%%%%%%%%%%%%%%%%%%%%%%%%%%
\section{Sectorial constraints and routed quantum circuits} \label{sec:routed framework}
\addcontentsline{toc}{section}{Sectorial constraints and routed quantum circuits}
%%%%%%%%%%%%%%%%%%%%%%%%%%%%%%%%%%%%%%%%%%%%%%

In this section, we review the basics for modelling operations that feature sectorial constraints. 

\subsection{Why routed circuits?}

The routed circuits framework \cite{vanrietvelde2020routed} was devised in response to situations  \cite{lorenz2020causal, Abbott_2020, Chiribella_2019, Kristj_nsson_2020} where the relevant Hilbert space of a system is a proper subspace of the tensor product of the spaces of each individual subsystem. For example, in the superposition of paths scenario, a particle described by a $d$-dimensional Hilbert space $P$ is sent down Alice's or Bob's transmission line depending on the logical value of a qubit $C$. Alice's Hilbert space $A=A^\textit{vac}_\star  \oplus A^\textit{par}_\star$ is a direct sum of a one-dimensional vacuum sector $A^\textit{vac}_\star$ and a $d$-dimensional particle sector $A^\textit{par}_\star$. Likewise, Bob's space can be written $B=B^\textit{vac}_\star  \oplus B^\textit{par}_\star$. The transformation can be represented as follows, where $\ket{\Omega}$ is the vacuum state:
\begin{equation} \label{superposition-paths}
\begin{split}
    \ket{0}_C\ket{\psi}_P & \rightarrow \ket{\psi}_A \ket{\Omega}_B \\
    \ket{1}_C\ket{\psi}_P & \rightarrow \ket{\Omega}_A \ket{\psi}_B
\end{split}
\end{equation}

It is easily checked that (\ref{superposition-paths}) defines a unitary transformation $U: C \otimes P \rightarrow (A^\textit{par}_\star \otimes B^\textit{vac}_\star) \oplus (A^\textit{vac}_\star \otimes B^\textit{par}_\star)$. However, one cannot provide a standard unitary quantum circuit representing this transformation in which $A$ and $B$ are each associated with one output wire, unless one extends $U$ to be defined on the physically irrelevant sectors $A^\textit{vac}_\star \otimes B^\textit{vac}_\star$ and $A^\textit{par}_\star \otimes B^\textit{par}_\star$, as well as a larger input space. One is  left with a dilemma: either accept a non-unitary diagrammatic representation of a scenario which should be physically be understood as unitary, or allow physically redundant degrees of freedom in the representation.

The dilemma is a direct consequence of a very fundamental fact about standard quantum circuit diagrams -- that putting two wires next to each other means taking the tensor product of the associated Hilbert spaces. When there are constraints that prevent certain subspaces of this tensor product space from being populated, this means including some redundancy in our representation. More generally, the problem with standard circuits is that they lack the resources to describe constraints on how transformations are allowed to act on certain relevant subspaces: they cannot explicitly represent \textit{sectorial constraints}. The very same deficiency leads to problems in completely different areas; for example, it makes it impossible for standard circuits to provide \textit{causal decompositions} of certain unitary transformations \cite{lorenz2020causal}.

The framework of routed quantum circuits solves this problem by supplementing standard quantum circuits with classical (more precisely, possibilistic) information representing sectorial constraints. We now describe that framework.

\subsection{Routed maps}

We start off by adding structure to the Hilbert spaces. A \textit{sectorized Hilbert space}\footnote{In \cite{vanrietvelde2020routed}, this was referred to as a `partitioned Hilbert space'. Here we are aiming for more precise terminology.} is a Hilbert space with a preferred decomposition into a set of orthogonal sectors (i.e.\ orthogonal subspaces), which we call a \textit{sectorization}. Not too formally, given a Hilbert space $A$ and a complete orthogonal family of projectors $\pi_A^i$, we can construct a sectorized version of the space by appealing to a preferred sectorization:
\begin{equation} \label{partioned-H-space-def}
    A^i = \bigoplus_{i \in \cz_A} \pi_A^i (A)
\end{equation}
Note that $A^i$ represents the whole sectorized space; its superscript does not stand for some specfic sector, but simply serves to remind us that the space is sectorized. We will represent specific sectors in the sectorization of the space using a star subscript $A_\star^i := \pi_A^i(A)$.

More formally, a sectorized space can  be defined as a tuplet including the original space $A$, the projectors $\pi_A^i$, and the possible values $\cz_A$ of $i$.

The tensor product of two sectorized Hilbert spaces $A^i$ and $B^j$ is rather natural, corresponding to the decomposition:

\begin{equation} \label{partioned-H-space-otimes}
    A^i \otimes B^j = \bigoplus_{i \in \cz_A, j \in \cz_B} A^i_\star \otimes B^j_\star
\end{equation}

That takes care of the Hilbert spaces; now we must address the transformations. We supplement the transformations with \textit{routes}, which encode sectorial constraints. A route is a \textit{relation} $\lambda$, a mathematical object which can be though of as a possibly multi- or empty-valued function; it can be represented as a Boolean matrix $\lambda_i^j$. Its Boolean element $\lambda_i^j$ determines whether a transformation is permitted to map the $i$th input sector to the $j$th output sector.

To state this more formally, consider a linear map $f: A \rightarrow B$ between Hilbert spaces which are sectorized to form $A^i$ and $B^j$ as described above. We say that $f: A \rightarrow B$ \textit{follows} a route $\lambda$ just in case
\begin{equation} \label{routedef}
    f = \sum_{ij} \lambda_i^j \cdot \pi_B^j \circ f \circ \pi_A^i
\end{equation}
As proven in \cite{vanrietvelde2020routed}, this is equivalent to the statement that for all $i$ and $j$, if $\lambda_i^j =0$ then $\pi_B^j \circ f \circ \pi_A^i=0$. Hence if $\lambda^j_i=0$ then $f$ cannot map any state with null support outside the $i$th sector of $A$, $\psi_A \in A_\star^i$, to a state with support in the $j$th sector of $B$, $B_\star^j$. Another equivalent formulation is that, given a matrix representation of $f$, the corresponding block in the block-decomposition provided by the projectors is zero whenever $\lambda_i^j=0$. We thus see how routes (and specifically, their zeroes) enforce sectorial constraints on linear maps.

Routes can be sequentially composed using matrix multiplication and parallel-composed using the cartesian product. Their compositions play well with the compositions of linear maps, as well as their adjoints: if $f$ follows $\lambda$ and $g$ follows $\epsilon$, then $g \circ f$ and $f \otimes g$ follow $\epsilon \circ \lambda$ and $\epsilon \times \lambda$ respectively, and $f^\dagger$ follows $\lambda^\top$.

We can now define a \textit{routed linear map} from $A^i$ to $B^j$ as a pair $(\lambda, f)$  for a route $\lambda$ and a linear map $f$ that follows it. Routed linear maps (hereon just `routed maps') can be sequentially and parallelly composed by composing their elements pairwise, and one can take their Hermitian adjoint by adjoining their elements. 

It will also be useful to define the notion of practical input or output spaces. The idea is that there can be elements $i$ of $\lambda$'s input space that are not connected by $\lambda$ to anything in the output space, i.e.\ satisfying $\lambda_i^j~=~0 \,\forall j$. This means that a linear map $f$ that follows $\lambda$ will necessarily be zero on the corresponding subspaces $A^i_\star$. As these $A^i_\star$'s are in practice never used by the routed map $(\lambda,f)$, we define the \textit{practical input space} of $(\lambda,f)$ as corresponding to the rest of the subspaces. 

Formally, given a routed map $(\lambda, f): A^i \rightarrow B^j$, $f$'s practical input space is given by 
\begin{equation}
    A^\textnormal{prac} := \bigoplus_i \  \alpha_i \cdot A_\star^i
\end{equation}
where $\alpha_i$ is given by
\begin{equation} \label{alphaj}
     \alpha_i := \sum_j \lambda_i^j.
\end{equation}

In (\ref{alphaj}), the sum is Boolean, i.e.\ given by $0+0=0$, $0+1=1$, $1+0=1$, $1+1=1$. $A^\textnormal{prac}$ is the orthogonal subspace of $A$ in which $f$ is not constrained by $\lambda$ to have null support. One defines the practical output space in an exactly analogously way, this time by taking the Boolean sum $\sum_i \lambda^j_i$.

\subsection{Routed circuits}

It is proven in \cite{vanrietvelde2020routed} that sectorized Hilbert spaces and routed maps form a  dagger symmetric monoidal category, which in turn implies that routed maps have a sound representation in terms of circuit diagrams.

For example, we can give a routed circuit diagram for the superposition-of-paths transformation discussed above. To this end, we will say that Alice and Bob have the \textit{sectorized} Hilbert spaces:
\begin{equation} 
\begin{split}
    A^i := A^0_\star \oplus A^1_\star \\
    B^j := B^0_\star \oplus B^1_\star
\end{split}
\end{equation}
where we have defined:
\begin{equation}
    \begin{split}
        A^0_\star & := A^\textit{vac}_\star \\
        A^1_\star & := A^\textit{par}_\star \\
        B^0_\star & := B^\textit{par}_\star \\
        B^1_\star & := B^\textit{vac}_\star 
    \end{split}
\end{equation}
The scenario can be represented as a routed transformation $(\delta, U)$, where $U$ is the transformation (\ref{superposition-paths}), and $\delta$ is the Kronecker delta function. Diagrammatically, this is represented by a box for $U$ and a floating label for $\delta$:
\begin{equation} \label{sup_of_channels_no_im}
    \input{sup_of_channels.tikz}
\end{equation}

In the diagram, wires represent sectorized Hilbert spaces; putting two wires next to each other represents the tensor product of those spaces as previously defined (likewise for transformations). We omit superscripts on wires with trivial (i.e.\ one-sector) sectorizations below, as in the rest of the paper, but it should be remembered that the mathematical object represented is technically a sectorized space. 

The role of the route $\delta$ is to enforce the constraint that $U$ has no support on its output space outside the one-particle sector $(A\otimes B)^{\rm prac} := (A^0_\star \otimes B^0_\star) \oplus (A^1_\star \otimes B^1_\star)$, which serves as its practical output space. Since such a `delta-route' simply matches up the values of the indices, we can equivalently write (\ref{sup_of_channels_no_im}) using the convenient shorthand of \textit{index-matching}:
\begin{equation} \label{sup of channels}
    \input{sup_of_channels_im.tikz}
\end{equation}

Now, in what sense does (\ref{sup of channels}) offer an improved representation of superposition of paths scenario? By matching up the indices, the diagram indicates the Hilbert space $(A \otimes B)^{\rm prac}$ upon restricting to which $U$ is unitary.\footnote{In more detail: the \textit{accessible space} \cite{vanrietvelde2020routed}, which can always be defined formally using the routes, at the top of our diagram coincides with the practical output space of the unitary. This means that the diagram unambiguously and formally defines a unitary transformation with this output space.} And the index-matching makes it immediately clear that sectors of the larger Hilbert space $A \otimes B$ in which there is not exactly one particle are irrelevant. The dilemma with which we introduced this section has been overcome: we have achieved a representation that makes clear the physical unitarity without encouraging us to consider irrelevant degrees of freedom.

More generally, routed linear maps $(\lambda, U)$ with the property that $U$ defines a unitary upon restriction to the practical input and output spaces are called \textit{routed unitaries}.\footnote{Or, in the terminology of \cite{vanrietvelde2020routed}, \textit{practical unitaries}.} These transformations are very important for us, since we will use them to define causal relations in the presence of sectorial constraints.

Finally, we introduce \textit{mixed} routed quantum circuits, in which the boxes become routed quantum \textit{channels}, and the wires become spaces of linear operators over Hilbert spaces. The routes then work in the same way and have the same meaning, forbidding input spaces that are in (i.e.\ have support only in) specific subspaces to be sent to other subspaces.\footnote{In \cite{vanrietvelde2020routed}, routes in the mixed case were made more expressive by having them also encode so-called \textit{coherence constraints}. This procedure, which involves a doubling of the spaces' and routes' indices, is unnecessary for our needs in this paper; we will therefore keep working with standard routes in the mixed theory as well, to avoid clutter, as was already done in \cite{vanrietvelde2021universal}. To translate the notations here to those of \cite{vanrietvelde2020routed}, one just has to replace $A^i$'s with $A^{ii'}$'s and $\lambda_i^j$'s with $\lambda_i^j \lambda_{i'}^{j'}$'s.}. To differentiate between unitary operators and unitary channels, we reserve calligraphic letters for the latter. For example,
\begin{equation}
    \input{sup_of_channels_chan.tikz}
\end{equation}
is a routed unitary channel.

Before moving on, a quick note on notation and terminology. Throughout this paper, we will stick to  using roman letters like $A$ to denote a Hilbert space. In a helpful abuse of notation, we will sometimes use the same letter to denote the system that that Hilbert space represents. Similarly, $A^i$ is used both to represent a sectorized Hilbert space and the corresponding \textit{sectorized system} -- that is, a system that suffers from sectorial constraints.

%Given a routed channel $(\lambda, \Phi): A^{i_1i_1'}_I \rightarrow A^{i_2i_2'}_O$, where $A^{i_2i_2'}_O$ is a copy of $A^{i_2i_2'}_I$, there is a particular choice of $\lambda$ that will be of interest to us. This is called a \textit{full-coherence delta route} and is given by $\delta_{i_1}^{i_2} \delta_{i_1'}^{i_2'}$. This route ensures that $\Phi$ is \textit{sector-preserving}, in the sense that any input states on $(A_I)^i_\star$ alone are mapped to $(A_O)^i_\star$, but it does nothing more: in particular, it enforces no constraints on the coherence between different sectors. By Theorem 6 of \cite{vanrietvelde2020routed}, $\Phi$ follows this route if and only if each of its Kraus operators $K_n$ in any Kraus representation is block diagonal in the sense that $K_n \in \bigoplus_{i_1i_2} \delta_{i_1}^{i_2} \ A^{i_1}_I \rightarrow A^{i_2}$.

\section{Causal influence through transformations}    \label{sec:first order}
What does it mean to say that some system $A$ exerts a causal influence on another system $D$? Let us assume, for the time being, that we can model $A$ as an input and $D$ as an output of some transformation.  In this case, an obvious idea is:

\begin{quote}
    \textit{$A$ influences $D$ just in case it is possible for information to flow from $A$ to $D$ through the transformation that connects them.}
\end{quote}

In the quantum case, this suggests we should define causal relations as signalling relations through quantum operations. However, in our view, not all quantum operations are suitable to define causal relations. This is because we assume that the existence of a causal relation should be independent of our state of knowledge, but general quantum transformations -- and the opportunities for signalling that they afford -- do depend on what we know.

For example \cite{barrett2020quantum}, if $A$ is the control input and $D$ is the target output of a quantum CNOT channel, then $A$ can signal to $D$ through this transformation. Hence we can reasonably say that $A$ is a cause of $D$. But then suppose that we have no idea about what happens at the target input, so we describe the situation with the channel obtained by preparing it in a maximally mixed state. Through \textit{this} channel, $A$ does \textit{not} signal to $D$. If we took signalling relations in general quantum channels (or instruments) to define causal relations, then we could legitimately say that $A$ is \textit{not} a cause of $D$. 

The upshot is that, if general quantum operations defined causal relations, then there could be disagreement about whether $A$ is a cause of $D$, without either of the options being objectively wrong: they would just correspond to different states of knowledge. This relativity is incompatible with our preferred perspective that causal relations are `out there' as objective features of the natural world, waiting to be discovered.\footnote{Or at least, as objective features relative to some division of the world into subsystems.} %\footnote{We note that this argument against defining causal relations with respect to non-unitary operations relied on an example in which a maximally mixed state that only serves to account for our ignorance and does not describe any objective property of nature is inserted into one part of a unitary. But if one believes that quantum pure states are ontic (meaning that two different pure states of the theory always correspond to two different states of nature), then the quantum channels obtained by inserting correct pure states into some input subsystems of a unitary do not depend on what we know. So one could remain consistent with our argument while defining causal relations in terms of a channel that arises from inserting pure states into input subsystems of a larger unitary, although one would have to refrain from doing so when mathematically the same channel arises from epistemic states. But we don't find this a particularly attractive proposal since (i) it makes the set of operations that bear causal relations depend on how those operations arise rather than just on their inherent mathematical form, and (ii) one then misses out on most of the appealing properties of causal relations defined in terms of unitaries that are described in the coming pages.}

To avoid this relativization of causal structure to states of knowledge, we assume that quantum causal relations correspond to signalling relations in a \textit{unitary} channel. This approach is consistent with the view that unitary processes (or at some suitable properties thereof) are objective features of the world.%\footnote{It is not, however, compatible with the view that non-unitary completely positive maps are part of the fundamental dynamics of the world.}

Such causal relations are the subject of this section. Firstly, we will focus on standard unitary channels whose input and output subsystems correspond to tensor factors of the overall Hilbert spaces, summarising the salient features of the framework presented in \cite{barrett2020quantum}. We will then generalize to the case of the routed unitaries described above.

\subsection{No-influence through an unrouted unitary channel}

We have argued that quantum causal relations should be understood in terms of the possibility of signalling through unitary transformations. But we have not spelled out precisely what this means, either by laying out exactly what resources one should be granted in attempting to send messages from $A$ to $D$, or by translating the idea into mathematics. In this subsection, we will offer six distinct notions of the lack of causal relation, which we call \textit{no-influence through an unrouted unitary channel}. 

Some of the notions correspond to quite distinct ideas of what it means to signal, and some are not obviously formulations of no-signalling at all. Remarkably, however, they all turn out to be equivalent. This is good news for our programme of defining causal relations as signalling relations through unitaries, since it means that there is one notion of no-influence that stands out as particularly natural, rather than a number of competing alternatives.

After defining no-influence, we will describe some of the most important consequences of the definition. In doing so, we will provide a compact summary of those features of the framework for quantum causal models developed in \cite{Allen_2017, barrett2020quantum} that are most relevant to our needs, as a preparation for their subsequent extension to systems suffering sectorial constraints.

An obvious candidate formulation of the claim that $A$ exerts no causal influence on $D$ through a unitary $U: A \otimes B \rightarrow C \otimes D$ is given by the following diagram, in which $\discard$ represents the trace operation \cite{Allen_2017}:
\begin{equation} \tag{U1} \label{U1}
    \input{NI-def.tikz}
\end{equation}

This says that the channel obtained by tracing out $C$ is equivalent to a channel that traces out $A$. Intuitively, the notion of no-influence being presented here is that if we only care about the non-effect, we may as well forget about the non-cause.

A closely related notion is that the channel obtained by tracing out $C$ should be independent of any local channel $\Phi$ applied to system $A$ before $U$:
\begin{equation} \tag{U2} \label{U2}
  \forall \Phi, \, \,  \input{NI_U2.tikz}
\end{equation}
 
This condition has a very clear operational meaning -- it says that no local interventions on a non-cause can be used to send signals to the non-effect.
 
 A third notion of no-influence from $A$ to $D$ is based on the intuition that if a (mixed) product state is fed into the channel, then the resulting reduced state at $D$ is independent of the choice of state at $A$:
 \begin{equation} \tag{U3} \label{U3}
 \forall \, \rho, \, \rho', \, \sigma, \, \,   \input{NI_U3.tikz}
\end{equation}
Note that this is a conceptually distinct notion of signalling to (\ref{U2}), since one might naively imagine a situation in which one can only signal from $A$ to $D$ via applying a local channel on $A$ to a \textit{nonseparable} state on $A \otimes B$.

While they have the advantage of being obviously related to signalling, all of the candidate no-influence relations so far make essential use of the trace operation $\discard$. This might give these conditions a somewhat anthropocentric feel, given that the trace is often taken to represent the ignorance of an agent. Given that we are trying to get away from the idea that causal relations have some essential connection to agents, one might prefer to give a no-influence relations at the level of pure unitary quantum theory. One way of doing this is to require the unitary to decompose in the following way.

\begin{equation} \tag{U4} \label{U4}
\exists \, V, \, W, \, \,  \input{NI_U4.tikz}
\end{equation}

The intuition here is that $A$ does not influence $D$ just in case there is some decomposition of the unitary into a unitary circuit in which there is no path of wires connecting $A$ and $D$. Here, a causal relation is understood as primarily a \textit{compositional} property of a unitary channel, rather than a signalling relation.

%Conditions (\ref{U1} - \ref{u4}) have all appeared in various places in the literature on causality in quantum foundations. Remarkably, they all turn out to be equivalent, as stated in Theorem \ref{equivalence-u1-u6}. Our only novel contribution for this section is to show they are also equivalent to two further statements involving the alebras of operators over $A$ and $D$. 

%The relevant algebras are the spaces of linear operators over $A$ and $D$, $\ca := \cl(A)$ and $\cd:=\cl(D)$ respectively. These are finite-dimensional Von Neumann algebras, closed under taking sums, matrix multiplications, and adjoints of the operator elements. The first of the novel conditions states that any local unitary operator $V \in \ca$ on $A$ acts trivially on $D$ once transformed via $\cu$ into the output space of the channel:

Another way to define no-influence without the use of the trace is to consider how the unitary channel $\cu:=U(\cdot)U^\dag$ acts on the local unitary operators on $A$. Specifically, we might require that for all unitaries $V\in \cl(A)$, there exists a unitary $V'\in \cl(C)$ such that:
\begin{equation} \tag{U5} \label{U5}
\input{NI_U5_v2.tikz}
\end{equation}

The intuition behind (\ref{U5}) is explained as follows. A local unitary intervention applied to $A$ before the implementation of $U$ results in a transformation $U (V \otimes I)$. By the unitarity of $U$, this is equivalent to $\cu(V \otimes I)  U = (V' \otimes I)  U$. So (\ref{U5}) corresponds to the intuition that $A$ is not a cause of $D$ just in case the result of applying a local unitary intervention on $A$ before the implementation of $U$ is equivalent to that of doing nothing before the implementation of $U$, and instead just unitarily operating on output systems other than $D$.

Our final notion of no-influence concerns the operator algebras $\ca := \cl(A)$ and $\cd:=\cl(D)$. Specifically, we require that the following commutation relation holds:
\begin{equation} \tag{U6} \label{U6}
    [U(\mathcal{A} \otimes I_B )U^\dagger, I_C \otimes \mathcal{D}] = 0
\end{equation}

In identifying causal relations with commutation relations, this last notion of no-influence corresponds to well-established intuitions about causality in quantum field theory.

%We note that (\ref{U1} -- \ref{u4}) have all appeared in the literature on quantum causality before -- our only novel contributions here are the last two notions of no-influence. 

Remarkably, every one of the intuitions discussed here leads to the precisely the same no-influence relation:

\begin{theorem}\label{equivalence-u1-u6}
Conditions (\ref{U1} - \ref{U6}) are all equivalent.
\end{theorem}

Theorem \ref{equivalence-u1-u6} is proven in Appendix \ref{appendix:equivalence-u1-u6}. This allows for a formal definition of no-influence through an unrouted unitary channel. 
\begin{definition} \label{definition:unroutedni}
    No-influence through an unrouted unitary channel. Consider a unitary transformation $U: A \otimes B \rightarrow C \otimes D$. We say that $A$ has no influence on $D$ through $U$, written $A \not\xrightarrow{U} D$, just in case conditions (\ref{U1} - \ref{U6}) hold. 
\end{definition}

We say that $A$ is a cause of $D$ through $U$, written $A \xrightarrow{U} D$, just in case it is not true that $A \not\xrightarrow{U} D$.

We thus have a notion of no-influence that is motivated by a number of distinct intuitions. It turns out that this notion leads to two very interesting and useful properties, to which we now turn.

\begin{proposition}[Time symmetry] \label{prop:time_sym_unrouted} $A$ is not a cause of $D$ through $U$ if and only if $D$ is not a cause of $A$ through $U^\dag$. That is,
\begin{equation} \label{eq:timesymunrouted}
    A \not\xrightarrow{U} D \Longleftrightarrow D \not\xrightarrow{U^\dagger} A ,
\end{equation}
\end{proposition}

This property is immediate from (\ref{U4}), as well as (\ref{U6}). The property feels quite intuitive in a time symmetric universe -- if the dynamics are reversible, and we define causal relations in terms of those dynamics, then it appears natural that those causal relations should be reversible too. 

However, remarkably, the dynamical time symmetry of a theory does not always imply that its causal structure is time symmetric, meaning that the latter is a distinctive feature of quantum theory. In particular, we note that classical causal relations, defined as dependencies in a reversible function, are not time-reversible. For example, in a classical CNOT gate, there is signalling from the control to the target. But in the time-reversed function (which is just the CNOT), there is no signalling from the target to the control. Proposition \ref{prop:time_sym_unrouted} thus provides a sense in which quantum theory actually feels much more natural than classical theory\footnote{Note, however, that time-symmetry of causal relations holds for the theory of classical \textit{symplectic transformations} \cite{friend2022private}.}.

The second useful property is the \textit{atomicity} of causal structure in unitary transformations \cite{barrett2020quantum, composableconstraints}.\footnote{The term `atomicity' is not used in \cite{barrett2020quantum}, even though the notion is introduced there (see Remark 4.4). The name was coined later in \cite{composableconstraints}.}

\begin{proposition}[Atomicity] \label{unrouted atomicity}
No-influence relations in a unitary transformation can be composed and decomposed in the following ways:
\end{proposition}
\begin{equation} \label{dag1unrouted} 
        (A \not\xrightarrow{U} E) \land (A \not\xrightarrow{U} F) \Longleftrightarrow (A \not\xrightarrow{U} E \otimes F)
\end{equation}
\begin{equation} \label{dag2unrouted} 
        (A \not\xrightarrow{U} F) \land (B \not\xrightarrow{U} F) \Longleftrightarrow (A \otimes B \not\xrightarrow{U} F)
\end{equation}
This proposition is proven in Appendix \ref{appendix:unrouted atomicity}.

Atomicity says that if a system $A$ cannot influence either system $E$ or system $F$ individually, it cannot influence the composite system $E \otimes F$. This is a remarkable property since there are pure states on $E \otimes F$ for which much (if not all) of the information is only present in the nonlocal correlations between the subsystems, and does not show up in the individual reduced density matrices of the subsystems. Thus, naively, one could imagine that $A$ could influence $E \otimes F$ without influencing its subsystems by only affecting the nonlocal correlations (indeed, this can happen for signalling relations in general quantum channels). Yet atomicity shows that this is impossible: the nonseparability of pure states does not lead to a similar sort of nonseparability at the level of causal structure.

Atomicity is also very useful since it enables us to represent the overall \textit{causal structure} of a unitary transformation using a directed graph between its individual, `atomic' input and output subsystems. This is a very important theme for us, so it is worth dwelling on. Consider a unitary transformation of the following form:
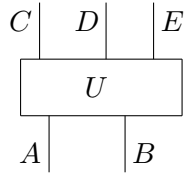
\begin{equation} \nonumber
    \input{atomicty_example_uni.tikz}
\end{equation}

We now know how to tell whether there is causal influence between any pair of input/output subsystems -- we just check using whichever of (\ref{U1}--\ref{U6}) is most convenient. We can thus write down six causal influence or no-influence relations, which can conveniently be represented with a directed acyclic graph with arrows representing the influence relations. For example, if our only  \textit{no}-influence relations are  $A \not\xrightarrow{U} D$, and $A \not\xrightarrow{U} E$, we can represent all six relations as follows:
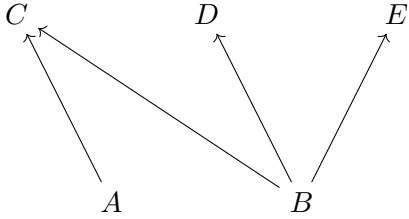
\begin{equation} \nonumber
    \begin{tikzpicture}
    \node (a) at (-2.5, 0) {$A$};
    \node (b) at (2.5, 0) {$B$};
    \node (c)at (-5, 5) {$C$};
    \node (d) at (0, 5) {$D$};
    \node (e) at (5, 5) {$E$};
    
    \draw[->] (a) to (c);
    \draw[->] (b) to (e);
    \draw[->] (b) to (d);
    \draw[->] (b) to (c);
    \end{tikzpicture}
\end{equation}

If we didn't know about atomicity, then we might think that there are still more facts about the causal structure that aren't captured by this graph. In particular, we might imagine that $A$ is a cause of the composite system $D \otimes E$. But if we bear atomicity in mind, this graph tells us that $A \not\xrightarrow{U} D \otimes E$, even though there is no node representing $D \otimes E$. Thus a directed acyclic graph between the five subsystems from the original decomposition of the unitary represents a much larger set of facts about the causal structure. 

More generally, atomicity implies that the causal relations among arbitrary subsets of the input/output subsystems are fixed by the causal relations among the individual subsystems. This justifies us in simply  \textit{defining} the causal structure of a unitary transformation as this directed acyclic graph:

\begin{definition}
 The causal structure of a unitary transformation is directed acyclic graph representing the causal relations between its individual input/output subsystems.
\end{definition}

While this notion of causal structure is good as far as it goes, it does not apply to situations where sectorial constraints are at play. In the next subsection, we generalize the notion to \textit{routed} unitaries, in order to accommodate them.

%since it means that the causal relations between arbitrary tensor products of inputs and outputs are all fixed by the no-influence relations between the individual `atomic' inputs and outputs considered in the original decomposition of the unitary transformation. This in turn imples that the causal structure of a unitary channel can be fully captured by a directed acyclic graph, where the input and output systems are the vertices. Further down the line, this same property enables one to represent any causal structure in the unrouted quantum causal models framework using a directed graph \cite{barrett2020quantum, barrett2021cyclic}.

\subsection{No-influence through a routed unitary channel} \label{sec:ni_ruc}

As (\ref{U2}) makes clear, no-influence can be understood as the independence of measurements on the non-effect on transformations on the non-cause. This suggests that if the possible transformations or measurements on subsystems are \textit{restricted}, different causal relations will arise. But our current definition of no-influence does not accommodate such restrictions; indeed, (\ref{U2}) quantifies over \textit{all} quantum channels on $A$. The unfortunate implication is that the existing no-influence condition will attribute the wrong causal relations to scenarios featuring restrictions on transformations and measurements.

As a concrete example, suppose that $A$ and $D$ are the control input and the target output respectively of a quantum CNOT gate. According to our current definition, $A$ is a cause of $D$. But now suppose that the only measurements one can perform on $D$ are measurements of the orthonormal basis vectors $\frac{\ket{0} \pm \ket{1}}{\sqrt{2}}$ (perhaps because of a superselection rule). In this case, all the possible measurements on $D$ are independent of any transformations on $A$. Then, presumably, we should say that $A$ is \textit{not} a cause of $D$. We need a more general notion of no-influence to accommodate this.

We can understand this restriction on measurements as the result of a sectorial constraint. Note that, due the Naimark dilation, any measurement on a system $D$ can be performed by implementing a channel with an ancillary output $D'$, and then measuring $D'$. Diagrammatically, this channel has the form
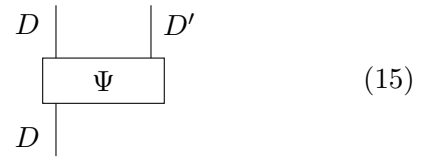
\begin{equation} \label{dilation channel}
\input{measurement_dilation_channel.tikz}
\end{equation}

Implicitly, (\ref{U2}) assumed that any measurements of this form on $D$ can be performed, since it requires that its reduced state should be independent of the channel $\Phi$ on $A$. Let us now restrict ourselves to the \textit{sector-preserving} measurements on a sectorized version of the system, $D^l$. These are the measurements that can be implemented using a channel of the form (\ref{dilation channel}) that maps each of $D^l$'s sectors to itself. In other words, sector-preserving measurements are the ones that can be performed using a routed channel of the form:
\begin{equation}
    \input{sp_channel_im_2.tikz}
\end{equation}

Suppose $D^l$ is a sectorized qubit with two sectors, spanned by $\frac{\ket{0} + \ket{1}}{\sqrt{2}}$ and $\frac{\ket{0} - \ket{1}}{\sqrt{2}}$ respectively. In this case, the measurements on $D^l$ that one can perform via measurements of $D'$ with this sort of channel are precisely the measurements of (probabilistic mixtures of) these two states. We can therefore accommodate the constraint from the previous example by restricting to sector-preserving channels.

More generally, we also want to consider sectorial constraints on the putative cause systems. We thus restrict to the sector-preserving channels $\Phi$ on the sectorized system $A^i$. These are the ones for which
\begin{equation}
    \begin{tikzpicture}
	\begin{pgfonlayer}{nodelayer}
		\node [style=map] (0) at (0, 0) {$\Phi$};
		\node [style=none] (1) at (0, 2) {};
		\node [style=none] (2) at (0, -2) {};
		\node [style=none] (3) at (0.75, 1.5) {$A^i$};
		\node [style=none] (4) at (0.75, -1.5) {$A^i$};
	\end{pgfonlayer}
	\begin{pgfonlayer}{edgelayer}
		\draw (2.center) to (1.center);
	\end{pgfonlayer}
\end{tikzpicture}

\end{equation}
is a valid routed channel.

We can therefore consider a causal relation between sectorized systems as the possibility of signalling using sector-preserving transformations on the cause and sector-preserving measurements on the effect. Since we are now dealing with sectorized systems, we can define the no-influence relations with respect to routed unitaries of the following form:
\begin{equation} \label{routed unitary}
    \input{routed_unitary.tikz}
\end{equation}

We can formulate the resulting notion of \textit{no-influence through a routed unitary} $(\lambda, U)$ from $A^i$ to $D^l$ as follows:

\begin{equation} \label{R1} \tag{R1}
  \forall \, \Phi, \, \Psi, \, \,   \scalebox{.8}{\input{NI_R1.tikz}}
\end{equation}
Applying this definition of no-influence to the example above, we indeed find that there is no influence from $A$ (which has a trivial sectorization) to the now-sectorized system $D^l$.

As in the case of unrouted unitary channels, this definition of no-influence turns out to be equivalent to a number of conceptually distinct formulations. We now lay these out, beginning with  a generalization of (\ref{U5}). This notion is once again motivated by the intuition that applying (now sector-preserving) local unitaries on the non-cause and then applying $U$ should be equivalent to applying $U$  and then some unitary that only acts trivially on the non-effect. 

However, since the non-effect is now a sectorized Hilbert space, we do not require that that this last unitary is confined to the $C$ wire. Instead, it can take the more general form of a family of unitaries on $C$ that is coherently controlled on the sector of $D$. More precisely, we can have something of the form $V' = \sum_l V_C^l \otimes \pi_D^l$, where each $V_C^l$ is a unitary on $C$, and each $\pi_D^l$ projects onto the $l$th sector of $D^l$. We can give this second notion of no-influence through a routed unitary the following diagrammatic representation:\footnote{We note that the equality here only holds between the unitary circuits, and not necessarily between the routes. That is why we allow a $D^l$ at the bottom right of the right side of the equation, which is index-matched with the $D^l$ at the top right, and a $D^n$ in the corresponding position on the left side, which may or may not be matched up with $D^l$ via the route $\lambda$.}

\begin{equation} \label{R2} \tag{R2}
\forall \,  V, \, \exists \, V', \, \,   \scalebox{.9}{\input{NI_R2.tikz}}
\end{equation}

Above, $V$ is any unitary operator on $A$ that is block diagonal in its sectorization. $\lambda^T$ denotes the transpose of $\lambda$. $W$ is a unitary matrix that maps each sector $D_\star^l$ to $\dot{D}^l_\star \otimes D^l_\star$, where $\dot{D}^l_\star$ is one-dimensional. Accordingly, if $D^l$ has $n$ sectors, then $\dot{D}^l$ is a sectorized Hilbert space with $n$ one-dimensional sectors. The map between $W$ and its adjoint on the RHS is a unitary operation $V'$ on the left wire coherently controlled by the value of $\dot{D}^l$ in the basis given by its sectorization.

A third notion of no-influence through a routed unitary is a generalization of the commuting-algebras formulation (\ref{U6}). For this, we need to consider \textit{sector-preserving} operator algebras, which are not generally factor algebras of the form $\cl(H)$. For a sectorized Hilbert space $A^i$, the corresponding sector-preserving algebra has the form:
\begin{equation} \label{spalgebra111}
    \ca := \bigoplus_i \cl(A_\star^i)
\end{equation}

$\ca$ can be embedded into an algebra of operators $\ca^*$ on our routed unitary's entire input space. $\ca^*$ is defined from $\ca$ using an `embedding map', defined in Appendix \ref{appendix:embedding}. The third notion of no-influence can then be written:
\begin{equation} \tag{R3} \label{R3}
    [U \ca^* U^\dagger, \cd^*] = 0
\end{equation}
where $\ca^*$ is the algebra for $A^i$ embedded into $U$'s input space, and $\cd^*$ is the algebra for $D^l$ embedded into $U$'s output space.

Finally, there is a last notion of no-influence which uses a special unitary transformation called the \textit{exchange gate}. We leave the formal definition of this gate to the next subsection, in which it will play a more central role. Suffice it to say for now that the exchange gate for some sectorized system can be intuitively regarded as the unitary acting on that system together with and unsectorized ancillary that allows them to exchange the maximum possible amount of information while respecting the sectorial constraints. Our final notion of no-influence through a routed unitary is that there should be no signalling from the ancillary input of the exchange gate on $A^i$ to the ancillary output of the one on $D^l$:
\begin{equation} \label{R4} \tag{R4}
  \scalebox{.9}{\input{NI_R4_v2.tikz}}
\end{equation}

Above, the route $\mu_{ij}:=\sum_{kl}\lambda_{ij}^{kl}$ simply restricts to $\cu$'s practical input space, and $\Phi$ is some quantum channel.

In the special case where the sectorization is trivial (i.e.\ involves just one sector), the exchange gate is just a swap. Hence, (\ref{R4}) is a generalization of (\ref{U1}).\footnote{Formally, the diagram obtained from (\ref{R4}) by assuming all routed and sectorizations are trivial (and therefore substituting swaps for exchanges) and (\ref{U6}) are slightly different, since the former includes ancillary systems such as $A^O$ that do not feature in the latter. But it is easily shown that each one can be derived from the other.}

Although motivated by distinct intuitions, each of these notions turns out to be equivalent, just like in the unrouted case.
\begin{theorem} \label{theorem:R1 - R3 eq}
(R1 - R4) are all equivalent.
\end{theorem}

Theorem \ref{theorem:R1 - R3 eq} is proven in Appendices \ref{appendix:proof-fg-equiv} and \ref{appendix: proof routed NI equivalence}. Thus there is a clear, natural choice for notion of no-influence when sectorial constraints are at play, which we capture as follows.

\begin{definition}
 Given a routed unitary transformation (\ref{routed unitary}), we say that $A^i$ has no influence on $D^l$ in $U$, written $A^i \not\xrightarrow{U} D^l$, just in case the conditions (R1 - R4) hold.
\end{definition}

We say that $A^i$ is a cause of $D^l$ through $U$, written $A^i \xrightarrow{U} D^l$, just in case it is not true that $A^i \not\xrightarrow{U} D^l$.

%The reader may surmise that this definition makes no mention of the route $\lambda$ associated with $U$ -- this is fine since changing the (valid) routed that one ascribes to $U$ does not affect the truth or falsehood of statements (R1 - R4).

The reader will notice that we have not provided generalizations of (\ref{U3}) or (\ref{U4}) for the routed case.
There does not appear to be any reasonable generalization of (\ref{U3}). One reason for this is that the practical input space of a routed unitary is not generally the tensor product of its subsystems, and, accordingly, starting with a product state $\rho_A \otimes \sigma_B$ that lies in the space and then changing $\rho_A$ to some $\rho_A'$ might take you out of that space. And anyway, this sort of switching should not be permitted since it cannot be brought about with a sector-preserving channel. We draw the moral that in the general case, causal relations are best thought of in terms of transformations rather than states. The question of whether (\ref{U4}) can be generalized, and shown to be equivalent to (\ref{R1} -- \ref{R4}), is left open for future work.

The next obvious question is whether this more general notion of no-influence leads to the same nice properties as its unrouted counterpart. It turns out that this is indeed the case: the no-influence relation through a routed unitary is time-symmetric and atomic.

\begin{proposition}[Time symmetry -- routed case]
Given the routed unitary (\ref{routed unitary}), $A^i$ is not a cause of $D^l$ in $U$ if and only if $D^l$ is not a cause of $A^i$ in $U^\dag$. That is,
\begin{equation} \label{eq:timesym} 
    A^i \not\xrightarrow{U} D^l \Longleftrightarrow D^l \not\xrightarrow{U^\dagger} A^i ,
\end{equation}
\end{proposition}

Time-symmetry follows immediately from (\ref{R3}).

For the atomicity property, consider a unitary transformation of the following form (expressed in the pure theory):
\begin{equation}\label{routed dag unitary}
    \input{routed_dag_unitary.tikz}
\end{equation}

Recall that, just like unsectorized Hilbert spaces, we can take a tensor product of sectorized Hilbert spaces as in \ref{partioned-H-space-otimes}. We then have the following proposition, proven in Appendix \ref{appendix:ni-compositionality}.

\begin{proposition}[Atomicity -- routed case] \label{ni-compositionality}
No-influence relations in the routed unitary (\ref{routed dag unitary}) can be composed and decomposed in the following ways:
\end{proposition}
\begin{equation} \label{dag1} 
        (A^i \not\xrightarrow{U} E^m) \land (A^i \not\xrightarrow{U} F^n) \Longleftrightarrow (A^i \not\xrightarrow{U} E^m \otimes F^n)
\end{equation}
\begin{equation} \label{dag2} 
        (A^i \not\xrightarrow{U} F^n) \land (B^j \not\xrightarrow{U} F^n) \Longleftrightarrow (A^i \otimes B^j \not\xrightarrow{U} F^n)
\end{equation}

Just like the unrouted case, this property allows us to represent the causal structure of a routed unitary as a directed acyclic graph between its input and output subsystems. Since we will define a fine-graining of this structure in the next subsection, called the `sectorized' causal structure, we call this the `unsectorized' causal structure of a routed unitary.

\begin{definition} \label{theorem:NIrouteddag}
The unsectorized causal structure of a routed unitary transformation is a directed acyclic graph representing the causal relations between its individual input and output subsystems.
\end{definition}

\subsection{Sectorized no-influence through a routed unitary transformation} \label{sec: trans fg}

Suppose we know that $A^i$ influences $D^l$ through a routed unitary. (\ref{R1}) then implies that it is possible for associated agents to send and receive messages by performing local, sector-preserving interventions on $A^i$ and $D^l$. However, one might also want to know whether it is possible to send messages via a restricted set of interventions on $A^i$ that only act nontrivially on a specific sector, such as $A^0_\star$. If so, one could further ask whether sending such messages is still possible if we additionally require that the local effect of the intervention on $D^l$ is only to modify the relative phases between the different sectors, and to do nothing to states that have support within only one sector. 

More generally, we can ask about not only whether one entire sectorized system influences another, but also whether there is influence between the systems' specific sectors and/or the relative phases between them. To tackle such questions, this section introduces \textit{sectorized} no-influence relations.

The essential tool here is the exchange gate, which we already mentioned in the previous subsection. Let us now describe its role and definition in detail, beginning with an intuitive account in terms of agents and their actions. Suppose that an agent Alice has unconstrained access to a system $A$. Alice wants to exploit, to the best of her ability, the possibilities for sending and receiving messages afforded by $A$. A natural strategy is to swap $A$ with an ancillary system, $A'$, of the same dimension. Then preparing $A'$ in her choice of initial state allows her to send arbitrary messages through $A$, and measuring the final state of $A'$ allows her to probe all the information that flowed into $A$.  Alice's use of the swap thus allows her to extract as much information as possible from $A$, and likewise insert as much information as possible into it.

In the presence of sectorial constraints, Alice is forbidden to use this strategy: the swap with an ancillary system is not sector-preserving on $A^i$. Intuitively, the point of the exchange gate is to provide her with the \textit{next best strategy}. In other words, the gate allows her to exchange the maximum amount of information possible between $A^i$ and an ancilla without violating the sectorial constraints. 

How can Alice do this? Taking $A^i$ to have $N$ sectors, one sensible approach is for her to keep in store not one, but $N$ ancillary systems, each corresponding to one of the sectors; she can then swap $A^i$ with the corresponding ancilla depending (coherently) on which sector is populated, e.g. swapping it with the first ancilla if it is in the first sector, etc. These ancillas will be called the \textit{sectorial ancillas}.

But swapping the individual sectors is not enough; one also needs to exchange the information contained in the relative phases between them. To this end, Alice will also need an $(N+1)$th ancilla of dimension $N$, on which a generalised CNOT gate is performed, whose control is the `which sector of $A^i$ is populated?' information. This $(N+1)$th ancilla will be called the `which-sector' ancilla $A_w$, as, when its input state was prepared in the computational basis, reading its output tells Alice about which sector of $A^i$ was populated (and decoheres this information). Alternatively, by preparing the input states of $A_w$ in the Fourier basis, Alice can introduce phase shifts between the sectors of $A^i$ through phase back-action.

Using both the sectorial ancillas and the which-sector ancilla is the best strategy for Alice: this is what the exchange gate represents. 

% We define the sectorized relations in terms of the exchange gate that featured in (\ref{R4}). We start by explaining what the gate does (a more thorough definition, that is useful for following proofs, is provided in Appendix \ref{appendix:xdeltadef}). 

% Intuitively, the exchange gate on a sectorized system $A^i$ and an ancilla $A'$ is the unitary transformation that exchanges the maximum amount of information between the two systems, whilst still being sector-preserving on $A$. If $A$ has $N$ sectors, then the ancilla factorizes into $N+1$ systems: $A'=A_1 \otimes ... \otimes A_N \otimes A_w$.  All the information in the $i$th sector $A_\star^i$ of $A^i$ is exchanged with the \textit{sectorial subsystem} $A_i$, while the information that isn't found within any particular sector is exchanged with the \textit{global subsystem} $A_w$. In particular, $A_w$ receives the information about which sector `happened', and affects the relative phase among the sectors.

Let us now turn to its mathematical definition.\footnote{An equivalent  definition, that is useful for following proofs, is provided in Appendix \ref{appendix:xdeltadef}.} The exchange gate acts on $A^i$ and $A'=A_1 \otimes ... \otimes A_N \otimes A_w$, and decomposes as a direct sum of $N$ terms, each acting on one sector. The $i$th term swaps the $i$th sector $A_\star^i$ of $A$ with $A_i$, increases the logical value of $A_w$ by $i$, and does nothing to the remaining sectorial wires. In other words, we have:
\begin{equation}
    \ket{m}_{A_w}\ket{\phi}_{\overline{A_i}} \ket{\xi}_{A_i} \ket{\psi_i}_A \xrightarrow{\texttt{EXCH}} \ket{m+i}_{A_w}\ket{\phi}_{\overline{A_i}}\ket{\psi_i}_{A_i} \ket{\xi}_A
\end{equation}
for a state $\ket{\psi_i}_A\in A_\star^i$ in the $i$th sector, a state $\ket{\phi}_{\overline{A_i}}$ on $\overline{A_i}:=\bigotimes_{j \neq i}A_j$, and the logical state $\ket{m}_{A_w}$. The exchange gate on a system with only one sector is just the swap. Hence the exchange gate is a generalization of the swap formally, as well as conceptually.

When writing the exchange gate as part of a circuit diagram, it is helpful to distinguish between $A'$ before and after the implementation of the exchange gate using the notations $A^O$ and $A^I$ respectively, and similarly for its subsystems. We can now rewrite the channel from (\ref{R4}) in terms of those subsystems:
\begin{equation} \label{sectorized rels} 
    \scalebox{.9}{\input{fine_grained_rels_v4.tikz}}
\end{equation}

This routed unitary is all we need to define the more fine-grained causal structure that we were looking for, as the exchange gate allows us to neatly differentiate between the causal influences that are mediated by each individual sector, and also by the `which-sector' information. As we already know how to define no-influence relations among the input/output subsystems of a routed unitary transformation, we can simply define a sectorized no-influence relation as the no-influence relation between a subsystem of $A'$ and $D'$ in the routed unitary above. For example, if $A_3^O \not\rightarrow D_w^I$ through the routed unitary in (\ref{sectorized rels}), then we take this to mean that the third sector of $A^i$ has no influence on which-sector information in $D^l$. If $A^i$ and $D^l$ have $N$ and $M$ sectors respectively, then we have a total of $(N+1)(M+1)$ sectorized no-influence relations.

Since they were defined in terms of our existing notion of no-influence through a routed unitary, the sectorized no-influence relations immediately inherit time-symmetry and atomicity. This means that we can define the sectorized causal structure as a directed acyclic graph, where a system with $N$ sectors is represented by $N+1$ vertices.
\begin{definition} \label{def:fg cs}
The sectorized causal structure of a routed unitary is a directed acyclic graph with a vertex for each sector of each input/output subsystem, and a vertex for the which-sector information in each system. The  arrows represent the causal relations between them.
\end{definition}

\begin{figure*}
\begin{equation} \tag{a}
    \input{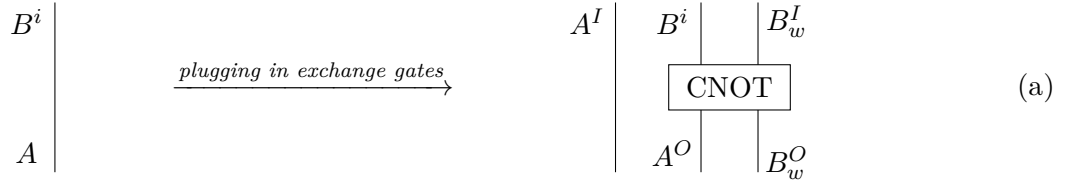} \qquad \ \ \ \ \  \xrightarrow{\textit{plugging in exchange gates}}  \ \ \ \ \  \qquad \input{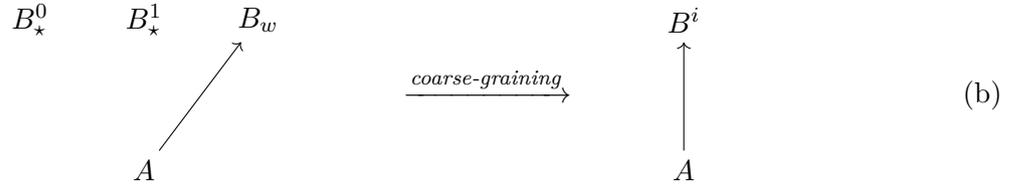}
\end{equation}
\newline
\newline 
\begin{equation} \tag{b}
    \begin{tikzpicture}
    \node (a) at (0, -2) {$A$};
    \node (b0) at (-3, 2) {$B_\star^0$};
    \node (b1) at (0, 2) {$B_\star^1$};
    \node (bw) at (3, 2) {$B_w$};
    \draw[->] (a) to (bw);
    \end{tikzpicture}  \ \ \ \ \ \ \ \ \ \ \  \xrightarrow{\textit{coarse-graining}} \ \ \ \ \  \ \  \  \begin{tikzpicture}
    \node (a) at (0, -2) {$A$};
    \node (b) at (0, 2) {$B^i$};
    \draw[->] (a) to (b);
    \end{tikzpicture}
\end{equation}  \caption{A routed unitary illustrating the need to include the which-sector ancilla in the exchange gate. The routed unitary on the left of (a)  is just an identity channel from a qubit $A$ with only one sector (so that we can omit an index) to a qubit $B^i$ sectorized into two one-dimensional sectors. The exchange gate on an unsectorized system is just the swap, and the exchange gate on a sectorized qubit is the CNOT controlled on that qubit. Hence putting in exchange gives us the right side of (a) where the control input and output to the CNOT is $A^O$ and $B^i$ respectively (the wires of the exhange gate corresponding to one-dimensional sectors are themselves one-dimensional, so they can be ignored). Evaluating the no-influence conditions on this second transformation gives us the more fine-grained sectorized causal structure on the left of (b), which yields the coarse-grained, unsectorized causal structure on the right when we combine the vertices associated with $B^i$. \newline \ \ \ \ Since they are both one-dimensional, there is obviously is no influence from $A$ to either of the sectors. Thus, if we omitted the which-sector information $B_w$, we would conclude that there is no influence at all from $A$ to $B^i$. But this would be wrong, since Alice can send a message to Bob by preparing $A$ in either $\ket{0}$ or $\ket{1}$, which Bob can then read by performing a (sector-preserving) computational basis measurement on $B^i$.
Therefore, not all the information in $B^i$ is encoded within its sectors: there is also information about which sector happened.} \label{fig:need which sector}
\end{figure*}

\begin{figure*}
\begin{equation} \tag{a}
    \input{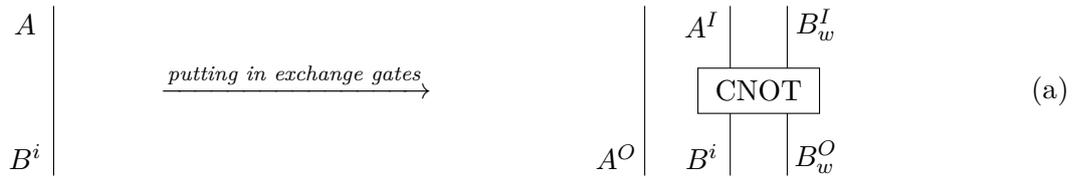} \ \ \ \ \ \ \ \ \ \  \xrightarrow{\textit{putting in exchange gates}}  \ \ \ \ \ \ \ \ \ \  \qquad \input{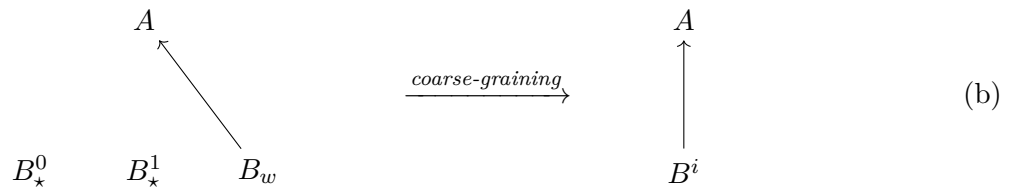}
\end{equation}
\newline
\newline 
\begin{equation} \tag{b}
    \begin{tikzpicture}
    \node (a) at (0, 2) {$A$};
    \node (b0) at (-3, -2) {$B_\star^0$};
    \node (b1) at (0, -2) {$B_\star^1$};
    \node (bw) at (3, -2) {$B_w$};
    \draw[->] (bw) to (a);
    \end{tikzpicture}  \ \ \ \ \ \ \ \ \ \ \  \xrightarrow{\textit{coarse-graining}}   \ \ \ \ \  \ \  \    \begin{tikzpicture}
    \node (a) at (0, -2) {$B^i$};
    \node (b) at (0, 2) {$A$};
    \draw[->] (a) to (b);
    \end{tikzpicture}
\end{equation}  \caption{A routed unitary that illustrates the need to consider influence through relative phases, hence the which-sector ancilla. The routed unitary on the left of (a)  is just the time reverse of that from Figure \ref{fig:need which sector}. From time-symmetry, it follows that the sectorized and unsectorized causal structures are obtained by reversing the arrows in Figure \ref{fig:need which sector}. Again, the which-sector information turns out to be essential for getting the right unsectorized structure; this time, because Bob can signal to Alice only by acting on the relative phase between the two sectors.} \label{fig:need which sector 2}
\end{figure*}

We also note that there is an unsectorized causal influence between an input and output sectorized system if and only if there is at least one corresponding sectorized causal relation. This follows from (\ref{R4}), atomicity, and the fact that the ancillary wires of the exchange gate are the tensor products of the $n+1$ wires that define the sectorized causal structure. This is equivalent to the following proposition, which makes clear the sense in which the sectorized causal structure is a fine-graining of the unsectorized one.

\begin{proposition} \label{cg from fg trans}
The unsectorized causal structure of a routed unitary transformation is obtained from the sectorized causal structure by combining all the vertices that correspond to the same sectorized system, while preserving the edge structure.
\end{proposition}

It is worth elaborating on what the sectorized causal relations mean, and why we need to consider the which-sector ancilla $A_w$.  Causal influence on a sector means the possibility of affecting the information within that sector. Now, since the total state space of the sectorized system is just the direct sum of its sectors, one might naively think that if none of its sectors are influenced by some potential cause, then the sectorized system itself is not influenced. But a moment's reflection shows that this is false, since a cause might be able to affect \textit{which} sector happens, and yet have no effect at all on the information contained within the sectors. And the question of which sector happens can indeed be settled with a sector-preserving measurement. 

All of this is particularly obvious if all the sectors are one-dimensional, so that they contain no information; an example is provided in Figure \ref{fig:need which sector}. This shows that accounting for the which-sector information via the which-sector ancilla is crucial for Proposition \ref{cg from fg trans} to hold, and therefore for our sectorized causal structure to deserve the status of a fine-graining of the unsectorized one. 

Similarly, the causal influence exerted by a sectorized system is not just the combination of that exerted by its sectors, since one can also change the relative phases between the sectors via a sector-preserving transformation. This is once again accounted for by the which-sector ancilla. An example is provided in Figure \ref{fig:need which sector 2}, which is just the time-reversed version of the previous example.

We briefly note that a number of alternative formulations of the sectorized no-influence relations are available. It is possible to associate each of the vertices in a sectorized causal structure with an algebra of operators -- in fact, a subalgebra of the sector-preserving algebra for the corresponding system. This algebra defines the class of channels and unitaries one quantifies over in expressions like (\ref{R1}) and (\ref{R2}), which then turn out to be equivalent to the no-influence relation provided in this subsection, assuming that any sectors involved have more than one dimension. This is discussed in Appendix \ref{appendix:proof-fg-equiv}.

\section{Causal structure in supermaps} \label{sec:higher order}

The no-influence relations introduced so far only allow us to consider a very limited range of causal structures. For it seems perfectly reasonable to consider causal structures, such as $A \rightarrow B \rightarrow C$, in which some systems act as both causes and effects. But in a unitary transformation, each subsystem can only act as a cause \textit{or} as an effect, depending on whether it is an input or an output of the unitary.

For an example of a scenario which might exhibit this causal structure, suppose that Alice and Charlie are separated not just by unitary channels, but by an intervention that can be freely chosen by a third agent, Bob. The situation is depicted below, where the gap in the wire represents Bob's ability to choose an arbitrary operation with which to fill it.

\begin{equation} \label{comb}
    \input{comb.tikz}
\end{equation}

Our existing notion of no-influence might help us determine whether $A$ influences $B$, and whether $B$ influences $C$.
But how do we tell whether $A$ influences $C$? Our existing notion of no-influence doesn't apply, since the systems are not the input and output of a single unitary transformation. 

This section extends our framework to deal with this more general sort of scenario. The diagram (\ref{comb}) can be formally interpreted as a quantum \textit{supermap} \cite{Chiribella_2008} \footnote{An alternative, equivalent representation is that of \textit{process matrices} \cite{oreshkov2012quantum}. See footnote \ref{foot: process matrices} for a brief comment on the connection between the two pictures, or Appendix A of \cite{vanrietvelde2022consistent} for an extensive conceptual and mathematical treatment of this connection.}. Taking this as a hint,  we start by characterizing the causal structure of supermaps, once again re-expressing compactly the central ideas of \cite{barrett2020quantum}. This involves reformulating these ideas in the language of supermaps, rather than in terms of the process matrices used in \cite{barrett2020quantum}. We then generalize to \textit{routed supermaps} \cite{vanrietvelde2021universal}, in which sectorial constraints can be accounted for, and in which a more fine-grained, sectorized causal structure can also be displayed.

\subsection{The causal structure of a standard supermap}

A quantum supermap transforms a set of quantum operations as an input into a single quantum operation as an output: it is an `operation on operations'. We will exclusive consider the deterministic supermaps defined precisely in \cite{Chiribella_2013}; the rough idea is as follows. A bipartite, deterministic quantum supermap on product channels, $\cs$, is a linear map of the form \begin{equation} \label{supermap form}
    \cs: \textnormal{Herm}(A^I \otimes B^I \rightarrow A^O \otimes B^O) \rightarrow \textnormal{Herm}(P \rightarrow F)
\end{equation}
where each of the italicised letters are Hilbert spaces, and $\textnormal{Herm}(X \rightarrow Y)$ is the space of Hermitian-preserving maps from $\cl(X)$ to $\cl(Y)$. To state a further requirement we consider an arbitrary collection of ancillary Hilbert spaces $\gamma_A^I, \gamma_A^O$, $\gamma_B^I$, $\gamma_B^O$, and an arbitrary pair of quantum channels:
\begin{equation} \begin{split}
    \cm_A: \cl(\gamma_A^I \otimes A^I) \rightarrow \cl(\gamma_A^O \otimes A^O) \\ 
    \cm_B: \cl(\gamma_B^I \otimes B^I) \rightarrow \cl(\gamma_B^O \otimes B^O)
\end{split}
\end{equation}

A bipartite deterministic supermap on channels is any linear map of the form (\ref{supermap form}) for which
\begin{equation}
    (\ci \otimes \cs \otimes \ci)[\cm_A \otimes \cm_B]
\end{equation}
is always a quantum channel, where the left and right $\ci$'s are the identities on $\textnormal{Herm}(\gamma_A^I\rightarrow \gamma_A^O)$ and $\textnormal{Herm}(\gamma_B^I\rightarrow \gamma_B^O)$ respectively. An $n$-partite, deterministic quantum supermap on product channels is then the obvious generalization of this idea. A general such supermap is depicted below, in which we consider an arbitrary factorization of its `past' $P$ and its `future' system $F$.
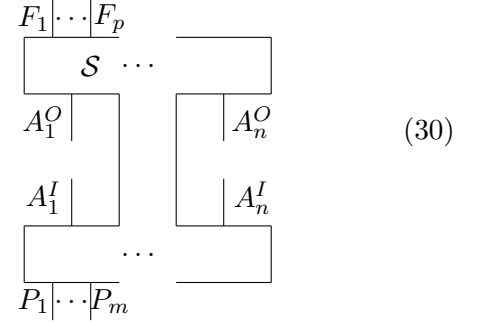
\begin{equation} \label{N-partite-supermap}
    \input{N-partite-supermap.tikz}
\end{equation}

Each node can be associated with an agent who chooses the intervention at that node. The commonly invoked intuition \cite{oreshkov2012quantum} is that each agent is in an isolated laboratory, the doors of which open only once to receive the input system associated with $A^I_l$, and then once more to eject the output system associated with $A^O_l$, leaving some time in between for the agent to perform an intervention.\footnote{If ancillas are used then these may be pictured as either residing in the lab the entire time, or as being received before the rest of the experiment and ejected afterwards.} These laboratories are what is represented by the nodes. Outside the labs these systems are allowed to interact, in a way described by the internal structure of the supermap.

Although we refer to agents for pedagogical purposes, we do not want the notion of agency to be essential to our framework. We should therefore always allow for the possibility that an agent does nothing at all, in which case it should be possible to regard the outgoing system as essentially the same as the ingoing one. For this reason, we assume here that $A_l^I$ and $A^O_l$ are of the same dimension for all $l$.

As in the case of transformations, we reserve the notion of causal structure for \textit{unitary} supermaps.\footnote{For supermaps this means either that it transforms $n$ unitary channels to another unitary channel, or that the associated process matrix (defined below) is unitary -- the two notions are equivalent \cite{araujo2017purification}.} Our reason for restricting to unitaries is the same as before -- we assume that unitary supermaps are the ontic supermaps that allow us to keep causal relations objective \cite{barrett2020quantum, Allen_2017}.

The objects that bear causal relations in a supermap are its nodes and its past/future subsystems. The relations can be defined in terms of the unitary channel obtained by pulling the outgoing wire of each node to the bottom of the page, and the ingoing wire to the top. This is called the \textit{process channel}\footnote{\label{foot: process matrices}The operator representing this channel via the Choi-Jamiołkowski isomorphism is commonly known as the \textit{process matrix} or \textit{process operator}, introduced in \cite{oreshkov2012quantum} and commonly used in the literature on indefinite causal order.}:

%It can be easily shown that, for a given input state global $\ket{0}_P$ if each agent performs a non-destructive measurement on their system, the probability of a particular set of outcomes $\{k_l\}_{l=1}^n$ obtaining is given by inserting the corresponding CP maps $\{\cm_i\}_{l=1}^n$ into the nodes and tracing out the resulting state -- this is called the \textit{generalized born rule}.
\begin{equation} \label{n-partite proc matrix}
    \input{n-partite_proc_matrix.tikz}
\end{equation}

This corresponds to each agent choosing a special intervention associated with each node: a swap onto an ancillary system of the same dimension as $A^I_l$. 

At this point, we already know what it means for an input to a unitary channel to influence some output. Hence we can define a \textit{direct cause} relation in terms of this influence relation. Specifically, the $i$th node is a direct cause of the $j$th node if and only if $A_i^O$ is a cause of $A_j^I$ in the process channel. We can also define direct cause relations involving the past/future subsystems: for example, the $i$th past system $P_i$ is a direct cause of the $j$th node if and only if $P_i$ is cause of $A_j^I$ in the process channel.

There is a nice intuition behind this definition. When each agent implements a swap, they prevent any information from flowing from the input systems of the nodes to the outputs. This means that any possibilities for signalling between agents at the nodes via their ancillas are entirely due to the internal structure of the supermap. In this case, the ability of one agent to signal to another is not dependent on, or mediated by, the other agents allowing information to flow through the node.

%The second point is that the definition is motivated a posteriori by its useful consequences, especially its operational consequences. In \cite{barrett2020quantum} the authors lay out a number of nontrivial operational constraints imposed by the direct cause relations of a supermap. The result is that the causal structure of a supermap serves as an effective compression of its operational consequences, which is just as it should be if causal structure is to be understood as properly scientific.

Moreover, when we define direct cause relations in this way, they inherit the atomicity property from the relation of no-influence through a unitary transformation. The direct causal relations between arbitrary subsets of the nodes and past/future subsystems are therefore fixed by the direct causal relations between the individuals. This enables us to define the causal structure of a supermap as a directed graph, where the vertices are the individual supermap nodes (since the $A_i^O$ and $A_i^I$ vertices can be merged together) and the past/future subsystems. Indirect cause relations are then defined as directed paths in this graph.

The causal structure of the supermap is easily obtained from the causal structure of its process channel:  

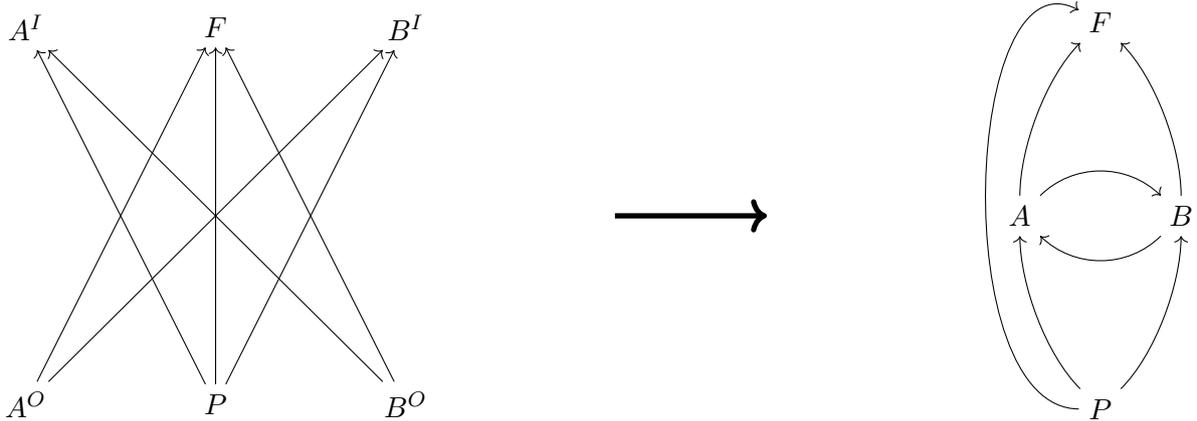
\begin{figure*}
    \centering
    \begin{equation} \nonumber
        \begin{tikzpicture}
        \node (ao) at (-5, -5) {$A^O$};
        \node (p) at (0, -5) {$P$};
        \node (bo) at (5, -5) {$B^O$};
        \node (ai) at (-5, 5) {$A^I$};
        \node (f) at (0, 5) {$F$};
        \node (bi) at (5, 5) {$B^I$};
        
        \draw[->] (ao) to (f);
        \draw[->] (p) to (f);
        \draw[->] (bo) to (f);
        \draw[->] (ao) to (bi);
        \draw[->] (bo) to (ai);
        \draw[->] (p) to (ai);
        \draw[->] (p) to (bi);
        \end{tikzpicture} \qquad  \qquad  \qquad  \begin{tikzpicture}
\draw[->, line width=0.7mm] (0,0) -- (4,0);
\end{tikzpicture} \qquad \qquad  \qquad 
        \begin{tikzpicture}[node distance={15mm}] 
\node (1) {$A$}; 
\node (6) [above right of=1] {};
\node (2) [above of=6] {$F$}; 
\node (5) [below right of=1] {};
\node (4) [above right of=5] {$B$}; 
\node (3) [below of=5] {$P$}; 

\draw[->] (1) to [out=45,in=135,looseness=1] (4); 
\draw[->] (4) to [out=225,in=-45,looseness=1] (1); 

\draw[->] (1) to [out=90, in =225, looseness=0.75] (2); 
\draw[->] (4) to [out=90, in=315, looseness=0.75] (2); 
\draw[->] (3) to [out=135, in=270, looseness =0.75] (1); 
\draw[->] (3) to [out=45, in=270, looseness=0.75] (4); 

\draw[->] (3) to [out=180, in =150, looseness=0.85] (2); 
\end{tikzpicture} 
    \end{equation}
    \caption{Moving from the causal structure of the process channel to the causal structure of the supermap. On the left-hand side, we have a possible causal structure of the process channel for a bipartite unitary supermap, with nodes $A$ and $B$ (in particular, this is what we get for the quantum switch \cite{Chiribella_2013, barrett2021cyclic}). To get the causal structure of the supermap, we just need to combine $A^I$ and $A^O$ into a single vertex representing the node $A$, and $B^I$ and $B^O$ into a single vertex representing $B$.  }
    \label{fig:supermap cs example}
\end{figure*}

\begin{definition}
The causal structure of a standard unitary supermap is a directed graph, obtained by combining the vertices corresponding to the same supermap node in the causal structure of the process channel, while preserving the edge structure. 
\end{definition}
An example of this procedure is provided in Figure \ref{fig:supermap cs example} for the causal structure of the quantum switch \cite{Chiribella_2013}.

We note that, while the causal structure of the process channel will always be acyclic, the process of combining vertices might induce cycles in the causal structure of the supermap, as in Figure \ref{fig:supermap cs example}. This implies that the vertices cannot be ascribed a partial order such that effects are always higher up than their causes. Accordingly, if a supermap has a cyclic causal structure, we say that it exhibits \textit{indefinite causal order} \cite{barrett2021cyclic}.

\subsection{The causal structure of a routed unitary supermap}

The above notion of the causal structure of a supermap is good as far as it goes, but it fails to accommodate sectorial constraints. Yet, as we aim to demonstrate with our analysis of purported implementations of the quantum switch, accounting for these constraints can be important for understanding the causal structure of a physical scenario. We need a more general framework.

Our approach will be to extend the framework sketched above so that it applies to supermaps acting on routed maps, which we call \textit{routed supermaps}. 
Defined in \cite{vanrietvelde2021universal}, routed supermaps are basically the same as unrouted supermaps, except that the input channels to each node of a routed supermap are required to follow a route associated with that node. 

Since we still do not want to rely on a notion of agency, we will assume that the ingoing and outgoing sectorized Hilbert spaces associated with the $l$th node, $A^{i_l}_{I, l}$ and ${}A^{i_l'}_{O, l}$, are copies of the same sectorized Hilbert space. By this we mean that they are sectorized into the same number of sectors, and that the dimensions of the sectors $A^{i_l}_{I, l\star}$ and $A^{i_l'}_{O, l\star}$ are the same for all $i_l=i_l'$. We further assume that each node only accepts sector-preserving channels; that is, the ones that follow the route $\delta_{i_l}^{i_l'}$.\footnote{The causal structure of unitary supermaps with more general routes is an interesting problem for future work.}

\begin{figure} 
    \centering
    \input{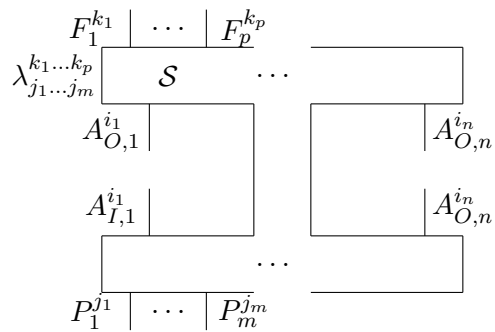} 
    \caption{An $n$-partite routed unitary supermap, represented in the pure theory. The $l$th node takes as inputs routed unitary transformations that map each $i_l$th sector to itself. Then the whole supermap outputs a routed unitary of the form $(\lambda, V)$, where $\lambda$ is some fixed route matrix, and $V$ is a unitary transformation that depends on the choice of inputs.}
    \label{routed supermap}
\end{figure}

As before, we assume that the routed supermaps that have causal structure are the routed unitary supermaps; that is, the ones that return routed unitary channels when a unitary channel is inserted into each one of its nodes. An example of the sort of supermap we are considering is illustrated in Figure \ref{routed supermap}, in which the matched indices represent the routes at the nodes.

How can we define direct causes in such a supermap? In the case of standard supermaps, we did this by inserting swaps into each node. But this trick won't work here, since the swap is not a sector-preserving intervention. Hence there is no guarantee that this will leave us with a routed unitary transformation -- in fact, there is no guarantee it will be a channel at all. Accordingly, inserting swaps does not necessarily leave us with an object with a causal structure that we know how to analyse. 

This is where the exchange gate introduced in Section \ref{sec: trans fg} once again comes in handy. As previously discussed, the exchange gate is the closest thing to the swap on a sectorized system that preserves its sectors. This suggests a promising strategy. We can form a \textit{routed process channel} by inserting exchange gates into each node:
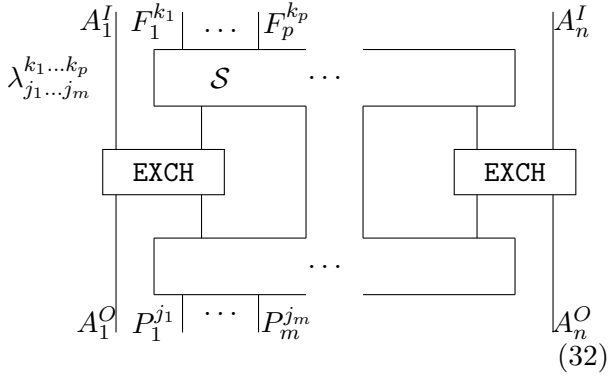
\begin{equation} \label{routed proc matrix}
    \input{N-partite_routed_proc_matrix.tikz}
\end{equation}

\begin{figure*} 
    \centering
    \input{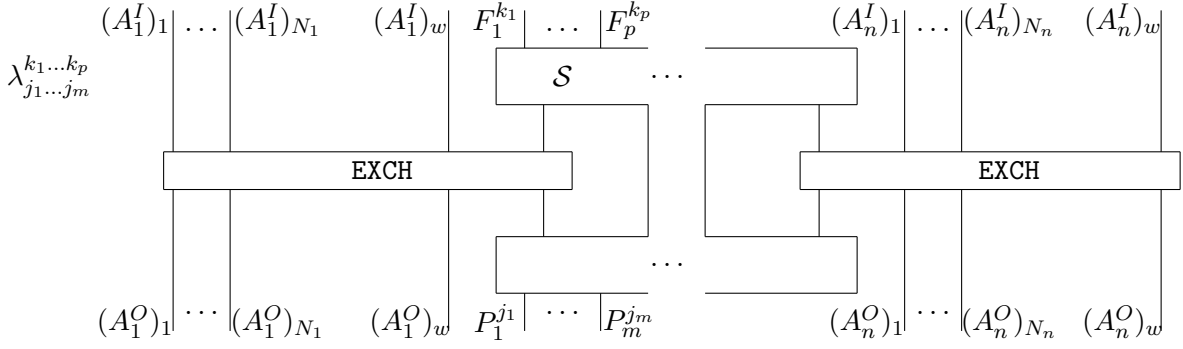}
    \caption{The routed process matrix with a fine-grained factorization of its input and output spaces. Direct causes between the sectorized relata of $\cs$ are defined in terms of no-influence relations between algebras over systems associated with the input and output wires in this circuit.}
    \label{fg factorization}
\end{figure*}

This is a routed unitary transformation, whose causal structure we know how to analyse. It reduces to the usual process channel for the special case of standard supermap  (which can be regarded as a routed supermap where all the sectorizations are trivial). If we define direct cause relations as signalling relations among the ancillary inputs and outputs of the exchange gates in this channel, we recover the intuition of a direct cause relation as the possibility of signalling even when agents extract and replace the maximum amount of information possible from their node.

This will be our approach. We say that the $l$th node is a direct cause of the $r$th node if and only if $A_l^O$ is a cause of $A_r^I$ in the routed process channel. We can similarly define direct causes involving past/future subsystems. For example, the $r$th past subsystem is a direct cause of the $l$th node if and only if $P_r^{j_r}$ is a cause of $A_l^I$ in the routed process channel. 

Recall that the atomicity of causal relations in a unitary channel meant that we could define the causal structure of an unrouted supermap as a directed graph, via the procedure sketched in Figure \ref{fig:supermap cs example}. A similar discussion applies in the routed case, leading to the following definition.

\begin{definition}
The unsectorized causal structure of a routed unitary supermap is a directed graph. The graph is obtained by combining the vertices corresponding to the same node of the supermap in the sectorized causal structure of the routed process channel, while preserving the edge structure. 
\end{definition}

As in the unrouted case, the process of combining pairs of vertices might lead to cycles, which we take to imply indefinite causal order.

Now recall that, for routed unitary transformations, the way the ancillary wires on the exchange gate factorized provided us with a way to define more fine-grained causal relations. We can provide a similar fine-graining of the causal structure of a routed unitary supermap. 

The first step is to write the process channel as in Figure \ref{fg factorization}, where we have decomposed the ancillary systems of each node's exchange gate into its subsystems. We then take the sectorized causal structure (from Definition \ref{def:fg cs}) for the decomposition of the routed unitary. (In this causal structure, for a past/future sectorized subsystem with $N$ sectors, there are $N+1$ vertices, whereas for a node with $N$ sectors, there are $2(N+1)$ vertices, since it is associated with the input and output wires  on its exchange gate.) Then, we just merge the input and output vertex for each ancillary system.

\begin{definition}
The sectorized causal structure of a routed unitary supermap is a directed graph. The graph is obtained from the sectorized causal structure of the routed process channel. It is obtained by combining each pair of vertices corresponding to the same sectorial ancilla, or to the which-sector ancilla, of the same node of the supermap, while preserving the edge structure. 
\end{definition}

The atomicity of causal relations through a routed unitary transformation then implies the following proposition, which justifies our characterisation of the sectorized causal structure as a fine-graining of the unsectorized one.

\begin{proposition} \label{theorem:FGfundamental}
The unsectorized causal structure of a routed unitary supermap is obtained from its sectorized structure by combining the vertices that correspond to the same node or input/output system of the supermap.
\end{proposition}

The distinction between the unsectorized and sectorized causal structures of a unitary supermap will be crucial for our analysis of alleged implementations of the switch. As we will soon argue, roughly speaking, the experiments only exhibit indefinite causal order at the unsectorized level. 

First though, we mention the time-symmetric nature of the causal structure of routed unitary supermaps. We can define the time-reversed version $\cs^\dag$ of a routed unitary supermap $\cs$ as the supermap satisfying
\begin{equation}
    \cs^\dag(\cu_1^\dag, \ldots, \cu_n^\dag) = \cs(\cu_1, \ldots, \cu_n)^\dag
\end{equation}
for any set of unitary channels (with ancillas) $\{\cu_k\}_k$ at the nodes. The routed process channel of $\cs^\dag$ is just the adjoint of the routed process channel for $\cs$, up to local unitary transformations applied to its unsectorized input/output subsystems. As explained in Appendix \ref{appendix:smap-timesym}, this implies the following.

\begin{proposition} \label{prop:smap-timesym}
The causal structure of the time-reverse of a routed unitary supermap is obtained from that of the original supermap by reversing the direction of the arrows.
\end{proposition}

%Since causal relations are not directly observable, definitions of causal structure are conventional, and in the final analysis they can only be justified by their relationships with detectable correlations. \cite{barrett2020quantum} provides such a justification for the case of definite causal order. Briefly, the key points are as follows. A unitary supermap has a causal structure represented by a DAG $G$ if and only the Choi-Jamiolkowski form of its process matrix takes the form of a product of commuting channels, where each channel is associated with one of the nodes and only depends on the parents of that node in the DAG. This can be used to prove that a variety of operational statements follow from certain structural features of $G$. The same operational statements are true of a non-unitary supermap which is \textit{compatible} with $G$, in the sense that the supermap admits a particular sort of purification in terms of a unitary supermap that has the causal structure $G$.

%****************************************************
\section{An application: the causal structure of purported implementations of the switch} \label{sec:switch}
%***************************************************

Lately, the quantum switch \cite{chiribella2009beyond, Chiribella_2013} has received considerable theoretical \cite{oreshkov2012quantum, paunkovic2020causal, barrett2021cyclic, Chiribella_2013, araujo2015witnessing, zych2019bell, moller2020gravitational} and experimental \cite{goswami2018indefinite, procopio2015experimental, rubino2017experimental, rubino2019experimental, cao2021experimental, nie2020experimental} attention. In particular, there has been an interesting debate \cite{oreshkov2019time, procopio2015experimental, paunkovic2020causal} about whether recent purported implementations of the quantum switch realize indefinite causal order. 

In this section and the next one, we apply our framework to offer a fresh perspective on the debate.
We provide a new model for the experiments called the \textit{routed switch}, which we argue better captures their causal structure than the standard quantum switch does.
% Our analysis makes three main contributions.
% Firstly, it offers a new distinction between weak and strong indefinite causal order, and suggests that the experiments \cite{procopio2015experimental, rubino2017experimental, rubino2019experimental} only realize indefinite causal order in the weak sense.
% Secondly, it indicates that the assumptions usually taken to motivate the view that the experiments realize indefinite causal order actually only suggest that they do so in this weak sense.
% Finally, it shows that the assumption that causal relations obtain between localized objects is not necessary to argue that they do not do so in any strong sense.
Our analysis allows us to conclude that there is a distinction to be made between a `weak' and a `strong' realization of indefinite causal order, with the current experiments \cite{procopio2015experimental, rubino2017experimental, rubino2019experimental} only realizing indefinite causal order in the weak sense. Importantly, to reach this conclusion, we do not need the assumption that causal relations should be defined as obtaining between objects that are localized in space and time. In fact, we argue using assumptions that were originally taken to motivate the view that the experiments do realize indefinite causal order, and show that these assumptions only point to a weak realization.
Thus, we push forward the debate by arguing that the conflicting \textit{a priori} assumptions usually thought to lead to different conclusions actually converge on the same one: that the experiments are not strong realizations of indefinite causal order.

\subsection{The experiments and the switch} \label{sec:rswitch reconstruction}

\begin{figure*} 
    \centering
    %\hspace{-2.0cm} 
    \input{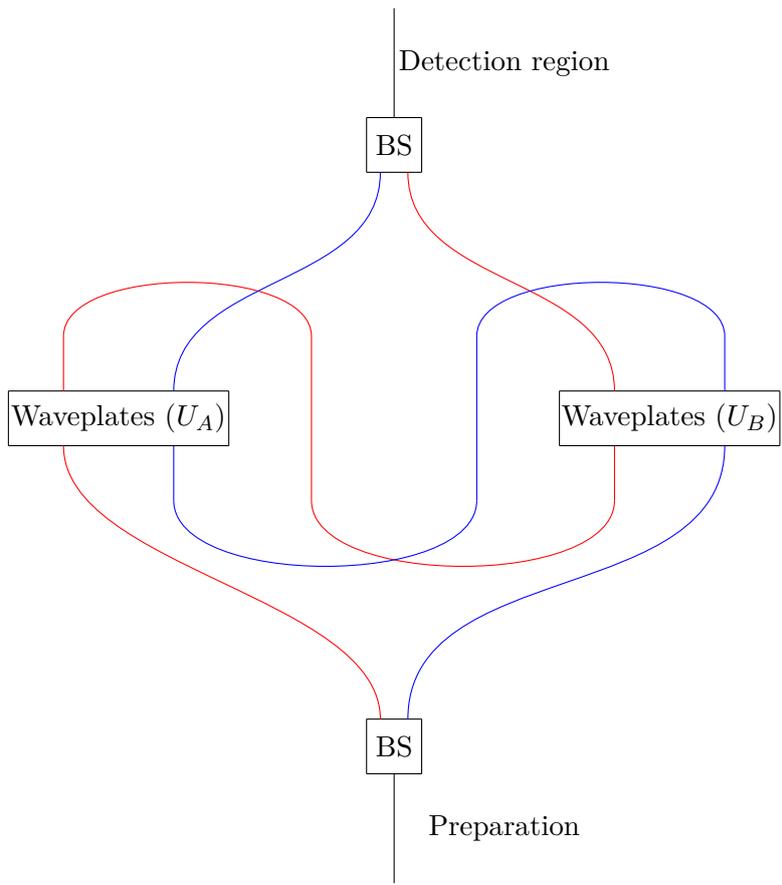}   \caption{The photonic experiments.}
    \label{fig:experiment_schematic}
\end{figure*}

Let us describe the relevant experiments. The ones \cite{procopio2015experimental, rubino2017experimental, goswami2018indefinite,  rubino2019experimental} that have excited the most controversy involve remarkably simple optical setups. 
The experimental setup common to \cite{procopio2015experimental, rubino2017experimental, rubino2019experimental} are depicted schematically in Figure \ref{fig:experiment_schematic}.
There are two sets of waveplates, each designed to implement a unitary transformation on the polarization degree of freedom of any photon passing through it.
We call these transformations $U_A$ and $U_B$, corresponding to the hypothetical agents Alice and Bob respectively.\footnote{In fact, the experiment in \cite{rubino2019experimental} involves two copies of this setup, where the ingoing photons are prepared in an entangled state. Accounting for this in the analysis does not lead to any substantially different conclusions.}

At the start of the experiment, a single photon is fired through a 50/50 beamsplitter, which sends it into a superposition of two paths. 
On the red path in the figure, the photon passes through Alice's waveplates first, and then through Bob's.
But on the blue path, the reverse is true: the photon passes through Bob's waveplates first, and then Alice's. The result is that the order in which the unitaries are applied to the photon's polarization  is \textit{coherently controlled} on its spatial degree of freedom.
More concretely, the unitary transformation on the overall Hilbert space $\ch^{\rm path} \otimes \ch^{\rm polarization}$ is given by the following function of $U_A$ and $U_B$:

\begin{equation} \label{switch def}
\begin{split}
        \texttt{SWITCH}(U_A, U_B) := & \ket{\color{red}{\rm red}}\bra{\color{red}{\rm red}}  \otimes U_BU_A \\ & \ \ + \ket{\color{blue}{\rm blue}}\bra{\color{blue}{\rm blue}}  \otimes U_AU_B
\end{split}
\end{equation}
In other words, the overall unitary transformation applied to the photon is precisely what one gets by modelling the experiment using the quantum switch supermap \cite{Chiribella_2013}. The isometric transformation applied to its polarization is obtained by inserting the state $\frac{\ket{\color{red}{\rm red}}+\ket{\color{blue}{\rm blue}}}{\sqrt{2}}$ to the left factor of $\texttt{SWITCH}(U_A, U_B)$.

At the end of the experiment, the two paths of the photons are recombined by another beamsplitter so that tests can be performed. In particular, causal witness tests \cite{araujo2015witnessing} are performed to verify that any bipartite unrouted supermap that accurately models the experiment must have a cyclic causal structure.

\subsection{The dynamical view vs. the spatiotemporal view}

The question is whether any of this implies that the experiments are physical realizations of indefinite causal order. We will reconstruct two opposing perspectives from the literature, which we call the \textit{dynamical view} and the \textit{spatiotemporal view}.\footnote{Note that, in providing this reconstruction, our primary objective is to articulate  what seem to us to be the most plausible detailed versions of the perspectives and arguments, in order to make progress in the debate. Our primary goal is not, therefore, to achieve perfect fidelity to the actual positions defended by the authors that inspired them. Although we think we have not strayed much from those actual positions, we leave it up to the reader to decide whether our eventual criticisms apply to those actual positions, or just to our particular attempt to reconstruct them in detail.} At the heart of the disagreement is how one should carve up the experiment into different \textit{causal relata}; that is, into different objects that bear the causal relations. 

\begin{figure}
    \centering
\begin{tikzpicture}[node distance={15mm}] 
\node (1) {$a$}; 
\node (5) [below right of=1] {};
\node (4) [above right of=5] {$b$}; 

\draw[->] (1) to [out=45,in=135,looseness=1] (4); 
\draw[->] (4) to [out=225,in=-45,looseness=1] (1); 
\end{tikzpicture} 
\qquad \ \ \ \  \ \ \ \
\begin{tikzpicture}
\node (x) {};
\node (a1) [below right of=x] {$a_1$};
\node (b2) [above right of=x] {$b_2$};
\node (b1) [right of=a1] {$b_1$};
\node (a2) [right of=b2] {$a_2$};

\draw[->] (a1) to (b2);
\draw[->] (b1) to (a2);
\end{tikzpicture}

    \caption{Intuitive depiction of the causal structure associated with the two- and four-relata views of alleged implementations of the switch. On the left-hand side, there is one relatum per gate, and each term in the superposition is associated with one of the arrows. On the right-hand side, there are two relata per gate, one for each time of implementation, and each term in the superposition is associated with one of the two islands in the graph.}
    \label{fig:intuitivedags}
\end{figure}
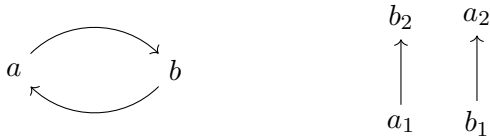

The \textit{dynamical view} is that the experiments realize indefinite causal order between two relata, one corresponding to  Alice's transformation, $U_A$, and the other to Bob's transformation, $U_B$. The intuition is depicted on the left side of Figure \ref{fig:intuitivedags}. 
More specifically, the dynamicist assumes that a causal relatum is\footnote{ \label{note:dynamicist} In fact, the dynamical view admits a number of possible variations, in which the causal relata are different. For example, they could be the sets of possible channels Alice and Bob could implement, or the (time-delocalized) systems that their transformations act on. Since these alternative relata are in one-to-one correspondence with the channel uses from the version of the view that we discuss, we have still have a total of two relata. And, by the same arguments we discuss below for the case of channel uses, these two relata can be held to give rise to the same causal relations and structures.} 
a single `use' of a channel, where `use' is somehow defined `operationally' \cite{araujo2014computational, procopio2015experimental, oreshkov2019time}. The challenge is then to say exactly just what this operational definition of the use of a channel could be. Unfortunately, this is far from straightforward in experiments that appear to realize indefinite causal order.

One approach defines the number of `uses' as the logical value of a hypothetical, auxiliary `flag' system affixed to the physical device responsible for implementing the channel \cite{araujo2014computational}. 
The device is hypothetically modified such that every time the channel is used, it raises the flag's logical value by one. 
Now, this definition might appear circular, since the definition of the modification seems to beg the question of when the channel is used.
But the idea is that we already know how to define channel use in the simple case of experiments with definite causal order, in which the channel can be considered as acting at a definite time. 
If we define the modification so that the flag correctly tracks the channel uses in these cases, then we more or less fix its behaviour in the cases where there is indefinite causal order.\footnote{There is some residual ambiguity in the flag's behaviour depending on (i) whether the raising of the flag is coherently or incoherently controlled on the use of the channel, and (ii) what the relative phase is if it is coherently controlled. But the precise definition one chooses does not matter much for the following discussion.}
We then take its behaviour in these conceptually less transparent scenarios to define the notion of channel use there.
Hence the flag definition of channel use provides a well-defined way of extrapolating our intuition from definite-order scenarios to indefinite-order scenarios.

This definition is `operational' in the sense that the flag could be constructed, and its  behaviour could be observed. 
However, one might hope that the definition of channel use could be made operational in another, perhaps stronger, sense.
In an experiment that can be described by a standard, acyclic quantum circuit, the circuit model makes clear what the operational significance of each channel is. 
That is, it tells the modeller which measurements one could perform to learn which channel was performed at a given point of circuit, or, conversely,  what implications a given channel has for various possible measurements.
One might hope for a similar characterization of the operational significance of channels used in an indefinite causal order, even though they cannot be represented as elements of a standard, acyclic circuit.

%In practice, the modification is defined such that it gives the obviously correct answer as to when the channel is used in the extreme and simple case where it is used at a definite time, and then the definition is extrapolated to cover cases where it is used at a superposition of times, as invariably happens in attempted tabletop realizations of indefinite causal order. However, as argued in \cite{oreshkov2019time}, this extrapolation from definite to superposed times is problematic, since, for example, there are many statements that are true of the logical states $\ket{0}$ and $\ket{1}$ but don't hold for arbitrary superpositions of them.

Considerable progress has been made in this direction in the bipartite and tripartite cases \cite{oreshkov2019time, wechs2023existence}. A key idea is that we can represent an experiment with a circuit that has a feedback loop; that is, a circuit with one of its outputs plugged into one of its inputs. (And we can relate this cyclic circuit to the standard, acyclic circuit model via a change in variables.) We can then take the individual mathematical transformations that build up the circuit to represent single uses of channels. This cyclic circuit can also be used to characterize the input and output subsystems of the transformations, and determine precisely which measurements tell us about these subsystems. We can thereby precisely characterize the operational significance of the channels themselves. We can then think of the use of channel as defined by the operational implications that we have carefully laid out.
 
Such operationally defined channels can act on  subsystems that are delocalised in time, and hence the channels themselves can be considered delocalized -- this is indeed the case for Alice's and Bob's channels in the photonic experiments. 

Once one assumes that the relata of causal relations are Alice's and Bob's operationally defined, time-delocalized channels, it seems fairly easy to argue that the experiments realize indefinite causal order.
For these channels are held to serve as inputs to the quantum switch, upon which they form an accurate model of the experiments.
Applying the framework for analysing the causal structure of standard supermaps sketched in Section \ref{sec:higher order}, one can show that the quantum switch has the cyclic causal structure on the right side of Figure \ref{fig:supermap cs example} \cite{barrett2021cyclic}. 
It seems to follow that the experiments themselves have a cyclic causal structure. 

In the actual experiments, a full tomography of the supermap is not performed, so it is not explicitly verified that the switch is a completely accurate model for the experiments (even if there are good theoretical reasons for believing as much).
However, causal witness tests \cite{araujo2015witnessing} are performed, which verify that \textit{if} any bipartite unrouted supermap accurately models the experiment, \textit{then} that supermap must have a cyclic causal structure.
So the dynamicist can argue that if the relata are two time-delocalized channels, the empirical data strongly suggests that there is indefinite causal order. 

We can now precisely formulate the dynamical point of view in argument form.

\begin{enumerate}
    \item The experiment consists of two operationally defined, time-delocalized channels.
	\item These channels serve as the relata of causal relations.
	\item  Moreover, these channels should be thought of as inputs to some bipartite unrouted supermap that accurately models the experiment.
	\item  The causal structure of the experiment is given by the causal structure of this supermap.
    \item Since the experiments certify a causal witness tests, it follows that they realize indefinite causal order.
\end{enumerate}

(1) and (2) are the central motivating assumptions of the dynamical view. (3) and (4) are auxiliary assumptions needed to make its conclusion (5) follow.

\begin{figure}
    \centering
    \hspace{-1.7cm} \input{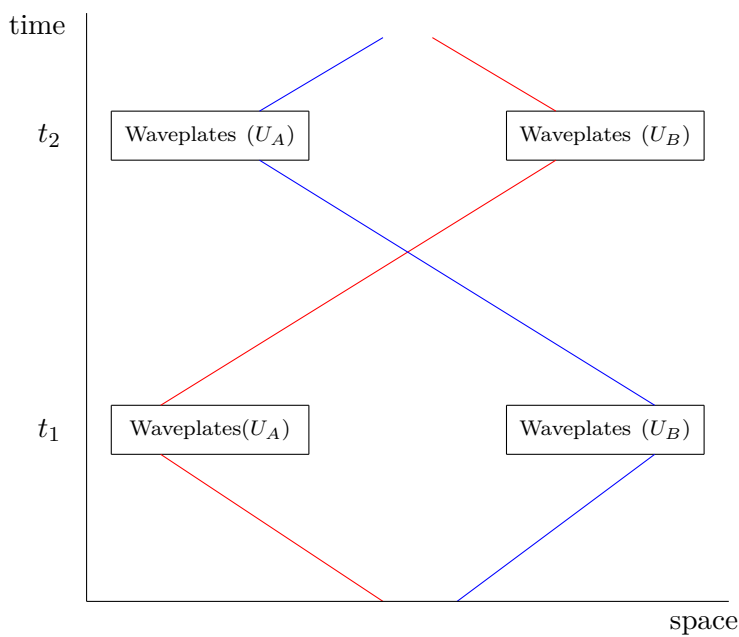}
    \caption{The spatiotemporal structure of the photonic expeirments.}
    \label{fig:experiment_spacetime}
\end{figure}

Let us now turn to a reconstruction of the rival perspective. The \textit{spatiotemporal view} \cite{paunkovic2020causal} is that the causal relata should be understood as spacetime events corresponding to the implementation of channels, so that they form a \textit{definite} causal order.\footnote{There exist variants of the spatiotemporal view in which the relata are not these spacetime events, but rather other objects in one-to-one correspondence with them. The approach in \cite{vilasini2017introduction, vilasini2020causality}, which takes signalling relations between localized systems as the most fundamental sort of causal relation, arguably falls into this category. The following discussion also applies to these variants (c.f.\ note \ref{note:dynamicist}).}  In the experiments from \cite{procopio2015experimental, rubino2017experimental, rubino2019experimental}, the photon enters Alice's gate at a different spacetime location depending on which of the superposed paths it is on, and likewise for Bob's, as illustrated in Figure \ref{fig:experiment_spacetime}.
So, even if we imagine that only two channels are implemented, each of them is implemented at two different spacetime events, which we call \textit{implementation events}. The four implementation events are taken to be the relata of causal relations. Assuming causal influences can only propagate inside the forward lightcone, it is not hard to see that this implies a definite causal order, as sketched intuitively on the right-hand side of Figure \ref{fig:intuitivedags}.
Note that the spatiotemporalist can consistently agree with the dynamicist on assumptions (1) and (3), while nevertheless rejecting (2) and (4).
%Due to the Minkowski structure of the spacetime in the lab, it is impossible to set things up so that the photon enters Alice's and Bob's gate at the same spacetime point regardless of the path, whilst maintaining that on one path it goes through Alice's first, and on the other it goes through Bob's first.

The dynamicist might argue that the different implementation events have no operational meaning \cite{paunkovic2020causal}. One attempt to rebut this is as follows. In principle, one could always check with a photon detector which of the two spacetime locations Alice's (or Bob's) channel was implemented at \cite{paunkovic2020causal}. Thus the two events are operationally meaningful, in the sense that they correspond to the two possible outcomes of this measurement. 

%The operationalist response to this argument \cite{oreshkov2012quantum, procopio2015experimental} can be understood as relying on a modified version of Leibniz's principles of the identity of indiscernibles. Leibniz argued that if two things could not even in principle be distinguished, then they should actually be argued as ontologically the same thing. Of course, Leibniz was not considering quantum mechanics, in which the process of distinguishing objects involve perform measurements that necessarily and sometimes radically alter their physical state. If one is interested in a particular sort of phenomenon in a quantum world, such as indefinite causal order, then one might want to require restrict the measurements that distinguish two things and thus prove their ontological distinctness to those that do not destroy the phenomenon that one is interested in. 

However, the dynamicist can object \cite{procopio2015experimental} that performing this measurement invariably destroys the interference between the paths. But it is only this interference that makes the causal witness tests that certify indefinite causal order possible.  
If the phenomenon that we are interested in is defined by the possibility of making this certification, it follows that it is impossible to measure which location the channel was implemented at \textit{without destroying the phenomenon of interest}. 
Conversely, assuming that the phenomenon of interest actually takes place, it is impossible to perform this measurement. 
By assuming that (a) this measurement is the only thing that could make the implementation events operationally meaningful, and (b) that measurements do not make something operationally meaningful if they destroy the phenomenon of interest, the dynamicist concludes that the implementation events are operationally meaningless.

But (a) is false: the operational content of the implementation events is not exhausted by measurements to find out which implementation event `happened' \cite{paunkovic2020causal}. In particular, \cite{paunkovic2020causal} proposes an experiment involving erasure in which the interference is not ultimately destroyed, and whose outcome is different depending on whether there are four implementation events or two (as in a gravitational realization of the switch \cite{moller2020gravitational}). Also, as we will argue below, the multiplicity of implementation events for each channel ensures that a larger range of operations can be performed by Alice and Bob than are typically accounted for.

Thus it cannot be maintained that the implementation events have no operational meaning.  But this does not prove the spatiotemporal view right, since it does not imply that the implementation events should be regarded as causal relata. After all, the dynamicist's operationally-defined channels also have operational meaning, so we have not demonstrated any sense in which the spatiotemporal view is superior. While the two  channels might each correspond to two operationally meaningful implementation events, the dynamicist can still legitimately maintain that the causal relata are those two channels. In that case, there are still only two relata, and they still exhibit an indefinite causal order.

We appear to have reached an impasse: unable to settle the dispute by appeal to operational meaningfulness, the dynamicist falls back on the \textit{a priori} assumption that causal relata correspond to channels, while the spatiotemporalist falls back on the \textit{a priori} assumption that causal relata are spacetime events.
The dynamicist's assumption seems to suggest there is indefinite causal order; the spatiotemporalist's assumption implies there is no such thing.

Here, we aim to break this impasse. 
%We offer an analysis of the photonic experiments described in \cite{procopio2015experimental, rubino2017experimental, rubino2019experimental}, suggesting that they only realize indefinite causal order in a weak sense. 
We will argue that the dynamicist's central assumptions (1) and (2) -- that there are two time-delocalized channels, and that these serve as causal relata -- do not in fact suggest that the experiments \cite{procopio2015experimental, rubino2017experimental, rubino2019experimental} should be understood as realizations of indefinite causal order in any strong sense. 
By the same token, we will show that the spatiotemporalist's central assumption -- that the causal relata should be localized in spacetime -- needs not be made in order to argue that the experiments are not strong realizations of indefinite causal order. 
Therefore, whichever starting assumptions one prefers, one should at most view the experiments as weak realizations of indefinite causal order.
We will make our case by motivating and analyzing a new model for the experiment called the \textit{routed switch}, to which we now turn.

\subsection{The routed switch and its causal structure} \label{sec:introducingrswitch}

One of our most important claims is the following. The quantum switch is both 
\begin{enumerate}[a)]
    \item an accurate model of the actual transformations experienced by the photon in the experiments; and 
    \item a bad supermap for reasoning about the causal structure of those experiments.
\end{enumerate}
%We noted in Section \ref{implementations_vs_realizations} that an experiment being an implementation of switch does not imply that it is a realization of indefinite causal order. Here, we argue that, although the switch provides an accurate model of the transformation experienced by the photon in the actual experiments, so that one could reasonably regard it as implemented by them, it nevertheless is not the best supermap for reasoning about their causal structure.

How can this be? If the switch tells an accurate story about the transformations that take place, why should we not trust what it says about causal structure? The answer is to be found in the intuition for causal influence sketched in Section \ref{sec:first order}. In a nutshell: causal influence is about how information \textit{can} flow, and not just about how it does flow. Therefore, one must take into account which transformations are \textit{possible}, rather than just those transformations that actually took place. This idea was reinforced by the definitions and examples of influence between sector-preserving channels from Section \ref{sec:first order}, where it was shown that the causal relations one attributes to a scenario can change due to a restriction on the set of transformations one calls `possible', even if the transformations that were actually performed are the same in both cases.

Moreover, to the extent that `indefinite causal order' is a concept of the foundations of  physics, the relevant notion of possibility here must be something like \textit{physical possibility}. 
Roughly, we should be concerned with what transformations could take place in keeping with the relevant physical laws, rather than with what transformations could take place given our choices, or given the particular tools we are using, or given our current level of technological sophistication.

\begin{figure}
    \centering
        \hspace{-1.7cm} \input{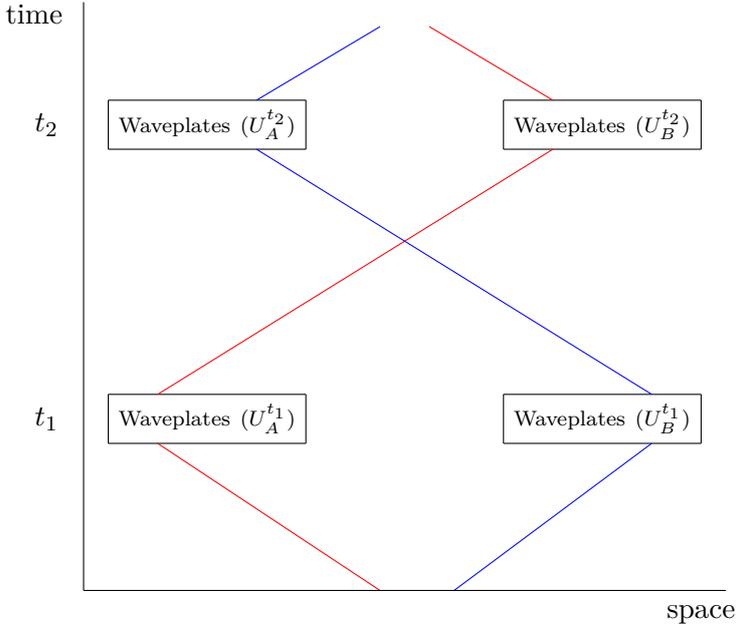}
    \caption{More general physically possible transformations, as implied by the spatiotemporal structure of the experiments.}
    \label{fig:experiment_spacetime2}
\end{figure}

And it turns out that the standard switch misses a wide range of physically possible transformations.
As the spacetime diagram in Figure \ref{fig:experiment_spacetime} makes clear, it is physically possible in the experiments from \cite{procopio2015experimental, rubino2017experimental, rubino2019experimental} for Alice to vary her transformation depending on the logical value of the control qubit. All she has to do is reach in after the first implementation event (when the photon enters her gate on the first path), and change her gate before the second implementation event (when the photon enters her gate on the second path). Bob could do the same thing, leading to the more general scenario depicted in Figure \ref{fig:experiment_spacetime2}. 

This possibility turns out to have important consequences for causal structure. Therefore, we generalize the switch to provide a model of the scenario described in Figure \ref{fig:experiment_spacetime2}. We want to allow transformations of the following form, where the agents can coherently control their choice of unitary based on the path.
\begin{equation} \label{routed switch inputs}
\begin{split}
      U_A &= \ket{\color{red}{\rm red}}\bra{\color{red}{\rm red}} \otimes U^A_{t_1} +  \ket{\color{blue}{\rm blue}}\bra{\color{blue}{\rm blue}}\otimes U^A_{t_2}\\
     U_B &= \ket{\color{red}{\rm red}}\bra{\color{red}{\rm red}}  \otimes U^B_{t_2} +  \ket{\color{blue}{\rm blue}}\bra{\color{blue}{\rm blue}}\otimes U^B_{t_1}
\end{split}
\end{equation}
(The subscripts refer to the time coordinates $t_1 < t_2$ of the different implementation events.) We do not, on the other hand, want to consider transformations that map the sector of $\ch^{\rm path} \otimes \ch^{\rm polarization}$ containing $\ket{\color{red}{\rm red}}$ to the one containing $\ket{\color{blue}{\rm blue}}$ or vice versa, since it is not physically possible, or even comprehensible, for the  agents to transform one path of the photon into another. So we only want to consider the transformations of the form (\ref{routed switch inputs}); that is, the sector-preserving transformations in the sectorization induced by the path. We allow these to serve as the inputs to a routed supermap called the routed switch, which is defined as follows. \footnote{More generally, the routed switch acts on unitaries with ancillas as     $\texttt{RSWITCH}(U_A, U_B) = \ket{\color{red}{\rm red}}\bra{\color{red}{\rm red}} \otimes (I_X \otimes U^{BY}_{t_2})(U^{AX}_{t_1} \otimes I_Y) \ \ + \ket{\color{blue}{\rm blue}}\bra{\color{blue}{\rm blue}} \otimes (U^{AX}_{t_2}  \otimes I_Y) (I_X \otimes U^{BX}_{t_1}$. This expression also allows us to calculate the action of the routed switch on arbitrary channels via their stinespring dilation.}

\begin{equation} \label{routed switch def}
\begin{split}
    \texttt{RSWITCH}(U_A, U_B) = &\ket{\color{red}{\rm red}}\bra{\color{red}{\rm red}} \otimes U^B_{t_2}  U^A_{t_1} \\ & \ \ + \ket{\color{blue}{\rm blue}}\bra{\color{blue}{\rm blue}} \otimes U^A_{t_2}  U^B_{t_1}
\end{split}
\end{equation}

\begin{figure} 
    \centering
    \input{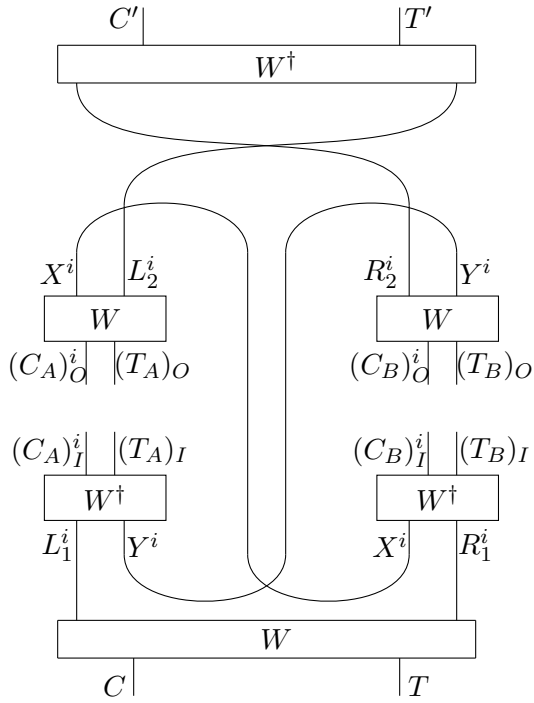}
    \caption{The routed switch. This is a bipartite supermap, in which we have factorized each node's ingoing and outgoing spaces into a sectorized control qubit and target system, e.g. $A_I^i=(C_A)^i_I \otimes (T_A)_I$. Thus each party has access to a sectorized copy of the control qubit. The sectorization is into one-dimensional sectors spanned by logical states , so the index matching on $(C_A)^i_I$ and $(C_A)^i_O$ restricts Alice to transformations of the form \ref{routed switch inputs}. $(\delta, W)$ is the superposition-of-paths routed unitary from (\ref{sup of channels}) (where we have relabelled $P\rightarrow T$). The wires bent into `cup' or `cap' shapes formally represent (the unnormalised) perfectly correlated entangled kets and bras respectively. When $i=0$, the particle exits through the left side of each $W$. This leads to the particle passing through Alice's node first, and then Bob's. The opposite is true for $i=1$. Inserting the transformations (\ref{routed switch inputs}) recovers the transformation (\ref{routed switch def}).
    %The past/future subsystems $P = C_p \otimes T_p$ and  $F = C_f \otimes T_f$ are tensor products of a control qubit and an $n$-dimensional target system. $L_1^k$ is an $(n+1)$-dimensional Hilbert space sectorized into an $N$-dimensional particle sector $L_{1\star}^1$ and a one-dimensional vacuum sector  $L_{1\star}^0$. $R_1^l$, $L_2^o$, $R_2$, $X^n$, and $Y^m$ all have the same structure and in each case the particle sector is associated with the superscript `1' and the vacuum with `0'. $A^i$ and $B^j$ are $2N$-dimensional Hilbert spaces with a preferred sectorization into two $N$-dimensional sectors that correspond to their order. $U:P \rightarrow (L_1^1 \otimes R_1^0) \oplus (L_1^0 \otimes R_1^1)$ is the `superposition of channels' unitary discussed in \cite{vanrietvelde2020routed}; it sends the state from $T_p$ to the particle sector of $L_1^k$ or $R_1^l$ depending on the logical value of $C_p$, it follows the route $\omega := (0, 1, 1, 0)$. $V$ is a unitary transformation that embeds the particle sector of its left input into the $1$ sector of its outputs, and the particle sector of its right input into the $2$ sector of its output. It follows the route $\epsilon$ which is given by $\epsilon^1_{10}=\epsilon^2_{01}=1$, $\epsilon^i_{kn}=0$ else. $\delta$ is the Kronecker delta function. Upside-down boxes denote adjoints, T denotes the transpose of a matrix. 
    }
    \label{fig:routedswitch}
\end{figure}

A circuit decomposition of the routed switch is provided in Figure \ref{fig:routedswitch}. Note that if we make the agents lose access to the control qubit, by requiring that $U^A_{t_1}=U^A_{t_2}$ and $U^B_{t_1}=U^B_{t_2}$, then we recover the standard switch, so the routed switch is strictly more general than the standard switch. Note also that the routed switch can in turn be recovered from even more detailed models underlying the standard switch, accounting for even more physical possibilities, such as its `causal box' \cite{portmann2017causal, vilasini2017introduction, vilasini2020causality}.\footnote{Should we then be using one of these more detailed models, since we have just empahsised the importance in accounting for physical possibilities? There are two reasons we stop at the routed switch. Firstly, the routed switch can be recovered from the more detailed models by assuming that agents cannot create or destroy particles, which is appropriate if one wants to think about the causal structure of the scenario as the possibilities for signalling afforded by the \textit{particular} photon that is fired through the beamsplitter at the start of the experiment. Secondly, as we will discuss below, the fact that the routed switch still takes time-delocalized channels as inputs makes its use compatible with the dynamicist's central assumptions, which in turn makes the conclusions of our analysis more significant.}

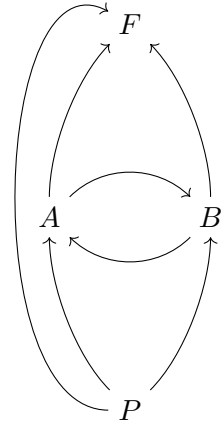
\begin{figure} 
\centering
\begin{tikzpicture}[node distance={15mm}] 
\node (1) {$A$}; 
\node (6) [above right of=1] {};
\node (2) [above of=6] {$F$}; 
\node (5) [below right of=1] {};
\node (4) [above right of=5] {$B$}; 
\node (3) [below of=5] {$P$}; 

\draw[->] (1) to [out=45,in=135,looseness=1] (4); 
\draw[->] (4) to [out=225,in=-45,looseness=1] (1); 

\draw[->] (1) to [out=90, in =225, looseness=0.75] (2); 
\draw[->] (4) to [out=90, in=315, looseness=0.75] (2); 
\draw[->] (3) to [out=135, in=270, looseness =0.75] (1); 
\draw[->] (3) to [out=45, in=270, looseness=0.75] (4); 

\draw[->] (3) to [out=180, in =150, looseness=0.85] (2); 
\end{tikzpicture} 
\caption{The unsectorized causal structure of the routed switch. This is the same as the causal structure of the standard switch (Figure \ref{fig:supermap cs example}). The presence of a cycle indicates that $A$ and $B$ cannot be embedded into a definite causal order.}
\label{fig:rswitch-cgdag}
\end{figure}

\begin{figure}
    \centering
\begin{tikzpicture} [scale=1.5]
% nodes
\node (aw) {$A_w$};
\node (a1) [below right of=aw] {$A_{\rm \color{red}{red}}$};
\node (b2) [above right of=aw] {$B_{\rm \color{red}{red}}$};
\node (x) [below right of=a1] {};
\node (y) [above right of=b2] {};
\node (b1) [above right of=x] {$B_{\rm \color{blue}{blue}}$};
\node (a2) [below right of=y] {$A_{\rm \color{blue}{blue}}$};
\node (bw) [below right of=a2] {$B_w$};
\node (p) [below of=x] {$P$};
\node (f) [above of=y] {$F$};

% edges
\draw[->] (p) to (a1);
\draw[->] (p) to (b1);
\draw[->] (a1) to (b2);
\draw[->] (b1) to (a2);
\draw[->] (a2) to (f);
\draw[->] (b2) to (f);
\draw[->] (aw) to [out=90, in=210, looseness=0.65] (f);
\draw[->] (bw) to [out=90, in=330, looseness=0.65] (f);
\draw[->] (p) to [out=30, in=270, looseness=0.6] (bw);
\draw[->] (p) to [out=150, in=270, looseness=0.6] (aw);
\draw[->] (p) to [out=180, in=180, looseness=1.25] (f);
\end{tikzpicture}
    \caption{The sectorized causal structure of the routed switch.  The directed path $P \rightarrow A_{\rm \color{red}{red}} \rightarrow B_{\rm \color{red}{red}} \rightarrow F$ corresponds to the branch where Alice's intervention comes first; $P \rightarrow A_{\rm \color{blue}{blue}} \rightarrow B_{\rm \color{blue}{blue}} \rightarrow F$ corresponds to the branch where Bob's comes first. As the lack of cycles indicates, once we break down the causal influences into those associated with each sector and the which-sector information, there is no need to ascribe an indefinite causal order to the routed switch.}  
\label{fig:rswitch-fgdag} 
\end{figure}
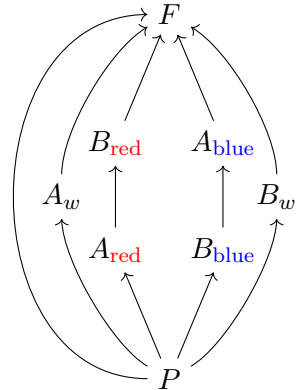

We now have a model of the experiment that accounts for a more appropriately broad range of physically possible transformations than the standard switch, which therefore tells a more trustworthy story about the causal structure of the experiments. So what is this story? Since the routed switch is a routed supermap, we can apply our framework to analyse causal structure at both the unsectorized and sectorized levels. The unsectorized causal structure, depicted in Figure \ref{fig:rswitch-cgdag}, is identical  to the causal structure of the standard switch uncovered in \cite{barrett2021cyclic}, which has an indefinite causal order, and is precisely the causal structure that the dynamicist is likely to attribute the scenario. However, the sectorized causal structure, depicted in Figure \ref{fig:rswitch-fgdag}, does admit a definite causal order.

%The inputs to the routed switch are two time-delocalized channels, which admit an operational definition in the sense of \cite{araujo2014computational} or \cite{oreshkov2019time}. Therefore, using the routed switch as a model for the experiments and their causal structure is consistent with the central assumptions (1) and (2) of the dynamical view; namely, that there are two such channels and they serve as causal relata. Moreover, applying the framework from Section \ref{sec:higher order} confirms the dynamicst's belief that there exists indefinite causal order between them. In fact, the unsectorized causal structure, depicted in Figure \ref{fig:rswitch-cgdag}, in which the entire nodes serve as relata, is identical  to the causal structure of the standard switch uncovered in \cite{barrett2021cyclic}, which is precisely the causal structure that the dynamicist is likely to attribute the scenario.

So we only have indefinite causal order at the coarse-grained, unsectorized level, that vanishes when we perform the sectorization. It is worth capturing this property with a general defintion.

\begin{definition} A routed supermap exhibits \textbf{strong indefinite causal order} if both its sectorized and unsectorized causal structures are cyclic. 
It exhibits \textbf{weak indefinite causal order} if its unsectorized structure is cyclic, but its sectorized structure is acyclic.
\end{definition}

Recall that, by Proposition \ref{theorem:FGfundamental}, the sectorized causal structure fixes the unsectorized causal structure. 
More precisely, the unsectorized directed graph is obtained by combining the sectorized vertices corresponding to the same unsectorized vertex.
Therefore, if a routed supermap exhibits weak indefinite causal order, then its indefinite causal order can be `explained away' as the result of coarse-graining a more detailed structure in which there is no indefinite causal order.
On the other hand, the indefiniteness of a routed supermap with strong indefinite causal order admits no such explanation, and accordingly must be regarded as a brute property of its causal structure.

The routed switch exhibits weak indefinite causal order.
One can easily check that Figure \ref{fig:rswitch-cgdag} is obtained from Figure \ref{fig:rswitch-fgdag} by combining the $A$ nodes and combining all the $B$ nodes, and thus one can understand the indefiniteness of the supermap as being entirely a consequence of this coarse-graining.

%Applying the routed switch to such scenarios, one sees that $U_A$ and $U_B$ act on \textit{time-delocalized systems} as described in \cite{oreshkov2012quantum}. For example, $U^A_1$ is acts on a Hilbert space associated with some time before the photon has entered either of the gates, whereas $U^A_2$ acts on space associated with some time after it has entered $U^B$. 

%\footnote{In fact, it is possible to arrange the apparatus in such a way that }

\subsection{So, are the experiments physical realizations of indefinite causal order?}

In the context of the experiments, the inputs to the routed switch are two time-delocalized channels, which admit an operational definition in the sense of \cite{araujo2014computational} or \cite{oreshkov2019time}. Therefore, we claim that using the routed switch as a model for the experiments and their causal structure is consistent with the central assumptions (1) and (2) of the dynamical view; namely, that there are two such channels and that they serve as causal relata (we will defend this claim in Section \ref{sec:conistent}).

And yet, the routed switch exhibits the formal property of weak indefinite causal order.
Since we take the routed switch to provide a good model for the causal structure of the experiments,  and, in particular, a better one than the standard switch, we take this to suggest that the experiments themselves are physical realizations of indefinite causal order only in the weak sense.
If this is right, then it seems that the dynamicist's central assumptions (1) and (2) do not actually suggest the experiments are realizations of indefinite causal order in anything more than in a weak sense.
By the same token, one needs not assume \textit{a priori} that causal relations obtain between objects that are localized in spacetime in order to argue that indefinite causal order is not strongly realized.
We conclude that there are  no compelling \textit{a priori} assumptions from which to argue that indefinite causal order is realized in more than a weak sense -- or at least, if there are, they have not yet been clearly articulated in the literature.

Now, if all this is correct, then there must be something wrong with the dynamicist's argument.
The issue is with propositions (3) and (4).
The fact that some time-delocalized channels and a supermap acting on them accurately model the transformations that take place in an experiment does not imply that they provide a useful model for the causal structure of the experiment.  This is because they may still not account for an appropriately broad range of physically possible transformations, or of possibilities for information flow.
Hence the combination of (3) and (4) should be rejected, and (5) is unsupported.

%Of course, the dynamicst's argument, as it was summarized with points (1 -- 5), did not actually rely on the use of the standard switch as a model, but only some bipartite unrouted supermap. As long as one uses such a supermap as a model for the causal structure, the passing of a causal witness test does indeed demonstrate indefinite causal order. But, for routed bipartite supermaps, such tests at most demonstrate weak indefinite causal order. So, given that one can also use a routed supermap as a model while maintain the dynamicist's central assumptions, the analysis also casts doubt on the dynamicist's reliance on causal witness tests in order to conclude indefinite causal order has been physically realized.

Essentially, the issue with the dynamicist's argument is that just because an experiment \textit{implements a supermap}, it does not automatically follow that it \textit{realizes its causal structure}.
And this is because the former just requires that the supermap models the actual transformations, while the latter requires that it also models the non-actualized but possible transformations.
For future discussions of the causal structure of an experiment, we therefore recommend that the questions of whether a supermap is implemented and whether its causal structure is realized should always be carefully distinguished.

We note that our arguments do not establish that all (potential) implementations of the switch other than \cite{procopio2015experimental, rubino2017experimental, rubino2019experimental} do not realize strong indefinite causal order. In particular, for an implementation in which it is physically impossible for the agents to access to the control qubit, there would be no reason to model it using the routed switch.

A gravitational implementation of the switch, such as the one proposed in \cite{moller2020gravitational}, would probably fall into this category. Another example is the photonic experiment in \cite{goswami2018indefinite}, in which Alice and Bob cannot change their actions depending on the order in a straightforward way. This reason for this is that the photon there has a very long coherence length. Hence there are no longer two implementation events for each agent, but rather elongated implementation \textit{regions} of spacetime, which are almost entirely overlapping. So Alice and Bob can no longer implement the routed switch by controlling their intervention on two distinct times of implementation. We recommend that any future debate over whether indefinite causal order has been realized yet should focus on this more challenging example.

%%%%%%%%%%%%%%%%%%%%%%%%%%%%%%%%%%%%%%%%%%%%%%%%%%%%%%%
\section{Possible objections to our analysis} \label{sec:rswitchconsequences}
%%%%%%%%%%%%%%%%%%%%%%%%%%%%%%%%%%%%%%%%%%%%%%%%%%%%%%

Here, we consider a number of possible objections to the analysis from the previous section. In particular, we discuss the criticisms that (i) our analysis only goes through relative to a particular notion of causal structure; (ii) we are wrong to claim that our approach is consistent with the central assumptions of the dynamical view; and (iii) there is something wrong with our formal definition of weak indefinite causal order. We will take each of these in turn.

\subsection{`The analysis relied on a particular view of causal structure'}

The first objection we consider is that our conclusion that the experiments are weak realizations of indefinite causal order is only true relative to our assumption that causal structure is to be understood as possibilities for information flow, as captured by some  appropriate routed supermap and our framework for analysing its causal structure.
There are other legitimate notions of causal structure, such as the possibilities for signalling between implementation events, or the partial order associated with a Minkowski manifold, for which indefinite causal order is not even weakly realized by the experiment \cite{paunkovic2020causal}.

We are happy to accept this point. 
We chose our notion because it is information-theoretic; because it makes causal structure of a physical scenario intrinsic to its quantum description; because it enables us to describe causal structures precisely and in great detail; and because it makes it seem initially plausible that indefinite causal order could be realized on a Minkowski spacetime. Nevertheless, undoubtedly other notions of causal structure that are more spatiotemporal and less information-theoretic than ours are also interesting and important in their own right, and they do indeed lead to a more forceful rejection of indefinite causal order. 
So we are happy for our conclusion that the `causal structure' of the experiments exhibits only weak indefinite causal order, rather than definite causal order, to be understood as true relative to our particular notion of causal structure.

%In fact, even given our preferred notion of causal structure, our conclusions are still relative to our particular choice of model, to which there are many alternatives. For example, the underlying causal box model of the switch also accounts for the physical possibilities we are concerned with, plus others, such as the creation and destruction of particles. And this model doesn't obviously exhibit indefinite causal order even in a weak sense. Again, we are happy to accept the conditional nature of our specific conclusion about how the causal structure is to be understood.

However, our conclusion that the experiments are at most weak, rather than strong, realizations of indefinite causal order seems to be more robust. For our analysis is compatible with the \textit{dynamicist's} notion of causal structure. Hence the analysis also suggests that the experiments are not strong realizations of indefinite causal order according to \textit{any} mainstream notion of causal structure.

\subsection{`The analysis is not actually compatible with the dynamicist's central assumptions'} \label{sec:conistent}

Another possible criticism is that we were wrong to claim that our analysis was compatible with the central assumptions of the dynamical view.
One way of defending this would be to point out that, although we do understand the experiment in terms of two time-delocalized channels (as per assumption (1)), we ended up considering a broader set of possible channels than those considered by the dynamicist.
However, as we have argued, this extra generality is needed in order to account for the physical possibilities for information flow, and hence to properly capture the causal structure.
So if the central assumptions of the dynamical view prohibit this extra generality, then we think they are wrong; while if they do not, then we think our analysis is consistent with them.

Another way to make this criticism would be to argue that our consideration of the individual sectors as relata in the sectorized causal structure can only be motivated by the spatiotemporalist's assumption that causal relata should be localized in spacetime.
After all, we did appeal to the existence of two distinct implementation events in order to justify our use of the routed switch, in which the sectorial constraints apply.

However, this is a misunderstanding of the argument. We were led to consider the sector-preserving channels (\ref{routed switch inputs}) on the basis that they account for a broader range of physically possible transformations than those accounted for by the switch. 
It is true that the physical possibility of these transformations is the \textit{consequence} of spatiotemporal considerations, but spatiotemporal considerations did not motivate our desire to account for them.
It is their physical possibility alone that motivated us to consider the full range of sector-preserving channels. 
Indeed, if their physical possibility were instead grounded in other, non-spatiotemporal considerations, then we would still be obliged to account for them.

This shows that our consideration of the sector-preserving channels and the routed switch is not motivated by any assumption of localization; it remains to show that our choice of causal relata does not rely on this either.
To see this, note that, as a routed supermap, the routed switch comes with a distinct sectorized causal structure which, by Proposition \ref{theorem:FGfundamental}, is strictly more detailed then its unsectorized counterpart.
We consider this structure, and its relations between individual sectors and which-sector informations, because they help us understand the causal structure at a deeper level -- in particular, how features of the unsectorized causal structure, such as cyclicity, come about in terms of the more detailed description.
But the fact that the sectorized relata provide this understanding does not depend on them being localized, and indeed the which-sector informations are actually \textit{de}localized. So even though the individual sectors are localized, our use of them as causal relata does not depend on any assumption that causal relata must be localized.

Another spin on this criticism would be to say that we should have characterized the dynamicist's assumption (2) as requiring that the time-delocalized  channels are the \textit{only} causal relata, in which case our analysis would not be consistent with it. If the assumption were so modified, then we would indeed reject it, since, as we have just argued, it is important to account for physically possible transformations, and doing so motivates the sectorized relata.

But perhaps there is still scope for the dynamicist to argue that there is something wrong or, at least misleading, about understanding the causal structure in terms of our sectorized relations and relata. This is the topic of the next subsection.

\subsection{`An appropriate physical notion of weak indefinite causal order has not been successfully formalized'}

We have relied on a formal notion that we chose to call `weak indefinite causal order'.
But perhaps the dynamicist could argue that our technical definition of this term fails to formalize any pre-formal, intuitive idea that its name might suggest. In that case, the routed switch exhibiting weak indefinite causal order in our formal sense would not obviously imply that the experiments are weak realizations of indefinite causal order in any interesting or relevant sense.

For example, the dynamicist could argue that our sectorized causal relations and relata are somehow illegitimate, or at least should be considered as conceptually secondary to the unsectorized relations  and relata. Then the lack of cycles in the sectorized picture would arguably not much weaken the sense in which indefinite causal order is obtained between the time-delocalized channels in the experiments.

One way to argue that the sectors are not legitimate relata is to argue that they have no operational meaning. But our response to this charge can be similar to the spatiotemporalist's, discussed in the previous section.
Alternatively, the dynamicist might deny that we can attribute our sectorized causal relations to the phenomenon of interest. For example, the arrow from $A_{\rm \color{red} red}$ to $B_{\rm \color{red} red}$ in Figure \ref{fig:rswitch-fgdag} means that the wire for the red sector on Alice's exchange gate signals to the wire for the red sector on Bob's exchange gate in the routed process channel for the routed switch. But if this is done experimentally, the Bob receiving information from Alice destroys the interference that makes the certification of indefinite causal order via a causal witness test possible. (This is because we end up effectively post-selecting on the sector in which the control qubit's state is $\ket{0}$.) Thus this alleged causal relation might be held to have no operational meaning in the context of the phenomenon of interest.

However, this argument backfires, since it would also prevent us from attributing the causal relations between the channels that lead the dynamicist to see the experiments as exhibiting indefinite causal order. For example, the arrow from $A$ to $B$ in the causal structure of the switch means that Alice can signal to Bob in the  process channel. But operationally witnessing this signalling would again destroy the interference. So the above argument would establish that this causal relation has no operational meaning, in which case we do not have a cyclic causal structure, and therefore we have a definite causal order. As long as one accepts our idea of causal relations as signalling relations through unitaries, attributing an indefinite causal order to this experiment necessarily involves considering possibilities for signalling which are not realized when the phenomenon of interest takes place. 

Of course, one could reject this idea of causal relations, and instead simply define `indefinite causal order' as the ability to pass a causal witness test. But since the routed switch can pass such a test, along with processes with a definite causal order and more parties,  this does not seem well-motivated. There is only a justification for interpreting a causal witness test as an indicator of indefinite causal order when one also assumes the correct model for the experiment is an unrouted supermap with a certain number of parties.

%Indeed, the very notion of a superposition of properties requires that it is legitimate to talk about properties associated with a state that would be destroyed were one to actually measure those properties. Accordingly, this sort of argument would also establish that we can't consider a particle in a superpositon of locations, since the measurement of location would destroy the superposition.

%%%%%% what if the operationalist says that the onlylegitmate way to attribute causal relations here is the verifcation of a causal witness?
%%%%%%.   ---- I guess that if you only consider witness certification, then its unclear why you attribute the computational advantages to a `superposition of causal orders'. You've effectively denied that its reasonable to attribute cause orders, even in an indefinite sense, to the scenario since verifying them would destroy the phenomenon.

%As a last resort, the dynamicist might accept the legitimacy of the sectorized relations, but take the coarse-grained, unsectorized structure to be somehow more fundamental. But this is quite problematic. Usually, a more fundamental structure fixes the less fundamental structure, and not the other way round. But here, the unsectorized structure is fixed by the sectorized structure. In light of this, it is unclear how the dynamicist can make this argument in anything but a very ad hoc way.

Another possible response from the dynamicist is that our sectorized causal structures are somehow problematic because an individual sector isn't the sort of thing that can bear causal relations. The argument might run as follows. When we partition a Hilbert space into different sectors, we are not thereby dividing a system into constituent parts. Instead, we are considering different, mutually exclusive ways that system could be. But a way that a system might or might not be should not qualify as a relatum of causal relations. Rather, the causal relata in experiment should be the constituent parts of the experiment that are `present' no matter what.

But this argument should be rejected, since it rules out a causal analysis of a classical version of the experiments which seems obviously reasonable. Imagine a classical version of the routed switch that incoherently controls the functions $f_{t_2}^B \circ f_{t_1}^A$ and $f_{t_2}^A \circ f_{t_1}^B$ on  a classical bit. We could perform an analysis similar to the one we have given for the quantum case, by using a classical version of the exchange gate. This analysis would also conclude that there is an indefinite causal order between $A$ and $B$, but that this disappears when we define more fine-grained causal relata via a sectorization into subsets of the classical variables. 

This analysis seems to be a fair reflection of the physical situation. True, it is possible that $A$ can affect $B$ on any particular run of the experiment, and it is possible that $B$ can affect $A$ on any particular run of the experiment. But, crucially, these possibilities turn out to be mutually exclusive: on any particular run of the experiment, causal influence between $A$ and $B$ propagates in only one direction. By letting the individual sectors of $A$ and $B$ define causal relata, we make this obvious but important physical fact clear at the formal level, by showing that their causal structure is acyclic. 

Far from being an unreasonable thing to do, here breaking down the causal relata into different sectors helps us to formally articulate the key conceptual points about the causal structure of the experiment: that there is indefinite causal order between $A$ and $B$, but only in a weak sense since on any particular run of the experiment causal influence passes only in one direction between them.  But if it were true that the different ways a system could be should never define causal relata, then we would not be able to perform this useful analysis. From this, we conclude that if there is a reason for forbidding the quantum sectors from defining causal relata, it is not because they correspond to mutually exclusive possibilities.

A better argument against performing the sectorization is based on the assumption that the quantum sectors, unlike the classical ones, are \textit{not} just mutually exclusive ways a system can be. This is evidenced by the fact that the relative phase between the sectors can be detected with a single measurement (and in fact is when indefinite causal order is verified with a causal witness test). But this relative phase is not associated within any particular sector -- it is a \textit{global} property of the sectors. From this, it is tempting to conclude that all sectors -- and the information they contain -- are somehow `present' on each run of the experiment. 
Applying this assumption to the photonic experiments, it seems we really do have causal influence in both directions between the gates on a single run, unlike in the classical version of the experiment. 

Now, this might seem to bring us towards thinking about sectors as constituent parts of a system, like we think of tensor factors.
And there is surely no problem with splitting up causal relata with the parts (commonly called `subsystems') associated with tensor factors.
This begs the question of why we can't split up the causal relata into different parts associated with sectors, as we have in our analysis. But the dynamicist might want to argue that they cannot be understood as \textit{independent} parts, so that we can't attribute to them their own causal properties (or at least, if we can then this picture should be regarded as secondary to the unsectorized one).

To support this, the dynamicist might appeal to the idea that the information in a system is not just the combination of the information within each sector. For this does not include the information in the coherence between sectors, which, by the global phase symmetry, cannot be reconstructed from the information present on each sector. Mathematically, the point is that the equation 

\begin{equation} \label{eq: sector sep}
     \pi_A^0 \ket{\psi}\bra{\psi}_A \pi_A^0 + \pi_A^1 \ket{\psi}\bra{\psi}_A \pi_A^1 = \ket{\psi}\bra{\psi}_A 
\end{equation}
does not hold for all $\ket{\psi}_A$, where $\pi_A^0$ projects onto a sector of $A$ and $\pi_A^1 := I -\pi_A^0$ projects onto the compliment sector. This is reminiscent of how the pure state of a bipartite system does not always separate onto a state for each factor; i.e. that 

\begin{equation} \label{eq: factor sep} {\rm Tr}_X(\ket{\psi}\bra{\psi}_{AX}) \otimes {\rm Tr}_A(\ket{\psi}\bra{\psi}_{AX}) = \ket{\psi}\bra{\psi}_{AX}\end{equation}
does not hold for all $\ket{\psi}_{AX}$. 

But the \textit{sectorial nonseparability} captured by (\ref{eq: sector sep}) is stronger than the usual notion of nonseparability across factors in the following sense. There are many pure states, called product states, for which (\ref{eq: factor sep}) does hold. But the only pure states for which (\ref{eq: sector sep}) holds are those for which one of the $\pi_A^i \ket{\psi}\bra{\psi}_A \pi_A^i = 0$. In other words, a pure state can only be understood as the combination of states on different sectors in the trivial case when there is only information associated with one of them. Intuitively, whenever both sectors are 
`present', there are relational properties between them (i.e. the coherence) that prevent us from understanding the pure state as a simple combination of them. So one could argue that, if the sectors do correspond to parts, they do not correspond to independent parts, that can be properly separated out either mathematically or conceptually. Rather, to the extent that they are all `present', they are always inextricably linked, or `entangled'. Given that the sectors are inextricably linked, it seems at least questionable that we can attribute them their own causal properties. For this reason, the dynamicist might either reject our sectorized perspective, or at least consider it as secondary to the unsectorized perspective from which there is indefinite causal order.

However, (\ref{eq: sector sep}) was given for pure states, and it does not generalize to mixed states. And according to our model, Alice's and Bob's time-delocalized systems are always each in an incoherent mixture of the two different sectors. This is because their systems are entangled, in the sense that if Alice's system is in the $t_1$ sector, then Bob's is in the $t_2$ sector, and vice versa. But the information associated with an incoherent mixture of two different sectors, unlike a superposition of them, can indeed be understood as the combination of the information associated with each sector. 

Still, in an \textit{improper mixture} \cite{d1966elementary, d2018conceptual} (that is, an incoherent mixture arising from tracing out one half of an entangled state), even if there is no coherence to be measured between two sectors, one can still induce local phase shifts between them that end up globally changing the entangled state; so sectorial nonseparability is still relevant in this sense. But Alice \textit{can} still do this in our model, by using the which-sector ancilla of the exchange gate. So where sectorial nonseparability is relevant, we account for it, but where we do not account for it, it is not relevant.

Moreover, we note that the argument from sectorial nonseparability requires an intuitive leap from the structure of the state space to causal structure. But the atomicity property has already shown us that such leaps are unwarranted, since state entanglement does not lead to an analogous entanglement of causal relations.  The general moral that we take from this is that causal structure in quantum theory does not always behave in the way one would naively expect from studying its state space. So even if it is surprising that there exists a natural and rigorous sense in which sectors can have their own causal relations, one's astonishment does not constitute an objection to them.

%****************************************************
\section{Discussion} \label{sec:discussion}
%****************************************************

\subsection{Summary}

In this paper, we built a framework that attributes a causal structure to scenarios where sectorial constraints are at play.
We started off with the assumption that the most basic sort of causal relation is a possibility for signalling afforded by a (routed or unrouted) unitary transformation.
We then proceeded to show that a mathematical formulation of this idea is equivalent to a number of other conditions, expressing distinct intuitions about what causal relations are.
These conditions imply that the causal structure of a (routed or unrouted) unitary transformation can be represented as a directed acyclic graph, due to the atomicity property \cite{barrett2020quantum}.
All of this indicates that there is a stand-out, natural notion of a causal relation in quantum theory, even when sectorial constraints are at play.
This notion is at the heart of our framework for causal structure.

When systems suffer sectorial constraints, we can break down their causal relations further. The resulting sectorized causal relations are between sectors and which-sector informations  in preferred partitionings of each system's Hilbert space.
These gives rise to the sectorized causal structure of a transformation, which can also be represented as a directed graph.
The sectorized causal structure is strictly more detailed than its unsectorized counterpart, and so can be regarded as a fine-graining.

We were able to use these notions of the causal structure of a transformation to define the sectorized and unsectorized causal structures of a (routed) supermap in a fairly straightforward way. 
The sectorized causal structure can be of considerable use in understanding scenarios which appear to display enigmatic causal properties, such as indefinite causal order. 
To illustrate this, we performed a sectorized and unsectorized analysis of recent photonic experiments that purport to implement the switch, using our more general routed switch as a model.
The analysis made clear that the cycles in the unsectorized causal structure of the experiment can be understood as the result of coarse-graining the more detailed, acyclic, sectorized causal structure. 
This provided an argument that the experiments realize indefinite causal order only in a weak sense.
Significantly, this is the first argument to this effect that did not start off by assuming that the objects that bear causal relations should correspond to local regions of spacetime.
Instead, the argument is based on acknowledging the impact that the sectorial constraints have on the causal structure.

\subsection{Possibility and causality}

Above all, this work demonstrates the importance of a simple dictum: \textbf{possibility is essential for causality}. More concretely, the point is that the question of which transformations are possible is of central importance for any information-theoretic account of causal structure. And the set of possible transformations might be nontrivial, in that it might not include all quantum operations on each of the systems of interest.

The reason the dictum matters so much is that failing to recognise it can lead one to attribute the wrong causal structure to a scenario. In the example sketched at the start of Section \ref{sec:ni_ruc}, ignoring the sectorial constraints on the system $D$ leads one to incorrectly state that it is influenced by $A$. A more pertinent example is provided by photonic implementations of the switch. There, the attribution of indefinite causal order can result from mistakenly thinking that just because the switch provides an accurate model of the actual transformations, its causal structure must reflect the causal structure of the experiments.

\subsection{Outlook}

There are plenty of avenues for further research. One pressing question is what the causal structure of a routed supermap can tell us about its compositional structure, and about the observable correlations it can generate. Future work will seek to lay out this out in detail, most likely by attempting to extend Theorems 4.10 and 8.3 from \cite{barrett2020quantum} to the routed case. A satisfactory answer to such questions, in addition to the contributions of this paper, would bring us most of the way to a fully-fledged theory of quantum causal models for sectorized systems.

Given that a part of the motivation for studying causal structure is to uncover the conceptual relationship between quantum theory and general relativity, it would be a useful step to investigate how frameworks for causal structure might generalize to the field-theoretic and relativistic cases. Most importantly, one could check how or whether the most striking features of causal structure in quantum theory are manifested in these more fundamental regimes, and determine whether any conceptually novel properties emerge.

A byproduct of our study of implementations of the quantum switch was the definition of a routed supermap of particular interest, the \textit{routed quantum switch}, which essentially models the situation where agents in a switch can coherently control their actions on the order they are in. Even though we restricted our use of it here to the discussion of a foundational question, it could also have more practical applications. In particular, as an extension of the switch it might hold similar or additional advantages for communication, computation, or thermodynamical procedures.

Another avenue is to find further physical applications of our framework. Any scenario featuring sectorial constraints and interesting causal features is likely to provide a worthwhile application; examples include the Aharonov-Bohm effect \cite{aharonov1959significance} (in which charge superselection appears to play a key role \cite{Erez_2010}), and the recently proposed del Santo-Dakić protocol \cite{Del_Santo_2018, Carrying, Massa_2019, faleiro2020}. It will also be interesting to see whether the understanding of causal structure in the presence of sectorial constraints can help one uncover new scenarios with exotic causal properties.

It is also likely possible to articulate a more systematic connection between routed maps and superselection rules, and hence also quantum reference frames \cite{bartlett2007reference}. Therefore, it might be possible to make some interesting general statements about the interplay between reference frames, superselection rules, and causal structure. It may be particularly interesting to explore the precise sense in which the causal structure one observes is dependent on one's quantum reference frame.

It would be interesting to provide an adaptation of our framework to the classical case. While a similar approach can be taken, there are some interesting differences, such as the possibility of determining all the information associated with the target input to the classical version of the exchange gate by measuring its ancillary output, and the fact that the `which-sector?' relata in the classical case never serve as causes, as a consequence of the absence of relative phases in classical theory.

Finally, it would be of interest to generalize the arguments we gave against the strong realization of indefinite causal order in \cite{procopio2015experimental, rubino2017experimental, rubino2019experimental} to the photonic experiment in \cite{goswami2018indefinite}. The obvious way of doing this involves the use of infinite-dimensional Hilbert spaces for Alice and Bob, formed by directly summing together a qubit for each time at which a branch of the photon passes through the waveplates. This goes beyond the scope of our framework as it stands, but one could provide an analysis within our framework of a version of the experiment in which the photon only passes through the waveplates at a very large, but still finite, number of times. This might provide an argument that \cite{goswami2018indefinite} does not realize strong indefinite causal order, which may be generalized even further to a wide range of experiments involving the superposition of various paths through spacetime.

\section*{Acknowledgements}

It is a pleasure to thank Giulio Chiribella, Hlér Kristjánsson, Ognyan Oreshkov, Nicola Pinzani, Tein van der Lugt, Matt Wilson, and two anonymous reviewers for helpful discussions and comments. AV is indebted to the organisers and participants of the 2021 Sejny Summer Institute -- in particular Timothée Hoffreumon, Natália Móller and Eleftherios Tselentis -- for discussions and ideas about implementations of the switch. We also thank Alexandra Elbakyan for her help to access the scientific literature. 

NO acknowledges funding from the UK Engineering and Physical Sciences Research Council (EPSRC). AV is supported by the EPSRC Centre for Doctoral Training in Controlled Quantum Dynamics.  This publication was made possible through the support of the grant 61466 ‘The Quantum Information Structure of Spacetime (QISS)’ (qiss.fr) from the John Templeton Foundation. The opinions expressed in this publication are those of the authors and do not necessarily reflect the views of the John Templeton Foundation.

\section*{Note Added}

Independent work appeared during the completion of the first version of this manuscript which provides a related fine-grained causal analysis of implementations of the switch \cite{vilasini2022embedding}.

\bibliography{references}

\appendix

\section{Proof of Theorem \ref{equivalence-u1-u6}} \label{appendix:equivalence-u1-u6}

The implication from (\ref{U1}) to (\ref{U2}) is obvious. That (\ref{U2}) implies (\ref{U3}) follows from noting that, for any density matrices $\rho$ and $\rho'$, one can always find a channel $\Phi$ such that $\rho = \Phi(\rho')$. Inserting $\rho' \otimes \sigma$ onto both sides of (\ref{U2}) then gives (\ref{U3}). The equivalence of (\ref{U1}) and (\ref{U3}) is proven in \cite{schumacher2005locality}, and the equivalence of (\ref{U1}) and (\ref{U4}) is proven in \cite{lorenz2020causal}. 

At this point, we have proven that conditions (\ref{U1}--\ref{U4}) are all equivalent. It remains to show that all of these are equivalent to (\ref{U5}) and (\ref{U6}).

It is obvious that (\ref{U4}) implies (\ref{U5}). That (\ref{U5}) implies (\ref{U6}) follows from the facts that $\ca:=\cl(A)$ is spanned by its unitary operators, and that the right side of (\ref{U5}) commutes with $I_C \otimes \cd$. 

Finally, to show that (\ref{U6}) implies (\ref{U2}), we define the channel $\tilde{\Phi}:=\cu \circ (\Phi \otimes \ci) \circ \cu^\dag$ on $C \otimes D$. We can then perform the following manipulations, in which the triangle labelled $`0$' represents the state $\ket{0}\bra{0}$. 
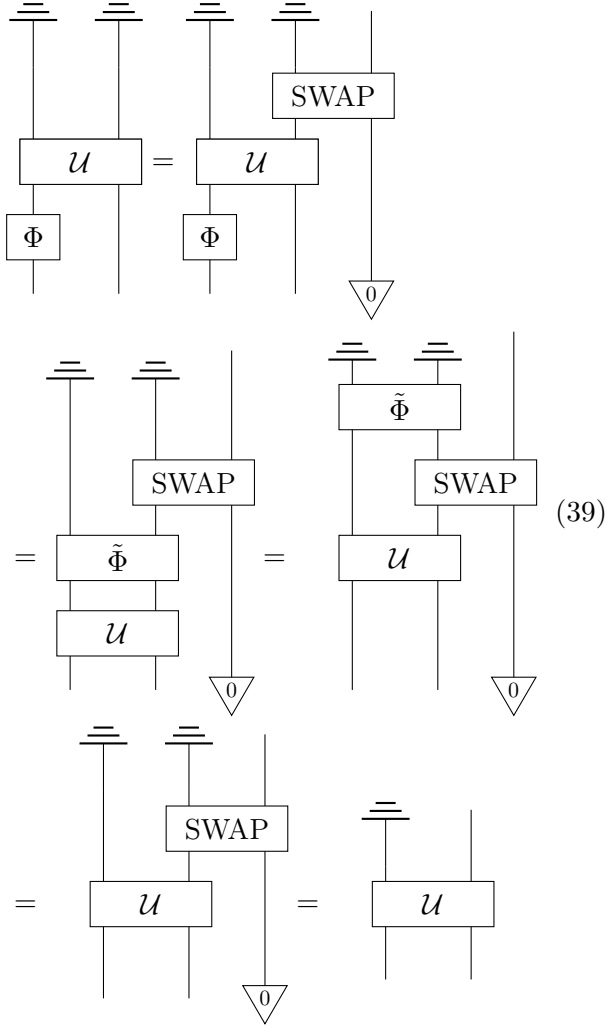
\begin{equation} \label{NI u6 implies u2}
\begin{split}
    & \input{ni_u6_implies_u2_1.tikz}  =  \input{ni_u6_implies_u2_2.tikz} \\
    & = \input{ni_u6_implies_u2_3.tikz}  = \input{ni_u6_implies_u2_4.tikz} \\
    & = \input{ni_u6_implies_u2_5.tikz}  = \input{ni_u6_implies_u2_6.tikz}
\end{split}
\end{equation}

To see why the third equality holds, note that since $\Phi$ is a channel on $A$, all of its Kraus operators are elements of $\ca$. It follows that the Kraus operators of $\tilde{\Phi}$ are elements of $U(\ca \otimes I_B)U^\dag$. By (\ref{U6}), we know these operators commute with all the operators in $\cd := \cl(D)$ that only act on $D$. In the manipulations above, the swap channel only acts on $D$ and its own ancilla, meaning that it will commute with $\tilde{\Phi}$.

%%%%%%%%%%%%%%%%%%%%%%%%%%%%%%%%%%%%%%%%%%%%%
\section{Proof of Proposition \ref{unrouted atomicity}} \label{appendix:unrouted atomicity}
%%%%%%%%%%%%%%%%%%%%%%%%%%%%%%%%%%%%%%%%
It follows quite simply from (\ref{U6}) that if $A \not\xrightarrow{U} E$, and $A \not\xrightarrow{U} F$, then $A \not\xrightarrow{U} E \otimes F$. Of course, if $A$ can signal to a subsystem then it can always signal to the composite system, so the converse also holds. We therefore have the first part of atomicity (\ref{dag1unrouted}), i.e. $(A \not\xrightarrow{U} E) \land (A \not\xrightarrow{U} F) \Longleftrightarrow (A \not\xrightarrow{U} E \otimes F)$. 

Taking into account the time symmetry property (\ref{eq:timesymunrouted}) leads to the second part of atomicity (\ref{dag2unrouted}), i.e.
$(A \not\xrightarrow{U} F) \land (B \not\xrightarrow{U} F) \Longleftrightarrow (A \otimes B \not\xrightarrow{U} F)$. Explicitly, we have:
\begin{equation} \label{atproof}
    \begin{split}
        (A \not\xrightarrow{U} F) \land (B \not\xrightarrow{U} F) & \Longleftrightarrow (F \not\xrightarrow{U^\dag} A) \land (F \not\xrightarrow{U\dag} B) \\
        & \Longleftrightarrow (F \not\xrightarrow{U^\dag} A \otimes B) \\
        & \Longleftrightarrow (A \otimes B \not\xrightarrow{U} F)
    \end{split}
\end{equation}
where in the first and third equivalences we have used time symmetry (\ref{eq:timesymunrouted}), and in the second equivalence we used the first part of atomocity, (\ref{dag1unrouted}).

%%%%%%%%%%%%%%%%%%%%%%%%%%%%%%%%
\section{The embedding map} \label{appendix:embedding}
%%%%%%%%%%%%%%%%%%%%%%%%%%%%%%%%%%

Consider two sectorized Hilbert spaces $A^i$ and $B^j$ and some subspace $H = \bigoplus_{i, j} \lambda^{ij} A^i_\star \otimes B^j_\star$, where $\lambda^{ij}$ is some route matrix. ($H$ might be the practical input or output space of some routed unitary.) A \textit{sector-preserving algebra} over $A^i$ is one that can be written as:
\begin{equation} \label{sector-preserving}
    \cx = \bigoplus_i \cx_i 
\end{equation}
where each $\cx_i$ is an algebra over $A^i_\star$. The point of the embedding map is to `embed' this algebra into an algebra over the larger space $H$.

The embedding map acts on operators from $\cx$. An operator from this algebra has the form $M = \bigoplus_i M_i$. The embedding map is defined as follows:
\begin{equation}
    M\xrightarrow{*} M^* := \bigoplus_{i, j} \lambda^{ij} M_i \otimes I_{B_\star^j}
\end{equation}
where $I_{B_\star^j}$ is the identity operator on $B_\star^j$. The codomain of $\xrightarrow{*}$ is given by:
\begin{equation}
    \cx \xrightarrow{*} \cx^* := \bigoplus_{i, j} \lambda^{ij} \cx_i \otimes I_{B_\star^j}
\end{equation}
where the $\cx_i$ are algebras over the $A_\star^i$ from (\ref{sector-preserving}).

We now prove that the embedding map has some properties that help with the proof of Theorem \ref{theorem:R1 - R3 eq}. Firstly, embedding map is invertible and linear. To see that it is invertible, note that any operator from $\cx^*$ can be uniquely written in the form $M^* = \bigoplus_{i, j} \lambda^{ij} M_i \otimes I_{B_\star^j}$, so we can easily define the inverse embedding map:
\begin{equation}
    M^* \xrightarrow{*^{-1}} M := \bigoplus_i M_i
\end{equation}

Now we prove that the embedding map is linear. Given the operators $M = \bigoplus_i M_i \in \cx$ and $N = \bigoplus_i N_i \in \cx$, consider some linear combination $R := aM +bN$. $R$ can be written $R=\bigoplus_iR_i$ where each $R_i = aM_i +bNi$. Applying the embedding map to a linear combination therefore has the following effect:
\begin{equation}
\begin{split}
    (aM + bN)^*  &= \bigoplus_{i, j} \lambda^{ij} (aM + bN)_i \otimes I_{B_\star^j}  \\
    & = \bigoplus_{i, j} \lambda^{ij} (aM_i + bN_i) \otimes I_{B_\star^j}  \\
    & = \bigoplus_{i, j} \lambda^{ij} (aM_i \otimes I_{B_\star^j}) +   \bigoplus_{i, j} (bN_i \otimes I_{B_\star^j}) \\
    & = aM^* + bN^*
\end{split}
\end{equation}
which proves linearity.
 
The combination of the linearity and the invertibility of the embedding map has the useful consequence that if a set of operators $\{M_k\}_k$ spans $\cx$, then $\{M_k^*\}_k$ spans $\cx^*$, which we now prove. Denote by $N^*$ an arbitrary member of $\cx^*$. $N^*$ can be mapped via  $\xrightarrow{*^{-1}}$ to $N \in \cx$, which by assumption can be written $N = \sum_k \alpha_k M_k$. We can apply $\xrightarrow{*}$ to both sides of this equation and use linearity to deduce $N^* = \sum_k \alpha_k M_k^*$.

%%%%%%%%%%%%%%%%%%%%%%%%%%%%%%%%%%%%%%%%%%%%%%%%%%%%%%%%
\section{The exchange gate} \label{appendix:xdeltadef}
%%%%%%%%%%%%%%%%%%%%%%%%%%%%%%%%%%%%%%%%%%%%%%%%%%%%%%%%%%%%%%

Given a sectorized Hilbert space $A^i$, the exchange gate is a unitary operator
\begin{equation}
    \texttt{EXCH}: A \otimes A^O \rightarrow A \otimes A^I
\end{equation}
where $A^O$ and $A^I$ are ancillary systems of the same dimension with the factorizations:
\begin{equation}
    \begin{split}
        A^O = A^O_1 \otimes \ldots \otimes A^O_N \otimes A^O_w \\
        A^I = A^I_1 \otimes \ldots \otimes A^I_N \otimes A^I_w 
    \end{split}
\end{equation}
where $N$ is the number of sectors. Each $A_i^O$ and $A_i^I$ has the dimension of the corresponding sector, and $A^O_w$ and $A_w^I$ are each of dimension N.

The exchange gate can be understood in terms of two more basic unitary operators. The first is called the `sectorial CNOT', or $\texttt{SCNOT}$, gate, and it acts only on $A$ and the which-sector ancilla. To define the $\texttt{SCNOT}$ gate, we first note for every sector $A^i_\star$ of $A^i$ we can write down a basis $\{\ket{i, j_i}\}_{j_i}$. We can then combine these to get a basis for $A$: $\{\ket{i, j_i}\}_{i, j_i}$. Then the gate $\texttt{SCNOT}$: $A \otimes A^O_w \rightarrow A \otimes A^I_w$ is defined as follows:
\begin{equation} \label{SCNOTdef}
    \texttt{\texttt{SCNOT}}\ket{i, j_i}_A\ket{k}_{A_w^O}= \ket{i, j_i}_A \ket{k+i}_{A^I_w} \ \ \ \ \ \forall\ket{i, j_i}_A \ \forall \ket{k}_{A^O_w}
\end{equation}
where $\{\ket{k}_{A^O_w}\}$ and $\{\ket{k}_{A^I_w}\}$ are logical bases for $A_w^O$ and $A^I_w$ respectively, and the addition is modulo $N$. In other words, the $\texttt{SCNOT}$ is a direct sum over the sectors of $A^i$ of operators which leave $A^i_\star$ alone send the ancillary basis $\{\ket{k}_{A^O_w}\}$ to $\{\ket{k+i}_{A^I_w}\}$.

The second gate is called the `routed swap' gate, or $\texttt{RSWAP}. \texttt{RSWAP}$: $A \otimes A^O_1 \otimes \ldots \otimes A^O_N \rightarrow  A \otimes A^I_1 \otimes \ldots \otimes A^I_N$ is a direct sum over the sectors of $A^i$ of unitary operators that swap $A^i_\star$ with the $i$th ancillary system. Let us make this more precise. We define $\texttt{SWAP}_i:= I_{A^i_\star \rightarrow A^I_i} \otimes I_{A^O_i \rightarrow A_\star^i}$, where $I$ represents the identity transformations between the Hilbert spaces denoted in the subscript. We denote by $I_{A^O_{\overline{i}} \rightarrow A^I_{\overline{i}}}$ the identity transformation from the tensor product of all except the $i$th ancillary input to the tensor product of all except the $i$th ancillary output. Then $\texttt{RSWAP}$ is given by:
\begin{equation} \label{RSWAPdef}
    \texttt{RSWAP} = \bigoplus_i \texttt{SWAP}_i \otimes I_{A^O_{\overline{i}} \rightarrow A^I_{\overline{i}}} 
\end{equation}

Then the exchange gate is defined as
\begin{equation} \label{eq:xdeltadef}
    \texttt{EXCH} := ( \texttt{RSWAP} \otimes I_{A^I_w} ) \circ (\texttt{\texttt{SCNOT}} \otimes I_{A^O_1 \ldots A^O_n}), \end{equation}
or, equivalently, as $\texttt{EXCH} :=(\texttt{\texttt{SCNOT}} \otimes I_{A^I_1 \ldots A^I_n}) \circ ( \texttt{RSWAP} \otimes I_{A^O_w} )$.

%%%%%%%%%%%%%%%%%%%%%%%%%%%%%%%%%%%%%%%%%%%%%%%%%%%%%%%%%%%%%%%%%%%
\section{A lemma for Theorem \ref{theorem:R1 - R3 eq}} \label{appendix:proof-fg-equiv}
%%%%%%%%%%%%%%%%%%%%%%%%%%%%%%%%%%%%%%%%%%%%%%%%%%%%%%%%%%%%%%%%%%%

In order to prove Theorem \ref{theorem:R1 - R3 eq}, we will first state and prove an important lemma. Since this will take a long time, we do it here, in a separate appendix to the rest of the proof. The lemma will show that our sectorized no-influence relation through a unitary channel -- which is always equivalent to (FG4) below -- is equivalent to three other conditions, except in the case where one of the relata is a one-dimensional sector.

\subsection{Stating the Lemma}

In order to state these conditions, we first associate the sectors and the which-sector information with algebras of operators. We associate the $i$th sector of $A^i$ with the algebra:
\begin{equation} \label{sectorial relatum}
    \ca^i:= \cl(A^i_\star) \oplus \textnormal{span}(I_{A^{\overline{i}}_\star})
\end{equation}
where $\cl(A^i_\star)$ is the algebra of linear operators on $A^i_\star$, $I_{A^{\overline{i}}_\star}$ is the identity operator on $A^{\overline{i}}_\star := \bigoplus_{i' \neq i} A_\star^{i'}$, and the span is over complex numbers (as it always will be subsequently). In other words, $\ca^i$ is the algebra of operators of the form $M_{A^i_\star} \oplus \alpha I_{A^{\overline{i}}_\star}$, where $M$ is an operator on $A^i_\star$, and $\alpha$ is a complex number. 

We also define an algebra corresponding to the which-sector information:
\begin{equation} \label{which sec? relatum}
    \ca^w := \textnormal{span}(\{ \pi^A_i \}_i)
\end{equation}
where the $\pi_i^A$ are the projectors used to define $A^i$'s sectorization: $\pi_i^A(A) =A_\star^i \ \forall i$. 

The algebras associated with $D^l$'s sectors and which-sector information are defined in a precisely analogous way. We now proceed to give conditions for sectorized no-influence relations (in the absence of one-dimensional sectors) which largely resemble the unsectorized no-influence relations but for the fact that they are defined in terms of these smaller algebras, rather than the full sector-preserving algebras of the form (\ref{spalgebra111})

%Since the sectors and which-sector information are in one-to-one correspondence with their associated algebras, one can talk about the causal relations as obtaining between the algebras themselves. In this appendix, we adopt this practice to make things less clunky. We therefore call the algebras defined above the \textit{sectorized relata} of causal relations.

 Given an algebra $\cx$ corresponding to one of $A^i$'s sectors or which-sector information, and given an algebra $\cy$ corresponding to one of $D^l$'s sectors or which-sector information, we are now ready to state the four conditions for no-influence between these two relata. In (\ref{FG1}) below, $\Phi$ is a channel with Kraus operators from $\cx$, and $\Psi$ is a channel with Kraus operators from $\cy \otimes \cl(* \rightarrow D^I)$\footnote{These are the operators of the form $\sum_k M_k \otimes \ket{\psi_k}$ for $M_k \in \cy$ and $\ket{\psi_k} \in D^I$.} for some ancillary space $D^I$. 
\begin{equation} \label{FG1} \tag{FG1}
\begin{split} 
    \scalebox{.8}{\input{NI_R1.tikz}} \\
     \forall \Phi \ \forall \Psi 
\end{split} 
\end{equation}

The second condition is given in the case that $\cy$ is a which-sector algebra (\ref{FG2a}), and separately in the case that it corresponds to a sector (\ref{FG2b}). In both cases, $V$ is any unitary operator from $\cx$. In (\ref{FG2a}) directly below, $W$, $\dot{D}$, and the routes all have the same meaning as in (\ref{R2}). $V'$ is a practical unitary controlled on $\dot{D}$ of the form $\sum_l V'_l \otimes \ket{l}\bra{l}_{\dot{D}}$, where $\{\ket{l}\}_l$ is the basis associated with $\dot{D}$'s sectorization. 
\begin{equation} \label{FG2a} \tag{FG2a} \begin{split}
     \input{NI_FG2a.tikz}  \\ 
     \forall V \ \exists V'
\end{split}
\end{equation}
Note that this equality is only required to hold for the transformations, and not necessarily the routes. 

On the other hand, in (\ref{FG2b}), $\dot{E}^u$ is a sectorized Hilbert space with two one-dimensional sectors, and $(L, \epsilon)$ is a routed unitary transformation that maps the specific sector that is associated with $D_\star^{q}$ associated with $\cy$ to ${D}^{q}_\star \otimes E^0_\star$, and maps the rest of the space $D_\star^{\overline{q}}$ to ${D}^{\overline{q}}_\star \otimes E^1_\star$, where $D_\star^{q}$ is the sector corresponding to $\cy$.
The route matrix $\epsilon_n^{pu}$ is equal to 1 if  $n=p=q$ and $u=0$, or if $n=p\neq q$ and $u=1$; otherwise, it is zero.
$V$ is again any unitary on $\cx$, and the controlled practical unitary labelled $V''$ is of the form $V''=   V_0'' \otimes \pi^D_q \otimes \ket{0}\bra{0}_E + V_1'' \otimes \ket{1}\bra{1}_E$,
where $V_0''$ acts on $C$, $V_1''$ acts on $C \otimes D$, $\pi^D_q$ is the projector in $D$ onto $D_\star^q$, and each $\ket{u}\bra{u}_E$ spans the sector $E_\star^u$. 

\begin{equation} \label{FG2b} \tag{FG2b} \begin{split}
  \input{NI_FG2b.tikz}  \\ 
     \forall V \ \exists V''  
\end{split}
\end{equation}
Again, the equation is not required to hold for the routes.

In (\ref{FG3}), $\cx^*$ is the algebra obtained by embedding $\cx$ into $U$'s practical input space. $\cy^*$ is the algebra obtained by embedding $\cy$ in $U$'s practical output space.
\begin{equation} \tag{FG3} \label{FG3}
    [U \cx^* U^\dagger, \cy^*] = 0
\end{equation}

(\ref{FG4}) is a statement of a no-influence relation through an \textit{unrouted} unitary transformation, conforming to Definition \ref{definition:unroutedni}. It is given in terms of the following unitary, which we denote by $S_U$:
\begin{equation} \label{fine-grained rels} 
    \input{fine_grained_rels_v2.tikz}
\end{equation}
Here, $T_{\textnormal{prac, in}}$ and $T_{\textnormal{prac, out}}$ are practical unitaries that embed $U$'s practical input and output space respectively into an unsectorized system. The routes in this diagram are given by $\mu_{ij}:= \sum_{kl} \lambda_{ij}^{kl}$ and $\gamma_{kl}:= \sum_{ij} \lambda_{ij}^{kl}$ . Also, in (\ref{FG4}), $X^O$ stands for the input subsystem related to $\cx$; either $A^O_i$ if $\cx=\ca^i$ or $A^O_w$ if $\cx=\ca^w$. Similarly,  $Y^I=D^I_l$ if $\cy=\cd^l$ or $Y^I=D^I_w$ if $\cy=\cd^w$.
\begin{equation} \label{FG4} \tag{FG4}
   X^O \not\xrightarrow{S_U} Y^I  
\end{equation}

%The definition of the unitary operator $W$ depends on the nature of the output relatum $\cy$. If $\cy$ is a `which-sector?' relatum of the form (\ref{which sec? relatum}), then it is defined as it was in (\ref{R2}) to map $D_\star^l$ to $\dot{D}^l_\star \otimes D^l_\star$ for all $l$, and $\dot{D}^l_\star$ is defined accordingly as consisting of the same number of sectors as $D^l$, but each of them one-dimensional. If, on the other hand, the $l$th `sectorial relatum', then, defining $D_\star^{\overline{l}}:= \bigoplus_{l' \neq l} D_\star^{l'}$, then $W$ maps  $D_\star^l$ to $\dot{D}^l_\star \otimes D^l_\star$ and $D_\star^\overline{l}$ to $\dot{D}^{\overline{l}}_\star \otimes D^\overline{l}_\star$, where $D^l$ is a qubit space with a sectorization into two one-dimensional sectors. $V'$ is again a unitary operation coherently controlled by the value of $\dot{D}^l$. The routes in the diagrams have the same meanings as in (\ref{R1}) and (\ref{R2}), except we have introduced $\epsilon$ whose definition again depends on the nature of $\cy$. If $\cy$ is a `which-sector?' relatum then $\epsilon_n^u=\delta_n^u$; otherwise $\epsilon_n^u$ is equal to 1 if $n=l$ and $u=0$ or if $n \neq l$ and $u=1$, and is 0 otherwise.

We define trivial sectorial relata as those algebras corresponding to sectors that are one-dimensional. We can now state the lemma:

\begin{lemma} \label{theorem:FGequiv}
Assuming neither $\cx$ nor $\cy$ correspond to trivial sectorial relata, conditions (FG1 -- FG4) are equivalent.
\end{lemma}
We will now prove the lemma.

\subsection{(FG2) is equivalent to (\ref{FG3})}

We start by proving that (FG2) $\Leftrightarrow$ (\ref{FG3}). For the rightward implication, it is immediate that the transformations on the right sides of (\ref{FG2a}) and (\ref{FG2b}) commute with $\cd^{w*}$ and $\cd^{l*}$ respectively. The implication then follows from the fact that $\ca^{w*}$ and each $\ca^{i*}$  are spanned by their unitaries. This in turn follows from the fact that $\ca^{w}$ and each $\ca^{i}$ are spanned by their unitaries, and from the linearity and invertibility of the embedding map. 

For the leftward direction, suppose first that $\cy=\cd^{w*}$. Then (\ref{FG3}) implies that, for any unitary $V \in \cx$, the unitary transformation in:
\begin{equation} \label{fg2-lhs}
    \input{FG2_LHS.tikz}
\end{equation}
commutes with each projector $Q_l \otimes \pi^D_l$, where $\pi^D_l$ is the projector onto $D_\star^l$, and $Q_l$ is the projector onto the largest orthogonal subspace of $C$, call it $c_l$, such that $c_l \otimes D_\star^l$ is part of $U$'s practical output space. This means that the transformation in (\ref{fg2-lhs}) can be written in the form $V' = \bigoplus_l V'_l$, where each $V'$ acts on $c_l \otimes D_\star^l$. It follows that it can be written as in the right-hand side of (\ref{FG2a}).

Now suppose $\cy=\cd^{l*}$. Set $H_l := c_l \otimes D^l$, and define $H_{\overline{l}}$ as the compliment of $H_l$ in $U$'s practical output space, $H_\textnormal{acc}$:
\begin{equation} \label{sectorization}
    H_\textnormal{acc} = H_l \oplus H_{\overline{l}}
\end{equation}
(\ref{FG3}) implies that any unitary $V \in \cu(\cx^*)$ commutes with the projector $Q_l \otimes \pi^D_l$ in $H_\textnormal{acc}$ onto $H_l$, and is therefore is block diagonal in this two-sector partitioning: $V=V_l \oplus V_{\overline{l}}$. (\ref{FG3}) implies that $V$ must commute with $(Q_l \otimes \cl(D_\star^l)) \oplus \textbf{0}_{\overline{l}}$, meaning that $V$ has the form $V= (V'_{c_l} \otimes I_{D_\star^l}) \oplus V_{\overline{l}}$. Any unitary operator of this form can be written as in the right-hand side of (\ref{FG2b}).

\subsection{(\ref{FG3}) implies (\ref{FG1})}

We now prove (\ref{FG3}) $\implies$ (\ref{FG1}). The reasoning is very similar to that in the proof that (\ref{U6}) implies (\ref{U2}), but it is carried out again explicitly here for the sake of completeness.

First we note that the transformation (although not necessarily the route) in the left-hand side of (\ref{FG1}) is equal to the one in
\begin{equation} \label{fg3 then fg1 1}
    \input{fg3_then_fg1_v2.tikz}
\end{equation}

%This is not completely trivial since $\cu$ is only a \textit{practical} unitary channel, so $\cu^\dag \circ \cu$ does not necessarily give the identity channel on the whole formal space. However, the equality of the transformations easily follows from the fact that they have the same action on all input states with $U$'s practical input state, and give 0 or `crash' for all terms outside it, as well as for all coherence terms between the accessible space and the rest of the space. 

(\ref{FG3}) implies the transformation in (\ref{fg3 then fg1 1}) is equal to the one in
\begin{equation}
    \input{fg3_then_fg1_2.tikz}
\end{equation}
since it entails that all Kraus operators of $\Phi$, once transformed by $\cu$ into operators acting on $U$'s output space, commute with all the Kraus operators of $\Psi$. Then we can trace out the channels after $\Psi$ to show that the transformation we have been discussing all along is equal to the one in:
\begin{equation}
    \input{fg3_then_fg1_3.tikz}
\end{equation}
This implies (\ref{FG1}).

\subsection{Sublemmas}

We state and prove two sublemmas required for the rest of the proof.

\begin{sublemma} \label{lemma:two scnots}
$A^O_w$ is a not cause of $D^I_w$ in (\ref{fine-grained rels}) if and only if $A^O_w$ is a not cause of $D^I_w$ in:
\begin{equation} \label{two scnots}
    \input{two_SCNOTS.tikz}
\end{equation}
\end{sublemma}

The `if' part is easily shown by partially tracing all the outputs except $D^I_w$ in (\ref{fine-grained rels}) and noting from the definition of the $\texttt{EXCH}$ gate that the \texttt{SCNOT} and \texttt{RSWAP} gates commute. The `only if' part follows by considering the following channel:
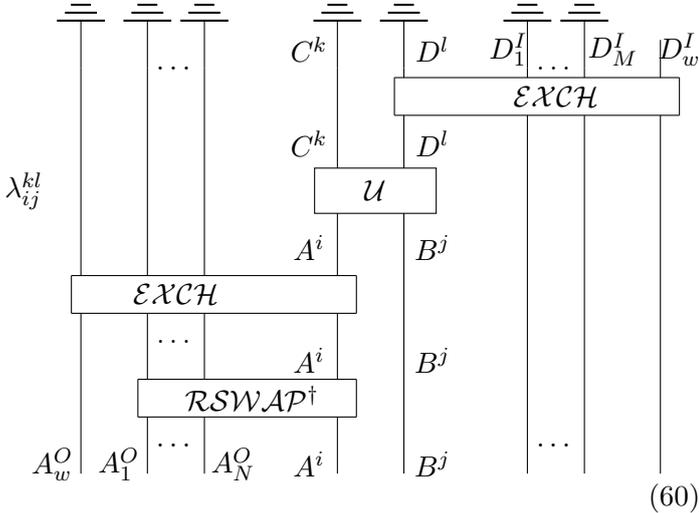
\begin{equation}
   \hspace{-1.0cm}  \input{fine-grained_rels.tikz}
\end{equation}
which can be simplified by noting that the $\mathcal{R} \cs \cw \ca \mathcal{P}$ on $D$ will be traced out and the one on $A$ will give the identity once combined with its adjoint. This leads to an expression from which one can trivially derive that $A^O_w$ is not a cause of $D^I$ in (\ref{two scnots}).

\begin{sublemma} \label{lemma:scnot-kruas} 
For any density matrix $\rho$
\begin{equation} \label{scnot state traced}
    \input{scnot_state_traced.tikz}
\end{equation}
and
\begin{equation} \label{scnot state}
    \input{scnot_state.tikz}
\end{equation}
have Kraus operators drawn from $\ca^{w}$ and $\cd^{w} \otimes \cl(* \rightarrow D^I_w)$ respectively.
\end{sublemma}

We denote the channels in (\ref{scnot state traced}) and (\ref{scnot state}) by $\cm$ and $\cn$ respectively. 
For any pair of states $\ket{\psi_i}$ and $\ket{\phi_i}$ both contained in $A^i_\star$, we have $\cm(\ket{\psi_i}\bra{\phi_i})=\ket{\psi_i}\bra{\phi_i}$. 
This means that not only does $\cm$ follow the delta-route depicted in (\ref{scnot state traced}) associated with the sectorization $H_\textnormal{acc}^\textnormal{in}=\bigoplus_i A^i_\star$, it also follows any similar delta-route obtained by also further sectorizing each sector $A^i_\star$ into a set of one-dimensional sectors. 
By Theorem 6 of \cite{vanrietvelde2020routed}, this implies that the Kraus operators of $\cm$ must be block diagonal in all of the sectorizations, implying that they are each of the form $\sum_i \alpha_i \pi^A_i$ for complex numbers $\alpha_i$, and hence a member of $\ca^{w}$. 

One can give a very similar argument that the Kraus operators $\cn$ must be block diagonal in the sense of belonging to $\bigoplus_l (\oplus_{m_l} \cl((D_\star^l)^{m_l}) \otimes \cl(* \rightarrow D^I_w)$ for any set of sectorizations $D_\star^l = \bigoplus_{m_l} (D_\star^l)^{m_l}$ of each $D^l_\star$ into one-dimensional sectors. This implies that each Kraus operator must be of the form $\sum_l \alpha_l \pi^D_l \otimes \ket{\xi_l}_{D^I_w}$, hence a member of $\cd^{w} \otimes \cl(* \rightarrow D^I_w)$.

\subsection{(\ref{FG1}) implies (\ref{FG4})}

Now we prove that (\ref{FG1}) implies (\ref{FG4}). We start by assuming $\cx=\ca^{w*}$ and $\cy=\cd^{w*}$. Given Sublemma \ref{lemma:scnot-kruas}, (\ref{FG1}) implies that, for any normalized density matrix $\sigma$,
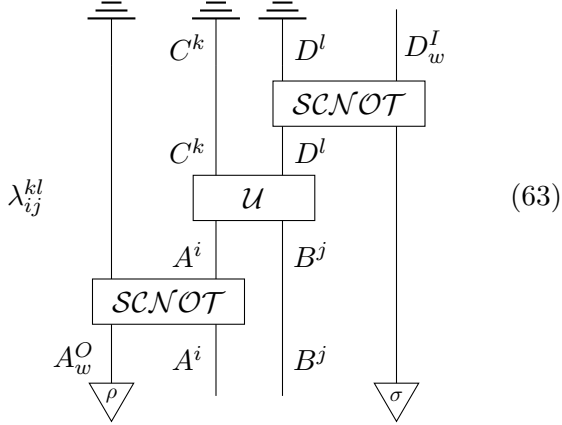
\begin{equation} \label{two scnots traced 1}
    \input{two_SCNOTS_traced_1.tikz} 
\end{equation}
is independent of the density matrix $\rho$.  Since $\rho$ is normalized, it follows that, for an arbitrary state $\rho'$ on $A^O_w$, the above is equal to:
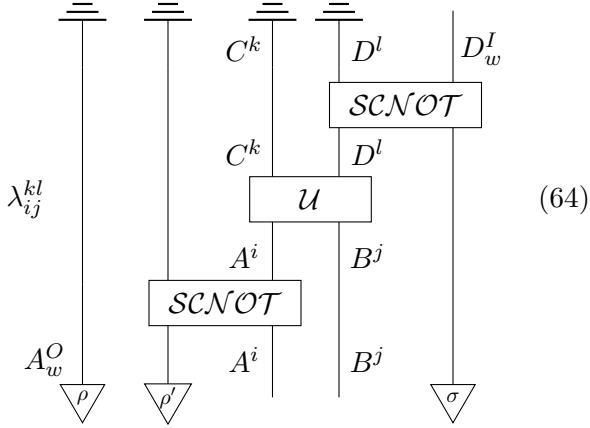
\begin{equation} \label{two scnots traced 2}
    \input{two_SCNOTS_traced_2.tikz}
\end{equation}
for any $\rho$ and all $\sigma$. We can use state tomography where we vary over $\rho$ and $\sigma$ but hold $\rho'$ fixed to deduce that:
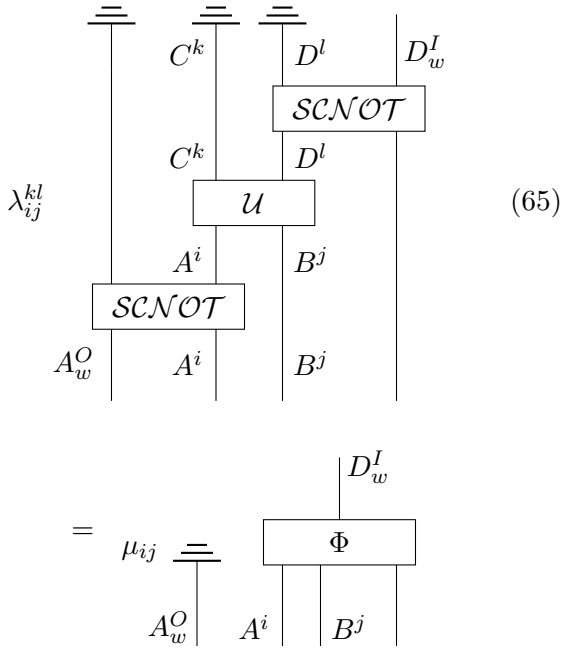
\begin{equation} \label{two scnots traced 3}
 \input{two_SCNOTS_traced_3.tikz}
\end{equation}
where $\mu$ is the matrix obtained by composing the route matrices on the left, and $\Phi$ is the following quantum channel:
\begin{equation}
    \input{phi.tikz}
\end{equation}

Given Sublemma \ref{lemma:two scnots}, (\ref{two scnots traced 3}) implies (\ref{FG4}).

The argument in the case that at least one of $\cx$ and $\cy$ corresponds to a sector is closely related to the first one. One just has to note that the \texttt{RSWAP} gate on $D^l$ (for example) can be written as a sequentially combination of commuting operators, on associated with each of $D^l$'s $M$ sectors:
\begin{equation}
    \texttt{RSWAP} = (I_{D_{\overline{M}}} \otimes R_M) \circ ... \circ ( I_{D_{\overline{1}}} \otimes R_1)
\end{equation}
where $I_{D_{\overline{l}}}$ is the identity from $\bigoplus_{l' \neq l} D^O_{l'}$ to $\bigoplus_{l' \neq l} D^I_{l'}$, and, for all $l$, 
\begin{equation}
    R_l := \texttt{SWAP}_l \oplus (I_{D_l} \otimes I_{D_\star^{\overline{l}}})
\end{equation}
where $I_{D_l}$ is the identity operator from $D^O_l$ to $D^I_l$, $I_{D_\star^{\overline{l}}}$ is the identity on $D_\star^{\overline{l}}=\bigoplus_{l' \neq l} D_\star^{l'}$, and SWAP$_l$ swaps the sector $D_\star^l$ with the ancillary wire $D^O_l$/$D^I_l$. Then the foregoing argument can be generalized to the case where at least one of $\cx$ and $\cy$ is a sectorial relatum by proving a sublemma similar to \ref{lemma:two scnots} where the \texttt{SCNOT} gates on any wires corresponding to the sectorial relata $\ca^i$ or $\cd^l$ are  replaced with the corresponding $R_i$ or $R_l$ -- these sublemmas are proved using similar arguments to the ones above. Then, maintaining the relevant substitutions, one repeats the argument encapsulated by equations (\ref{two scnots traced 1} -- \ref{two scnots traced 3}), which, importantly, does not depend on any special properties of \texttt{SCNOT}.

\subsection{(\ref{FG4}) implies (\ref{FG3})}

Finally, we prove that (\ref{FG4}) implies (\ref{FG3}), assuming the relata are nontrivial. The argument varies depending on whether the non-cause and non-effect are sectors or which-sector information. However, it will suffice to prove the claim in the two cases where both relata are of the same kind, before indicating how the arguments are to be adapted in the other two cases.

%F def :=\sum_{m, n=0}^{D-1}\textnormal{exp}(\frac{2\pi imn}{D})\ket{m}\bra{n}

We start by assuming that the no-influence relation is between the which-sector informations, so our algebras are $\cx=\ca^{w*}$ and $\cy=\cd^{w*}$. Let $F$ denote the quantum Fourier transform operator on the ancilla input $A^O_w$ to the \texttt{SCNOT} gate on $A$. Consider the unitary operator on $A$ obtained by inserting the state $F\ket{-n}= \frac{1}{\sqrt{N}} \sum_{m=0}^{N-1} \textnormal{exp}(\frac{-2\pi imn}{N}) \ket{m}$ into the ancilla input of the \texttt{SCNOT} gate, and ignoring the output state on $A^I_w$. This can be written as:
\begin{equation} \label{Vn}
    V_n = \sum_m \textnormal{exp}(\frac{2\pi imn}{N}) \pi^A_m
\end{equation}
where each $\pi^A_m$ is the projector onto $A^m_\star$.\footnote{More precisely, the point is that $\texttt{SCNOT}(I_A \otimes F\ket{-n}_{A_w^O}) = V_n \otimes F\ket{-n}_{A_w^I}.$} Denoting by $P_m$ the projector onto the largest  orthogonal subspace $b_m$ of $B$ such that $A^m_\star \otimes b_m$ is inside $U$'s practical input space, $V_n$ is embedded into the practical input space as follows (see Appendix \ref{appendix:embedding}):
\begin{equation}
    V_n^* := \sum_{m=0}^{N-1} \textnormal{exp}(\frac{2\pi imn}{N}) \pi^A_m \otimes P_m
\end{equation}

Since, by the unitary of $F$, $\{  \frac{1}{\sqrt{N}} \sum_{m=0}^{N-1} \textnormal{exp}(\frac{2\pi imn}{N}) \ket{m}\}_{n=0}^{N-1}$ spans any computational basis state $\ket{n}$ for $A$, it is clear that $\{V_n^* \}_{n=0}^{N-1}$ spans any projector $\pi^A_n \otimes P_n$, and thus that the same set spans $\ca^{w*}$. It will therefore suffice to show that $\{\cu(V_n^*) \}_{n=0}^{N-1}$ commutes with $\cd^{w*}$, for any $V_n^*$.

By Sublemma \ref{lemma:two scnots}, (\ref{FG4}) implies that
\begin{equation}
    \input{two_scnots_traced_4.tikz}
\end{equation}
is independent of $\rho$. Define $W_n:=\cu(V_n^*)$ and $\cw_n$ as the associated unitary channel, and set $\mathcal{P}_\textnormal{acc}:=P_\textnormal{acc}(\cdot)P_\textnormal{acc}$ for the projector $P_\textnormal{acc}$ onto $U$'s practical output space. Then, by equating the cases where $\rho=F\ket{n}\bra{n}F^\dag$ and $\rho=F\ket{0}\bra{0}F^\dag$, one can easily show that:
\begin{equation} \label{fg4 then fg3 one}
 \input{fg4_then_fg3_one.tikz}
\end{equation}
where we choose $\tau$ to be $\tau= \ket{0} \bra{0}$, and for all density matrices $\sigma$ on $U$'s practical output space.

Consider the sectorization of the practical output space of $U$ given by $H_\textnormal{acc} = \bigoplus_l (c_l \otimes D^l_\star$), where each $c_l$ is an orthogonal subspace of $C$. In the following, we consider states in this space, where a subscript $l$ on a state vector indicates that it is wholly contained in the corresponding sector $c_l \otimes D_\star^l$. Suppose, for contradiction, that $\exists \ket{\psi_0}: \ W_n\ket{\psi_0}= \sum_{l=0}^{M-1} a_l \ket{\phi_l}$, where at least one of the complex numbers $a_m$ for $m \neq 0$ is non-zero. If we substitute $\sigma= \ket{\psi_0}\bra{\psi_0}$ into (\ref{fg4 then fg3 one}), it is easy to show that the resulting state on the left-hand side is $\sum_{l=0}^{M-1} |a_l|^2 \ket{l}\bra{l}$, whereas the right-hand one is $\ket{0}\bra{0}$, violating the equality. Hence $\cu(V_n^*)$ maps $c_0 \otimes D^0_\star$ to itself. Of course, the choice of the $0$th sector was arbitrary, meaning that $\cu(V_n^*)$ maps each of the $M$ subspaces to itself, and is thus is block diagonal in the sectorization $H_\textnormal{acc} = \bigoplus_l c_l \otimes D^l_\star$. Hence it commutes with $\cd^{w*}$. Since this is true for all $n$, and $\{W_n \}_{n=0}^{D-1}$ spans $\ca^{w*}$, it follows that $[U \ca^{w*} U^\dag, \cd^{w*}]=0$.

Now, suppose instead that both relata correspond to the zeroth sector, so our algebras are $\cx=\ca^{0*}$ and $\cy=\cd^{0*}$. As discussed in the previous subsection, one can derive a version of Sublemma \ref{lemma:two scnots} for these relata in which the \texttt{SCNOT} channels in the sublemma are replaced with channels $\mathcal{R}_0$ corresponding to $R_0$. This, together with (\ref{FG4}), implies that the following is independent of the choice of unitary channel $\cv=V(\cdot)V^\dag$:
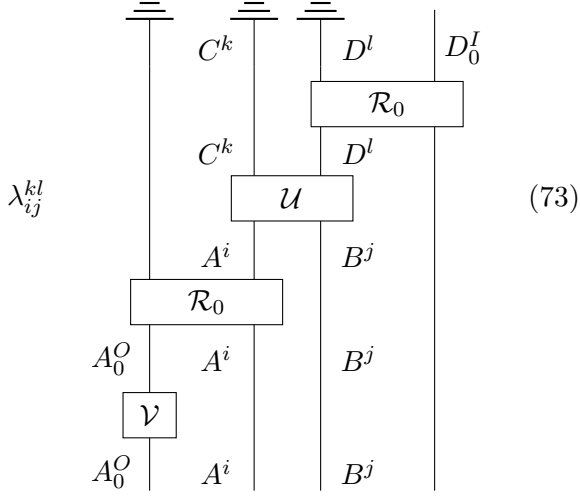
\begin{equation}
    \input{two_r0s_traced_one.tikz}
\end{equation}
from which it follows that:
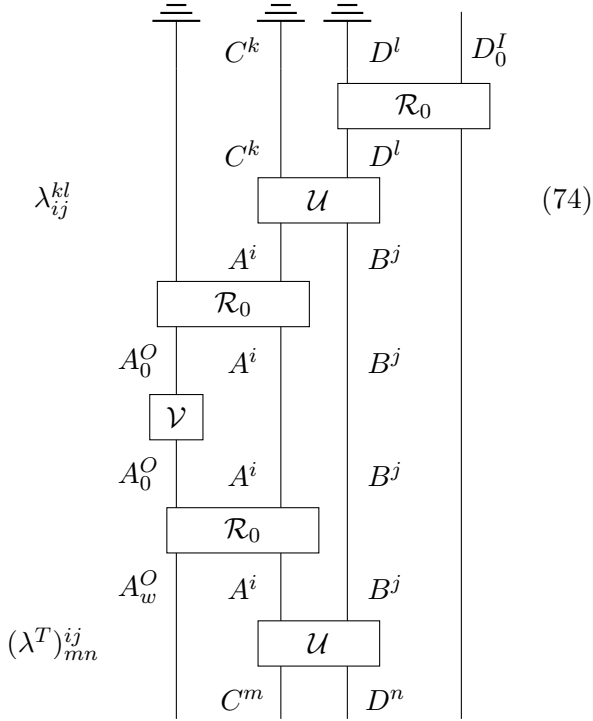
\begin{equation} \label{two r0s traced two}
    \input{two_r0s_traced_two.tikz}
\end{equation}
is also independent of $\cv$. 

Suppose that $\ket{\psi}$ is an eigenstate of the unitary operator $V$ on $A^O_0$. We define a unitary operator $V'$ on $A$:
\begin{equation} \label{V' form}
    V' =  V_0 \oplus \lambda_\psi^V I_{\overline{0}}
\end{equation}
where $\lambda_\psi$ is the eigenvalue associated with $\ket{\psi}$, $V_0$ is a unitary operator of the form $V$ on the sector $A^0_\star$, and $I_{\overline{0}}$ is the identity operator on $A_\star^{\overline{0}}$.

We further define $W_{V'} := \cu(V'^*)$ and denote by $\cw_{V'}$ the corresponding unitary channel. By inserting $\ket{\psi}\bra{\psi}$ into the input $A^O_0$ in (\ref{two r0s traced two}) and equating the  general case to the special case where $V$ is the identity, we deduce that:
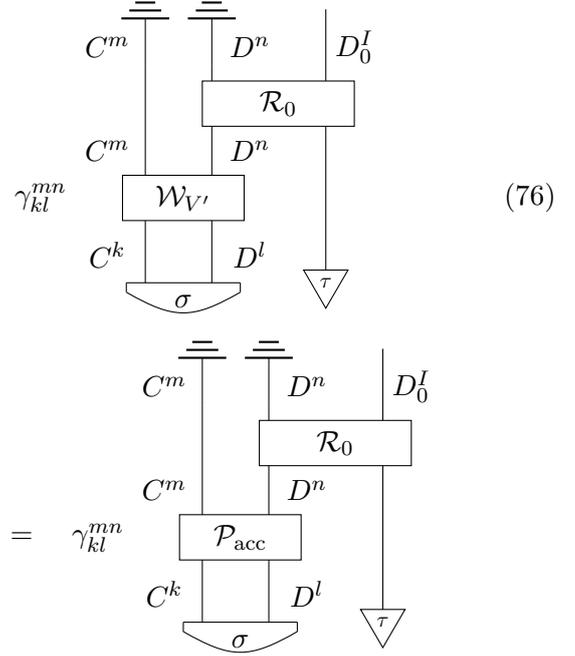
\begin{equation} \label{fg4 then fg3 two}
    \input{fg4_then_fg3_two.tikz}
\end{equation}
for arbitrary states $\sigma$ and $\tau$, where the route $\gamma_{kl}^{mn}:=\sum_{ij}\lambda_{ij}^{kl} (\lambda^T)_{mn}^{ij}$. 

We will proceed to show that $W_{V'}$ commutes with $\cd^*$, starting by showing that it is block diagonal in the sectorization (\ref{sectorization}) of $U$'s practical output space for $l=0$.  Consider some states $\ket{\psi_0}$ and $\ket{\phi_0}$ in $H_0 := c_0 \otimes D_\star^0$, and $\ket{\phi_{\overline{0}}}$ in the compliment space $H_{\overline{0}}:= \bigoplus_{l\neq0} c_l \otimes D_\star^l$. Suppose, for contradiction, that $W_{V'}\ket{\psi_0}= a\ket{\phi_0} + b \ket{\phi_{\overline{0}}}$ for $b \neq 0$. Then, substituting $\sigma = \ket{\psi_0}\bra{\psi_0}$ into both sides of (\ref{fg4 then fg3 two}) gives that $|a|^2 \textnormal{Tr}_{C} \ket{\phi_0}\bra{\phi_0}_{CD^I_0} + |b|^2 \tau_{D^I_0} = \textnormal{Tr}_{C} \ket{\psi_0}\bra{\psi_0}_{CD^I_0}$. It is always possible to choose a $\tau_{D^I_0}$ so that this equality is violated. Hence $W_{V'}$ maps $H_0$ to itself. Then unitarity implies that $W_{V'}$ is block diagonal in the sectorization $H_{\textnormal{acc}}=H_0 \oplus H_{\overline{0}}$.

We can therefore write 
\begin{equation} \label{W exp}
    W_{V'} = W_{V'}^0 + W_{V'}^{\overline{0}}
\end{equation}
where the operators on the right-hand side act on $H_0$ and $H_{\overline{0}}$ respectively. Now, by considering arbitrary choices of $\sigma$ contained in $\cl(c_0 \otimes D_\star^0)$, we can use tomography to deduce from (\ref{fg4 then fg3 two}) that
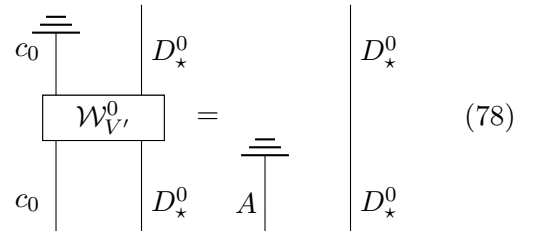
\begin{equation}
    \input{hellot.tikz}
\end{equation}
to which we can apply the essential uniqueness of purification to deduce that $W_{V'}^0 = T_{c_0} \otimes I_{D_\star^0}$ where $I_{D_\star^0}$ is the identity on $D_\star^0$ and $T_{c_0}$ is some unitary on $c_0$. Substituting this into (\ref{W exp}), we can see that $W_{V'}$ commutes with $\cd^{0*}$.

%%%%%
We now want to show that the unitary operators of the form $V'$ from (\ref{V' form}) span $\ca^{0}$. We will do this by explicitly showing how one can construct any operator from $\ca^{0}$ by taking linear combinations of operators of the form $V'$. 

In the case where $V$ is the identity operator (up to global phase) on $A^O_0$, $V'$ is the identity (up to global phase) on $A$. When $V$ is not the identity (up to global phase), it has at least two distinct eigenvalues. We can therefore obtain the following two operators from $V$, assuming that $A_0^O$ has a dimension greater than one (i.e. that the relatum $A_\star^0$ is nontrivial).
\begin{equation} \label{span}
    \begin{split}
       & V_0 \oplus e^{i\alpha} I_{\overline{0}} \\
        & V_0 \oplus e^{i\beta} I_{\overline{0}}
    \end{split}
\end{equation}
for $e^{i\alpha} \neq e^{i\beta}$. Subtracting the bottom operator from the top then gives us the operator $\textbf{0}_0  \oplus (e^{i\alpha}-e^{i\beta}) I_{\overline{0}}$, which of course spans all operators of the form
\begin{equation}\label{span2}
  \textbf{0}_0  \oplus aI_{\overline{0}}
\end{equation} 
for arbitrary complex numbers $a$. These operators together with those in (\ref{span}) span the operators of the form:
\begin{equation}\label{span2.5}
    V_0 \oplus aI_{\overline{0}}
\end{equation} 
for unitary $V_0$.

Note that since we can start with any unitary $V$, we can end up with any unitary $V_0$ in this expression. The operators of this form for unitary $V_0$ span the linear operators of the form:
\begin{equation}\label{span3}
    M \oplus aI_{\overline{0}}
\end{equation}
(as is easily seen from the fact that the unitary operators on a Hilbert space span the linear operators on that space). But these are precisely the operators in $\ca^0$. This shows that the operators of the form $V'$ span $\ca^0$. As discussed in Appendix \ref{appendix:embedding}, this implies that operators of the form $V'^*$ span $\ca^{{0}*}$. (\ref{FG3}) follows.
%%%%%

The cases where the two relata are different in kind are proven by simply mixing arguments for the two cases in which they are the same. For example, if $\cx=\ca^{w*}$ and $\cy=\cd^{l*}$, then one repeats the argument from the first case that one can induce a set of unitary transformations of the form $V_n$ from (\ref{Vn}) on $A$ by putting in states to the ancilla input of the \texttt{SCNOT} gate of the form $F\ket{n}$, and that the associated embedded unitaries $V_n^*$ span $\ca^{w*}$. Then one applies the argument above to show that (\ref{FG4}) implies that these operators, once transformed by $\cu$ to the practical output space, must commute with $\cd^{l*}$, implying that $[\cu(\ca^{w*}), \cd^{l*}]$. Upon making a similar adaptation for $\cx=\ca^{i*}$ and $\cy=\cd^{w*}$, the theorem is proved.

%%%%%%%%%%%%%%%%%%%%%%%%%%%%%%%%%%%%%%%%%%%%%%%%%%
\section{Proof of Theorem \ref{theorem:R1 - R3 eq}} \label{appendix: proof routed NI equivalence}
%%%%%%%%%%%%%%%%%%%%%%%%%%%%%%%%%%%%%%%%%%%%%%%%%%
We will prove Theorem \ref{theorem:R1 - R3 eq} by assuming Lemma \ref{theorem:FGequiv}, which is proven independently in Appendix \ref{appendix:proof-fg-equiv}. 
 
Given the routed unitary (\ref{routed unitary}), denote by (fg) the proposition that a relation of no-influence holds between each sectorized relatum associated with $A^i$ and each sectorized relatum associated with $D^l$. We will show that (\ref{R2}) is equivalent to (\ref{R3}), that (\ref{R3}) and (\ref{R4}) are each equivalent to (fg), that (\ref{R3}) implies (\ref{R1}) and that (\ref{R1}) implies (\ref{R4}).

We start with (\ref{R2}) $\Leftrightarrow$ (\ref{R3}). The rightward direction follows from the fact that the right side of (\ref{R2}) commutes with $\cd^*$, and the fact that $\cu(\ca^*)$ is spanned by its unitaries. 
For the leftward direction, consider an arbitrary unitary $V \in \ca$, and the associated unitary $V^*$ embedded in $U$'s practical input space (see Appendix \ref{appendix:embedding} for the definition of the embedding map). 
Consider also the decomposition of $U$'s practical output space of the form $H_\textnormal{acc} =\bigoplus_l c_l \otimes D_\star^l$, where each $c_l$ is a subspace of $C$, and consider the projectors $Q_l \otimes \pi^D_l$ onto each term.
(\ref{R3}) implies that $\cu(V^*)$ commutes with all of these projectors.
%If $V^*$ is the embedded unitary (see Appendix \ref{appendix:embedding}), then (\ref{R3}) implies that $\cu(V^*)$ commutes with all projectors of the form $ Q_l \otimes \pi^D_l$ where $Q_l$ is a projector on $C$ onto the largest subspace of $C$, $c_l$, such that $c_l \otimes D_\star^l$ is contained in $U$'s practical output space. 
%we note that (\ref{R3}) implies that any unitary $\cu(V^*)$ such that $V \in \ca$ commutes with all projectors of the form $ Q_l \otimes \pi^D_l$ where $Q_l$ is a projector on $C$ onto the largest subspace of $C$, $c_l$, such that $c_l \otimes D_\star^l$ is contained in $U$'s practical output space. 
Hence $\cu(V^*)$ is block diagonal in the sectorization $H_\textnormal{acc} = \bigoplus_l c_l \otimes D^l_\star$, and may be written $\cu(V^*) = \bigoplus_l W_l$. (\ref{R3}) also implies that each $W_l$ commutes  with an abritrary operator of the form $I_{c_l} \otimes M_{D^l_\star}$, which implies that each $W_l$ can be written $\ \cu(V^*) = W'_{c_l} \otimes I_{D^l_\star}$. Hence $\cu(V^*) = \bigoplus_l W'_{c_l} \otimes I_{D^l_\star}$, from which (\ref{R2}) follows. 

The fact that (\ref{R3}) $\implies$ (fg) follows simply from Lemma \ref{theorem:FGequiv}, and the fact that the algebras associated with the  sectors and which-sector information (see Appendix \ref{appendix:proof-fg-equiv}) are subalgebras of the algebra associated with the sectorized system. For the converse direction, note that the union of all those subalgebras spans the algebra associated with the sectorized system. In other words, denoting the union of the algebras associated to the multi-dimensional sectors and which-sector information of $A^i$ as $S_A$, $S_A$ spans $\ca$. Similarly, $S_D$ spans $\cd$. It follows from $S_A$ spanning $\ca$ that $S_A^*$ spans $\ca^*$ (see Appendix \ref{appendix:embedding}). By (fg) and  Lemma \ref{theorem:FGequiv}, we have $[\cu(S_A^*), S_D^*]=0$. (\ref{R3}) follows. This proves that (\ref{R3}) $\Leftrightarrow$ (fg).

For (fg) $\Leftrightarrow$ (\ref{R4}), consider the transformation in (\ref{sectorized rels}). By considering $U$'s practical output space  $\bigoplus_{kl}\sum_{ij} \lambda_{ij}^{kl}C_\star^k \otimes D_\star^l$ as a single Hilbert space, and similarly with its practical input space, one can think about (\ref{sectorized rels}) as an \textit{unrouted} unitary transformation. The fact that (fg) $\Leftrightarrow$ (\ref{R4}) is then straightforward consequence of Proposition \ref{unrouted atomicity}.

The fact that (\ref{R3}) $\implies$ (\ref{R1}) is given by exactly the same argument as the proof that (\ref{FG3}) $\implies$ (\ref{FG1}), except that now $\Phi$ and $\Psi$ can now be arbitrary sector-preserving quantum channels as in (\ref{R1}). One just needs to note that since $\Phi$ is sector-preserving, its Kraus operators are drawn from $\ca$, and since $\Psi$ is sector-preserving, its Kraus operators are drawn from $\cd \otimes \cl(*\rightarrow D')$. Both of these facts follow from Theorem 6 of \cite{vanrietvelde2020routed}. 

The argument for (\ref{R1}) $\implies$ (\ref{R4}) is very similar to the one encapsulated by equations (\ref{two scnots traced 1} -- \ref{two scnots traced 3}) upon replacing $\cs \cc \cn \co \ct$ with exchange channel $\ce \cx \cc \ch$; in fact it is slightly simpler since there is no need for Sublemma \ref{lemma:two scnots}.

%%%%%%%%%%%%%%%%%%%%%%%%%%%%%%%%%%%%%%%%%%%%%%%%%%%%%%%%%%%%%%
\section{Proof of Proposition \ref{ni-compositionality}} \label{appendix:ni-compositionality}
%%%%%%%%%%%%%%%%%%%%%%%%%%%%%%%%%%%%%%%%%%%%%%%%%%%%%%%%%%%%%
To prove (\ref{dag1}), note that the practical output space of the routed unitary (\ref{routed dag unitary}) can always be written in the form
\begin{equation}
    H_{\textnormal{acc}} = \bigoplus_{lmn} \omega^{lmn} D^l_\star \otimes E^m_\star \otimes F^n_\star
\end{equation}
where $\omega$ is a route matrix. Consider the sector-preserving algebras $\ce:= \bigoplus_m \cl(E_\star^m)$ and $\cf:= \bigoplus_n \cl(F_\star^n)$ over $E^m$ and $F^n$ respectively. Then the sector-preserving algebra for the composite system $E^m \otimes F^n$ is $\cx:=\bigoplus_{mn} \cl(E^m \otimes F^n) =\ce \otimes \cf$.

In Appendix \ref{appendix:embedding}, we defined a map $\xrightarrow{*}$ that embeds sector-preserving algebras into a practical output space. The embedded algebras  $\ce^*$, $\cf^*$ and $\cx^*$ have the form:
\begin{equation} \label{algebra-dsums}
    \begin{split}
        \ce^* &= \bigoplus_{lmn} \omega^{lmn} \ \ce^*_{lmn} \\
        \cf^* &= \bigoplus_{lmn} \omega^{lmn} \ \cf^*_{lmn} \\
        \cx^* &= \bigoplus_{lmn} \omega^{lmn} \ \cx^*_{lmn}
    \end{split}
\end{equation}
where we have defined:
\begin{equation}
    \begin{split} \label{dsum-comps}
        \ce_{lmn}^* :&= I_{D_\star^l} \otimes \cl(E^m_\star) \otimes I_{F^n_\star} \\
        \cf_{lmn}^* :&= I_{D_\star^l} \otimes I_{E^m_\star} \otimes \cl(F^n_\star) \\
        \cx_{lmn}^* :&= I_{D_\star^l} \otimes \cl(E^m_\star \otimes F^n_\star) \\
    \end{split}
\end{equation}
where $I_{D_\star^l}$ is the identity operator on $D^*_\star$, etc. 

Now, consider an arbitrary operator $M_{lmn} \in \cx^*_{lmn}$. Since the product operators on the space $E^m_\star \otimes F^n_\star$ span $\cl(E^m_\star \otimes F^n_\star)$, it is easily deduced from the algebra forms (\ref{dsum-comps}) that $M_{lmn}$ can be written 
\begin{equation}
    M_{lmn} = \sum_i \alpha_i (N_i^{lmn} \circ R_i^{lmn})
\end{equation}
where $N_i^{lmn} \in \ce_{lmn}^*$ and $R_i^{lmn} \in \cf_{lmn}^*$ for each $i$.

Since any operator in $\cx^*$ is just a direct sum of operators each from some $\cx_{lmn}^*$, it follows that any $M' \in \cx^*$ can be written in the form:
\begin{equation}
\begin{split}
    M' &= \bigoplus_{lmn} M_{lmn} \\ 
    &= \bigoplus_{lmn} \Big( \sum_i \alpha_i (N_i^{lmn} \circ R_i^{lmn}) \Big)
\end{split}
\end{equation}
where all $N_i^{lmn} \in \ce_{lmn}^*$ and all $R_i^{lmn} \in \cf_{lmn}^*$. If we define a new set of operators of the form $\tilde{N}_i^{lmn} := {N}_i^{lmn} \oplus ( \bigoplus_{l'm'n' \neq lmn} \textbf{0}_{l'm'n'})$, where $\textbf{0}_{l'm'n'}$ is the zero operator in $\ce_{l'm'n'}^*$, and similarly define $\tilde{R}_i^{lmn}$, then the direct sum above turns into a regular sum:
\begin{equation} \label{cx^*-comm-form}
    M' =  \sum_{lmni} \alpha_i (\tilde{N}_i^{lmn} \circ \tilde{R}_i^{lmn})
\end{equation}
where each $\tilde{N}_i^{lmn} \in \ce^*$ and each $\tilde{R}_i^{lmn} \in \cf^*$.

Now, suppose that $(A^i \not\xrightarrow{U} E^m) \land (A^i \not\xrightarrow{U} F^n)$. By (\ref{R3}), this means that 
\begin{equation} \label{commutes}
   [\cu(\ca^*), \ce^*]=[\cu(\ca^*), \cf^*]=0
\end{equation}

But (\ref{cx^*-comm-form}) shows that each operator from $\cx^*$ can be written as a sum of sequential compositions of operators from $\ce^*$ and $\cf^*$. It follows that $[\cu(\ca^*), \cx^*]=0$, implying that $A^i \not\xrightarrow{U} E^m \otimes F^n$.

Conversely, suppose $A^i \not\xrightarrow{U} E^m \otimes F^n$, and thus $[\cu(\ca^*), \cx^*]=0$. It is easily deduced from the expressions (\ref{algebra-dsums}) and (\ref{dsum-comps}) that $\ce^*$ and $\cf^*$ are subalgebras of $(\ce \otimes \cf)^*=\cx^*$. This means that $[\cu(\ca^*), \ce^*]=0$ and $[\cu(\ca^*), \cf^*]=0$, so we have $(A^i \not\xrightarrow{U} E^m) \land (A^i \not\xrightarrow{U} F^n)$. This proves (\ref{dag1}).

The other part of atomicity then follows from time symmetry. More precisely, (\ref{dag2}) then follows from (\ref{eq:timesym}) and (\ref{dag1}) via essentially the manipulations in equation (\ref{atproof}) from Appendix \ref{appendix:unrouted atomicity} we replace the Hilbert spaces with sectorized Hilbert spaces.

\section{Proof of Proposition \ref{prop:smap-timesym}} \label{appendix:smap-timesym}

To show that the causal structure of a general routed unitary supermap is time symmetric, let us start by explaining why it is so for the special case of an unrouted unitary supermap.
The process channel for an unrouted supermap is
\begin{equation}
    \cs^\dag(\texttt{SWAP}, \ldots, \texttt{SWAP}) = \cs(\texttt{SWAP}, \ldots, \texttt{SWAP})^\dag
\end{equation}
since the swap gate is self-adjoint. In other words, the process channel for the adjoint of the supermap is just the adjoint of the process channel for the original supermap.
But we already know, from Proposition \ref{eq:timesymunrouted}, that the causal structure of a unitary \textit{channel} is time symmetric.
It then follows from the definition of the causal structure of a unitary supermap in terms of its process channel that it is also time symmetric.

In summary, the fact that the swap gate is self-adjoint implies that the causal structure of an unrouted unitary supermap is time symmetric. If the exchange gate were self-adjoint, then we would be able to make an essentially identical argument. But it isn't: although \texttt{RSWAP} is self-adjoint, the other gate needed to define the exchange gate, \texttt{SCNOT}, is not (see Appendix \ref{appendix:xdeltadef} for the definition of \texttt{EXCH} in terms of these two unitaries). This apparently raises the possibility of time asymmetry in the routed case.

But in fact, $\texttt{SCNOT}$ has a weaker property than self-adjointness, which is enough to secure time symmetry. Specifically, it is equal to its own adjoint up to a local unitary applied to both its ancillary input and output. Defining the negation unitary $U = \sum_j\ket{-j}\bra{j}$, we have
\begin{equation} \label{localunis}
    \texttt{SCNOT} = ( I_A \otimes U_{A^I})\texttt{SCNOT}^\dag (I_A \otimes U_{A^O}  )
\end{equation}

To prove (\ref{localunis}), we note that $\texttt{SCNOT}^\dag=\sum_{ij} \pi_A^i \otimes \ket{j}\bra{j+i}= \sum_{ij} \pi_A^i \otimes \ket{j-i}\bra{j}$, and calculate the action of the right-hand side to find
\begin{equation}
    \ket{i}\ket{j} \rightarrow \ket{i}\ket{-j} \rightarrow \ket{i}\ket{-j-i} \rightarrow \ket{i}\ket{j+i}
\end{equation}

Now, as (\ref{R1}) makes clear, changing a unitary channel by composing it with another unitary that is both
\begin{enumerate}[a)]
    \item  local (meaning that it that acts only on one of the original unitary's input or output subsystems); and 
    \item sector-preserving
\end{enumerate} 
does not alter its causal relations. The negation unitary $U$ above is trivially sector-preserving since it acts on systems that only have one sector. Therefore, the process channel
\begin{equation}
    \cs(\texttt{EXCH}, \ldots, \texttt{EXCH})
\end{equation}
for a routed unitary supermap $\cs$ has the same (sectorized and unsectorized) causal structure as 
\begin{equation}
    \cs(\texttt{EXCH}^\dag, \ldots, \texttt{EXCH}^\dag).
\end{equation}
But by the time symmetry of the causal structure of routed unitary channels, $\cs(\texttt{EXCH}^\dag, \ldots, \texttt{EXCH}^\dag)$ has the causal structure obtained by reversing the arrows in the causal structure of $\cs(\texttt{EXCH}^\dag, \ldots, \texttt{EXCH}^\dag)^\dag$ -- the process operator for the time-reversed supermap $\cs^\dag$. It follows from the definition of the causal structure of routed unitary supermaps in terms of their process channels that the causal structure of $\cs^\dag$ is obtained from the causal structure of $\cs$ by reversing the direction of the arrows.

\end{document}